\documentclass[twocolumn,twocolappendix]{aastex63}
\usepackage{mathtools}
\usepackage{natbib}
\usepackage{float}
\usepackage{blindtext}
\usepackage{isomath}
\usepackage{hyperref}
\shorttitle{X-ray scaling relations for the ESZ selected sample}
\shortauthors{Lovisari et al.}
\begin{document}

\title{X-ray scaling relations for a representative sample of Planck selected clusters observed with XMM-Newton}

\correspondingauthor{Lorenzo Lovisari}
\email{lorenzo.lovisari@cfa.harvard.edu}

\author[0000-0002-3754-2415]{Lorenzo Lovisari}
\affiliation{Center for Astrophysics $|$ Harvard $\&$ Smithsonian, 60 Garden Street, Cambridge, MA 02138, USA}

\author[0000-0002-4962-0740]{Gerrit Schellenberger}
\affiliation{Center for Astrophysics $|$ Harvard $\&$ Smithsonian, 60 Garden Street, Cambridge, MA 02138, USA}
\author[0000-0003-0302-0325]{Mauro Sereno}
\affiliation{INAF - Osservatorio di Astrofisica e Scienza dello Spazio di Bologna, via Piero Gobetti 93/3, I-40129 Bologna, Italia}
\affiliation{INFN, Sezione di Bologna, viale Berti Pichat 6/2, 40127 Bologna, Italy}
\author[0000-0003-4117-8617]{Stefano Ettori}
\affiliation{INAF - Osservatorio di Astrofisica e Scienza dello Spazio di Bologna, via Piero Gobetti 93/3, I-40129 Bologna, Italia}
\affiliation{INFN, Sezione di Bologna, viale Berti Pichat 6/2, 40127 Bologna, Italy}
\author{Gabriel W. Pratt}
\affiliation{AIM, CEA, CNRS, Universit\'e Paris-Saclay, Universit\'e Paris Diderot, Sorbonne Paris Cit\'e, F-91191 Gif-sur-Yvette, France}
\author[0000-0002-9478-1682]{William R. Forman}
\affiliation{Center for Astrophysics $|$ Harvard $\&$ Smithsonian, 60 Garden Street, Cambridge, MA 02138, USA}
\author{Christine Jones}
\affiliation{Center for Astrophysics $|$ Harvard $\&$ Smithsonian, 60 Garden Street, Cambridge, MA 02138, USA}
\author[0000-0002-8144-9285]{Felipe Andrade-Santos}
\affiliation{Center for Astrophysics $|$ Harvard $\&$ Smithsonian, 60 Garden Street, Cambridge, MA 02138, USA}
\author[0000-0002-3984-4337]{Scott Randall}
\affiliation{Center for Astrophysics $|$ Harvard $\&$ Smithsonian, 60 Garden Street, Cambridge, MA 02138, USA}
\author[0000-0002-0765-0511]{Ralph Kraft}
\affiliation{Center for Astrophysics $|$ Harvard $\&$ Smithsonian, 60 Garden Street, Cambridge, MA 02138, USA}



\begin{abstract}
We report the scaling relations derived by fitting the X-ray parameters determined from analyzing the XMM-Newton observations of 120 galaxy clusters in the Planck Early Sunyaev-Zel'dovich sample spanning the redshift range of 0.059$<$$z$$<$0.546.  We find that the slopes of all the investigated scaling relations significantly deviate from the self-similar predictions, if self-similar redshift evolution is assumed. When the redshift evolution is left free to vary, the derived slopes are more in agreement with the self-similar predictions.
Relaxed clusters have on average $\sim$30$\%$ higher X-ray luminosity than disturbed clusters at a given mass, a difference that, depending on the relative fraction of relaxed and disturbed clusters in the samples (e.g. SZ vs X-ray selected), have a strong impact in the normalization obtained in different studies. Using the core-excised cluster luminosities reduces the scatter and brings into better agreement the $L$-$M_{tot}$ and $L$-$T$ relations determined for different samples. $M_{tot}$-$T$, $M_{tot}$-$Y_X$, and $M_{tot}$-$M_{gas}$ relations show little dependence on the dynamical state of the clusters, but the normalizations of these relations may depend on the mass range investigated.
Although most of the clusters investigated in this work reside at relatively low redshift, the fits prefer values of $\gamma$, the parameter accounting for the redshift evolution, different from the self-similar predictions. This suggests an evolution ($<$2$\sigma$ level, with the exception of the $M_{tot}$-$T$ relation) of the scaling relations. 
For the first time, we find significant evolution ($>$3$\sigma$) of the $M_{tot}$-$T$ relation, pointing to an increase of the kinetic-to-thermal energy ratio with redshift. This is consistent with a scenario in which higher redshift clusters are on average more disturbed than their lower redshift counterparts.
\end{abstract}



\section{Introduction} \label{sec:intro}
X-ray and Sunyaev-Zel'dovich (SZ) surveys are two independent probes of the same physical component in galaxy clusters: the hot gas filling the space between galaxies. However, these surveys have a different dependence on the gas density: the X-ray emission scales with the square of the electron gas density, while the SZ effect scales linearly. Due to that, the SZ experiments detect a larger fraction of disturbed systems than the X-ray surveys which detect more centrally peaked and relaxed galaxy clusters (\citealt{2016MNRAS.457.4515R}, \citealt{2017MNRAS.468.1917R}, \citealt{2017ApJ...843...76A}, \citealt{2017ApJ...846...51L}, \citealt{2019A&A...628A..86B}). This is an important fact, because relaxed and disturbed clusters populate a different region of the residual space with respect to the best-fit $L$-$M_{tot}$ and $L$-$T$ relations, with relaxed (disturbed) objects having, on average, an X-ray luminosity higher (lower) than the mean  (e.g. \citealt{2009A&A...498..361P}; for example, see Figs. 2 and 4 right panels). This offset is probably associated with the strength of cool-cores that boost the cluster X-ray luminosity. Mergers also likely contribute to the scatter, because the total masses can easily be incorrectly estimated when the clusters are not in Hydrostatic Equilibrium (HE), as happens during cluster mergers.
Thus, a different sampling of the galaxy cluster population leads to observed relations that differ both in slope and normalization for the different samples.  Moreover,  different trends in X-ray luminosity are shown to be correlated with other X-ray observables, e.g. temperature, inducing significant covariance between cluster properties (e.g. \citealt{2016MNRAS.463.3582M}, \citealt{2019arXiv190610455S}, \citealt{2019NatCo..10.2504F}, \citealt{ser+19_hscxxl}). Therefore, the comparison between studies with different cluster selection is very challenging. We also note that the fraction of relaxed and disturbed systems may evolve with redshift which further complicates the comparison between local and distant samples, if the relative fraction of relaxed and disturbed systems in the sample is unknown.

Since SZ surveys are thought to be very close to being mass-selected, and as such, unbiased, the different fraction of relaxed and disturbed systems in X-ray and SZ surveys also raises concerns about the representativeness of the X-ray selected samples which are often used to define our current understanding of cluster physics and as calibration samples for numerical simulations or cosmological studies.

The Planck Early Sunyaev-Zeldovich (ESZ, \citealt{2011A&A...536A...8P}) cluster catalog is a good reference set for characterizing mass-selected cluster samples, for studies of structure formation including comparison with theory and simulations, as well as for cosmological tests. With the exception of one candidate, all the ESZ clusters have been independently confirmed (e.g. \citealt{2011A&A...536A...8P}, \citealt{2014A&A...571A..29P}).  Most crucially, compared to ground-based SZ surveys, which have observed only a few thousand square degrees, Planck's all-sky ($|$b$|$$>$15$^{\circ}$) cluster survey provides a large statistical sample spanning a broad mass range, including the rare, very massive clusters.

In \cite{2017ApJ...846...51L} several morphological parameters were derived to investigate the difference between the dynamical state of the clusters in SZ and X-ray surveys. The comparison between the Planck Early Sunyaev-Zeldovich (ESZ, \citealt{2011A&A...536A...8P}) selected clusters with the REXCESS sample (\citealt{2007A&A...469..363B}), an X-ray selected cluster sample, indicated that the Planck clusters are, on average, less relaxed and have a lower fraction of cool core systems. This result confirmed the prediction by numerical simulations (e.g. \citealt{2005ApJ...623L..63M}) and previous findings by \cite{2016MNRAS.457.4515R,2017MNRAS.468.1917R} and  \cite{2017ApJ...843...76A}, and likely reflects the tendency of X-ray surveys to preferentially detect clusters with a centrally-peaked morphology, which are more X-ray luminous at a given mass, and on average more relaxed.

In the near future, eROSITA (\citealt{2012arXiv1209.3114M}) will provide catalogs with a large number of galaxy groups and clusters which, for the reasons outlined above, may not be representative of the whole cluster population. Thus, to fully exploit the entire eROSITA sample of detected clusters to constrain cosmological parameters, we need to take into account the different selection effects, including morphology, and Malmquist or Eddington biases.  While several methods (see e.g. \citealt{2006MNRAS.372..578P}, \citealt{2009ApJ...692.1033V}, \citealt{2009A&A...498..361P}, \citealt{2010MNRAS.406.1773M}, \citealt{2015A&A...573A.118L}, \citealt{2017MNRAS.469.3738S}) have been proposed to account for Malmquist and Eddington biases, little has been done to incorporate and account for the morphology bias which is expected to be important for eROSITA (and for X-ray survey data in general). In fact, most of the clusters detected with eROSITA will have too few photons to derive gas density and temperature profiles (and thus too few to determine mass profiles). Therefore the total masses will be mostly estimated using scaling relations (e.g. using the $L$-$M_{tot}$ relation). Hence, cosmological studies will rely on  the solid understanding of the scaling properties, for both relaxed and disturbed  clusters. In addition, knowing the selection function may not be sufficient, if the scaling relations are determined using a sample which is not representative of the whole cluster population.

In this paper we derive the X-ray scaling relations for a representative sample of Planck selected clusters and we compare them with  ones derived using X-ray selected samples and for a sample of clusters detected with the South Pole Telescope (SPT,  \citealt{2011PASP..123..568C}). Moreover, we investigate the scaling properties for subsamples of relaxed and disturbed clusters to highlight the impact on the relations of a different fraction of regular and dynamically active systems.

This paper is structured as follows. In Sect. 2 we describe the sample and the XMM-Newton data reduction. The determination of the X-ray properties  and the methodology for quantifying the scaling relations is described in Sect.  3. We present our results in Sect. 4 and discuss them in Sect. 5. Section 6 contains summary and conclusions. 

Throughout this paper, we assume a flat $\Lambda$CDM cosmology with $\Omega_m$=0.3 and $H_0$=70 km s$^{-1}$ Mpc$^{-1}$. Log is always base 10 here. Uncertainties are at the 68$\% $ c.l. Several studies determined the $L$-$M_{tot}$ and $L$-$T$ relations using the luminosities in the 0.5-2 keV band instead of the 0.1-2.4 keV band used in this work. To compare these relations with our results, we corrected the normalizations, assuming a scaling factor of 1.62 obtained assuming an unabsorbed APEC model in XSPEC \citep{1996ASPC..101...17A} for a cluster temperature of 5 keV, abundance of 0.3 solar, and a redshift of 0.2. The scaling factor only changes by a few percent by varying these input parameters. 

\section{Data}
Our sample contains 120 galaxy clusters  observed with XMM-Newton and originally selected from the Planck Early Sunyaev-Zeldovich (ESZ, \citealt{2011A&A...536A...8P}). As described in \cite{2017ApJ...846...51L}, these are the ESZ clusters for which $R_{500}$ was completely covered by XMM-Newton observations, allowing the estimation of the morphological parameters. $R_{500}$ was estimated using iteratively the $M_{500}$-Y$_X$  relation given in \cite{2010A&A...517A..92A}. As shown in \cite{2017ApJ...846...51L}, the mass and redshift distributions of these systems are representative of the whole ESZ sample of 188 galaxy clusters. 

Observation data files (ODFs) were downloaded from the XMM-Newton archive and processed with the XMMSAS (\citealt{2004ASPC..314..759G}) v16.0.0 software for data reduction. The initial data processing, to generate calibrated event files from raw data, was done by running the tasks {\it emchain} and {\it epchain}. We only considered single, double, triple, and quadruple events for MOS (i.e. PATTERN$\le$12) and single events for pn (i.e.  PATTERN==0) and we applied the standard procedures for bright pixels and hot columns removal (i.e. FLAG==0), and pn out-of-time correction.
All the data sets were cleaned for periods of high background due to the soft protons, following the two-step procedure extensively described in \citet{2011A&A...528A..60L}.
The point-like sources were detected with the {\it edetect-chain} task and visually inspected before excluding the regions with point sources from the event files.
The background event files were cleaned by applying the same PATTERN selection, flare rejection, and point-source removal as for the corresponding target observations.

\section{X-ray quantities}
\subsection{Luminosities}
The X-ray luminosities have been derived within two different apertures: 0-1$R_{500}$ and 0.15-1$R_{500}$ (referred to as core-excised luminosity). The total count rates within the different apertures, have been derived  by integrating the surface brightness (SB) derived in the 0.3-2 keV band and then converted into the 0.1-2.4 keV band (hereafter $L_X$) and bolometric (i.e. 0.01-100 keV band, hereafter $L_{bol}$) luminosities with XSPEC, using the best fit spectral model estimated in the same aperture. Uncertainties take into account both the statistical factors and the uncertainties in the derivation of $R_{500}$.  The relative errors  were estimated via Monte Carlo realizations by randomly varying the  observational data points of the SB profiles to determine a new best fit. The randomization was performed assuming a Gaussian distribution with mean and standard deviation equal to the observed uncertainties.

\subsection{Temperatures}\label{kTsubsect}
Spectroscopic temperatures have been obtained by fitting the spectra with an APEC thermal plasma model with an absorption fixed at the total (neutral and molecular, see \citealt{2013MNRAS.431..394W}) $N_H$ value estimated using the SWIFT online tool\footnote{http://www.swift.ac.uk/analysis/nhtot/index.php}, with the exception of a few clusters (marked with a star in Table \hyperlink{foo}{A1}), which were found to have a significantly different absorption than the value indicated from the tool.
All MOS and pn spectra were fitted simultaneously in the full (i.e. 0.3-10 keV) energy band, with temperature and abundance linked, while the normalizations were left free to vary to account for the different cross-calibration between the detectors (see e.g. \citealt{2015A&A...575A..30S}). The modeling of the background is described in \cite{2019MNRAS.483..540L}. 

The radial temperature profiles have been derived by extracting the spectra from successive annular regions created around the X-ray peak. We required a minimum width of 30$^{\prime\prime}$ to ensure that the redistribution fraction of the flux is at most about 20$\%$ (\citealt{2009ApJ...699.1178Z}) and a S/N$\ge$50 to ensure an  uncertainty of $\sim$10$\%$ in the spectrally resolved temperature (and consequently in the fitted temperature profiles). We also required the source-to-background count rate ratio to be higher than 0.6 to reduce the systematic uncertainties in the temperature measurements (see \citealt{2008A&A...486..359L} for more details). The number of obtained bins per cluster is listed in Table \hyperlink{foo}{A1}. The profiles are then deprojected, using the method presented in \cite{2006ApJ...640..710V}. The implementation has been done following \cite{2017MNRAS.469.3738S}, where  the parameters of the deprojected temperature profile are determined using a Markov Chain Monte Carlo.  The input temperature profiles are projected along the line of sight at every radius that has a measurement and compared to the observed profile, until convergence.
Then, the global and core-excised temperatures, used in the scaling relations, were determined by integrating the deprojected profiles along the line of sight (starting from the center and 0.15R$_{500}$, respectively) weighted by the emission measure, and accounting for the detector response. We verified that these temperatures are in good agreement with the temperatures derived using a single spectral extraction: a linear fit gives a slope of 0.97$\pm$0.03 and an intrinsic scatter of $\sim$6$\%$. It should be noted that not only uncertainties in the 3D temperature profile parameters, but all MCMC chains are used for the following steps (e.g., calculating the total mass), to assure the covariance of parameters is taken into account.

\subsection{Masses}
\begin{figure*}[t!]
\figurenum{1}
\hbox{
\centering
\includegraphics[width=1\textwidth]{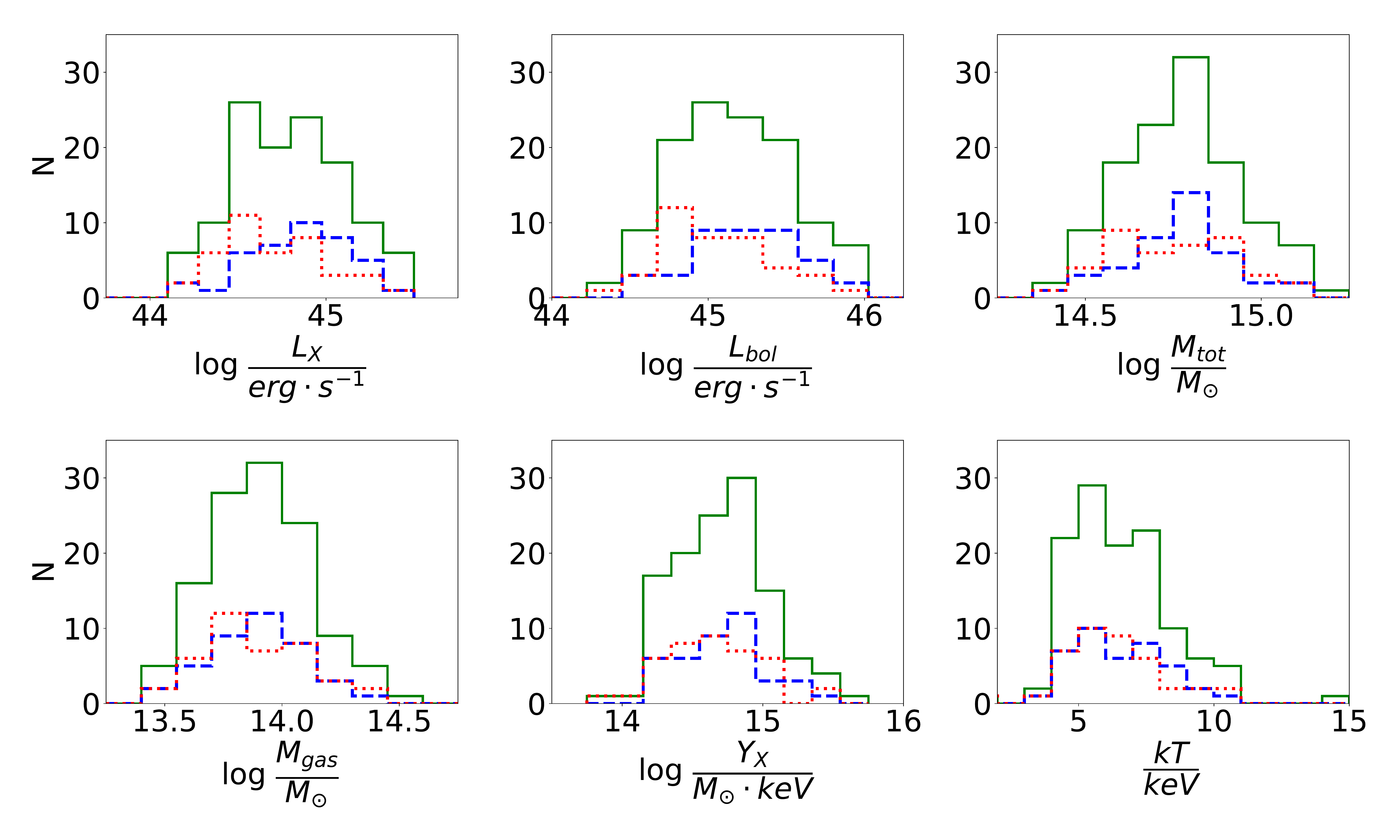}
}
\caption{We show the distribution of soft band luminosity (0.1-2.4 keV) $L_{X}$ ({\it top-left}), bolometric luminosity (0.01-100 keV) $L_{bol}$ ({\it top-center}), total mass $M_{tot}$ ({\it top-right}), gas mass $M_{gas}$ ({\it bottom-left}), $Y_X$=$M_{gas}$$\times kT$ ({\it bottom-center}), and temperature $kT$ ({\it bottom-right}), of the ESZ sample of  galaxy clusters observed with XMM-Newton. The distribution of the full sample is shown in green, while the distribution of the most relaxed (1/3 of the total) and most disturbed (1/3 of the total) clusters is shown in blue and red, respectively.}
\label{fig:parhisto}
\end{figure*}

The total cluster masses can be obtained by solving the HE equation. Assuming spherical symmetry, the total cluster mass $M$ within a radius $r$ is given by
\begin{equation}\label{eq:HE}
M(<r)=-\frac{r k_BT}{G\mu m_p} \left\{ \frac{d\ln{\rho}}{d\ln{r}} + \frac{d\ln{T}}{d\ln{r}}  \right\}
\end{equation}
where $k_B$ and $G$ are the Boltzmann and gravitational constants, and $\mu$ is the mean particle weight in units of the proton mass $m_p$. The observational inputs needed for this calculation are the density profiles obtained in \cite{2017ApJ...846...51L} and the temperature profiles obtained as discussed in Sect. \ref{kTsubsect}.
We solved Eq. \ref{eq:HE} for the radii of the spectral extraction regions (temperature measurements) and fited an NFW model up to the outer regions  (\citealt{1997ApJ...490..493N}) for the mass profile with the relation from \cite{2013ApJ...766...32B} between the Dark Matter concentration $c_{500}$ and $R_{500}$ as a prior, which constrains the posterior distribution of $R_{500}$ to reasonable values (see \citealt{2017MNRAS.469.3738S} for more details). The mass of the gas is then calculated by integrating the density profile within $R_{500}$ estimated from Eq. 1, and, together with the cluster temperature, it is used to determine the $Y_X$(=$M_{gas}\times kT$) and  $Y_{X,exc}$(=$M_{gas}\times kT_{exc}$) parameters. The derived properties for individual clusters are listed in Table \hyperlink{foo}{A1}, and their distribution is shown in Fig. \ref{fig:parhisto}.

\section{Fitting the scaling relations}
We investigated the following relations: $L$-$M_{tot}$, $L$-$T$, $M_{gas}$-$T$, $M_{tot}$-$T$, $M_{tot}$-$Y_X$, and $M_{tot}$-$M_{gas}$. We fitted the relations using both the soft band (0.1-2.4 keV) and bolometric luminosities (0.01-100 keV), as well as  both the global or core-excised properties, when appropriate. The full uncertainty covariance  matrix between the X-ray properties was computed and used  for  the analysis of the scaling relations.
For each set of parameters (X,Y), we linearly fit our data  as:
\begin{gather}\label{eq:lira}
\log{\left(\frac{Y}{C1}\right)}=\alpha+\beta\log{\left(\frac{Z}{C2}\right)}+\gamma\log{{F_z}}\pm\sigma_{Y|Z}\\
\centering{\log{X}=\log{Z}\pm\sigma_{X|Z}}
\end{gather}
where $Z$ is the intrinsic cluster property (i.e. the `true' quantity), $\alpha$ denotes the normalization, $\beta$ the slope, $\gamma$ the evolution with redshift, $\sigma$ the intrinsic scatter\footnote{
The intrinsic scatter manifests as a data distribution around a relation. Therefore, the smaller the intrinsic scatter value is, the closer the data distribution is to strict linearity. In LIRA, the intrinsic scatter refers to the probability of the  variable of interest (X when $\sigma_{X|Z}$ is considered, or Y in the case of $\sigma_{Y|Z}$) given the latent property Z (e.g. the true cluster mass).} in the two variables X and Y, and $F_z=E_z/E_{z,ref}(z_{ref}=0.2)$. $E_z=H_z/H_0=[\Omega_m(1+z)^3+\Omega_{\Lambda}]^{0.5}$ indicates the dependence on the evolution of the Hubble constant at redshift $z$.
The pivot points, C1 and C2, have been chosen to be roughly the median values of the sample and they are summarized in Table \ref{tab:pivot}. For each relation, we also provide the predicted slope $\beta_{self}$ and redshift evolution $\gamma_{self}$ in the case where gravity is the dominant process, a scenario which is referred as self-similar model (see e.g \citealt{2012MNRAS.421.1583M}).

\begin{table}[t!]
\centering
\caption{Self-similar values and pivot points used in the scaling relations in the form of Y$\propto$ E$_z^{\gamma}$ X$^{\beta}$. In the second and third columns we provide the predictions from the self-similar scenario for the redshift evolution $\gamma_{\rm self}$ and scaling relation slope $\beta_{\rm self}$, respectively. C1 and C2 values are the pivot points used in Eq. \ref{eq:lira}.}
\label{tab:pivot}
\begin{tabular}{l c c l l }
\hline\hline
relation (Y,X) & $\gamma_{\rm self}$ & $\beta_{\rm self}$ & C1 & C2  \\
\hline
$L_X$-$M_{tot}$ & 2 & 1 & 5$\cdot10^{44}$ erg/s & 6$\cdot10^{14}M_{\odot}$ \\            
$L_X$-$T$ & 1 & 3/2 & 5$\cdot10^{44}$ erg/s & 5 keV \\                  
$L_{bol}$-$M_{tot}$ & 7/3 & 4/3 & 1$\cdot10^{45}$ erg/s & 6$\cdot10^{14}M_{\odot}$ \\            
$L_{bol}$-$T$ & 1 & 2 & 1$\cdot10^{45}$ erg/s & 5 keV \\          
$M_{tot}$-$T$ & -1 & 3/2 & 6$\cdot10^{14}M_{\odot}$ & 5 keV \\    	      
$M_{tot}$-$Y_X$ & -2/5 & 3/5 & 6$\cdot10^{14}M_{\odot}$ & 5$\cdot10^{14}M_{\odot}$ keV \\ 
$M_{tot}$-$M_{gas}$ & 0 & 1 & 6$\cdot10^{14}M_{\odot}$ & $10^{14}M_{\odot}$ \\  
\hline 
\end{tabular}
\end{table}

We fit the data with a linear relation of the variables in log space using the R-package LIRA\footnote{\label{lira}LIRA (LInear Regression in Astronomy) is available from the R archive network at \url{https://cran.r-project.org/web/packages/lira/index.html}}. LIRA is based on a Bayesian method that can deal with heterogeneous data and correlated errors, and allow normalizations, slopes and scatters (and relative uncertainties) to be fitted simultaneously (see \citealt{2016MNRAS.455.2149S} for more details). As a default, all the important parameters are left free to vary, and central values and uncertainties are summarized in Table \ref{table:bestfit} (and a visual recap of all $\beta$, $\gamma$, and $\sigma$ is presented in Sect. 6, see Fig. \ref{fig:slopes} and \ref{fig:scatter}). The impact of freezing some of the parameters is discussed in  Appendix \ref{sect:fitcomp} where, for comparison, we also provide  the central values obtained using the routine {\it LINMIX} by \cite{2007ApJ...665.1489K} and the best-fit values from {\it BCES} by \cite{1996ApJ...470..706A}, and we discuss the resulting differences. To allow the reader to reproduce our results, in Appendix \hyperlink{foo3}{C}, we provide the LIRA commands to be used in the different cases investigated in this paper.

In Table \ref{table:bestfit}, we also provide an estimate of the goodness of the fit, computed as:
\begin{equation}\label{eq:chi2like}
C_\text{gof}=\frac{1}N_\text{cl}{}\sum_i^{N_\text{cl}} \frac{(y_i-\alpha-\beta x_i-\gamma \log F_z)^2}{\delta_{y,i}^2+\sigma_{Y|Z}^2+\beta^2(\delta_{x,i}^2+\sigma_{X|Z}^2)-2\beta \delta_{xy,i}}
\end{equation}
where $\delta_x$ and $\delta_y$ denote the statistical uncertainties and $\delta_{xy}$ is the uncertainty covariance. This term gives an idea of the goodness-of-fit but it does not follow a (reduced) $\chi^2$ statistic. In fact, the intrinsic scatters $\sigma_{X|Z}$ and $\sigma_{Y|Z}$ are estimated in the regression procedure and are not known a priori. Furthermore, $C_\text{gof}$ is computed for the mean values of the parameter posterior distribution and not for the maximum likelihood parameters.

\section{Results}
To investigate the impact of the cluster dynamical state on the scaling relations, we used the morphological information (i.e. centroid-shift and concentration parameter) from \citet{2017ApJ...846...51L} to select the most relaxed ``R" (1/3 of the total) and most disturbed ``D" (1/3 of the total) clusters in the ESZ sample. Their distribution is also shown in  Fig. \ref{fig:parhisto}. In the following we discuss the individual scaling relation results for the full ESZ sample and for the subsamples of relaxed and disturbed clusters. 

\subsection{L$_X$-M$_{tot}$}\label{sect:LM}
\begin{figure}
\figurenum{2}
\centering
\includegraphics[width=0.5\textwidth]{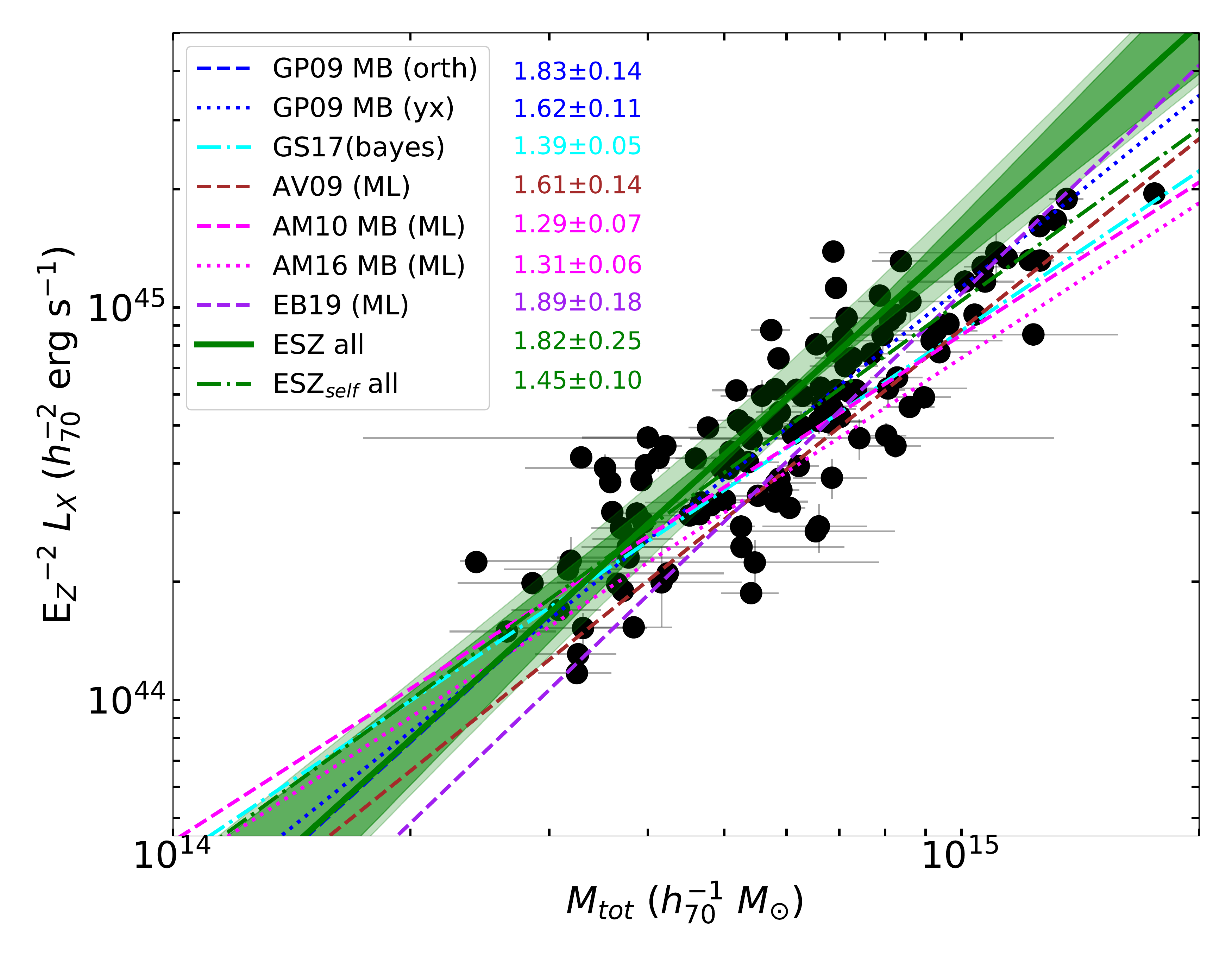}
\caption{Comparison between the $L_X$-$M_{tot}$ relation determined in this work (green lines) with the bias-corrected relations obtained with well-studied samples selected in both X-ray and SZ. $L_X$ are the soft band (0.1-2.4 keV) luminosities. The full green line represents the fitted scaling relation with the time evolution free to vary in the fit. The dark and light shaded areas represent the 1$\sigma$ statistical error and scatter, respectively. The dash-dotted green line represents the fitted relation assuming self-similar redshift evolution.  The slopes of the relations derived from the X-ray selected samples, with the exception of GP09, are flatter than those from relations derived using SZ derived samples (i.e. ESZ and EB19). For the acronyms in the legend see Sect. \ref{sect:LM}.}
\label{fig:LM}
\end{figure}

In Fig. \ref{fig:LM} we show the results for the $L_X$-$M_{tot}$ relation. The relation is corrected for the Eddington bias (see \citealt{2016MNRAS.455.2149S} for more details), but not for the Malmquist bias, which is negligible\footnote{In a forthcoming paper by Andrade-Santos et al. we will show that, indeed, the Malmquist bias is not important for  the  present SZ selected  samples.} when fitting the X-ray properties of an SZ selected sample. 
The fitted relation is shown with a solid green line, while the dark green shaded area encloses the 1$\sigma$ confidence region around the median scaling relation. We also show the bias-corrected relations derived for well known X-ray selected samples: REXCESS (\citealt{2009A&A...498..361P}, GP09), HIFLUGCS (\citealt{2017MNRAS.469.3738S}, GS17), 400d (\citealt{2009ApJ...692.1033V}, AV09), and the flux-limited samples of massive clusters by \citet[AM10]{2010MNRAS.406.1773M} and \citet[AM16]{2016MNRAS.463.3582M}. Moreover, we show the recent result from \citet[EB19]{2019ApJ...871...50B} who investigated the X-ray properties of a sample of SPT clusters spanning the redshift range from $z$=0.2 to $z$=1.5.

We find a relation steeper than the prediction of the self-similar scenario (i.e. $\beta$$>$1), which is probably the result of the combined effect of gas cooling, AGN feedback, and subcluster mergers. We also find mild evidence for negative redshift evolution (i.e. $\gamma$$<$1) in agreement with \cite{2015MNRAS.450.3675S}. This might be a sign of additional radiative cooling and uniform (pre-)heating at high redshift. 

There is good agreement between the slopes determined using Planck and SPT selected samples, despite the different distributions of cluster masses and redshifts (i.e. the SPT selected clusters used by EB19 are on average at higher redshift and lower mass than the ones used in our work). However, there is a normalization offset on the order of $\sim$45$\%$ at $M_{tot}$=6$\times$10$^{14}$M$_{\odot}$ and $z$=0.2 between the two relations, which reduces to $\sim$23$\%$ when self-similar redshift evolution (i.e. $\gamma$=2) is assumed.  A large offset in the scaling relations is also observed when comparing our results with those derived from the X-ray selected samples, with the exception of the REXCESS sample, which instead agrees extremely well with our results. However, as noted by AM16, a straightforward comparison between the different studies is difficult because the total masses have been derived using different methods (e.g. AV09 and GS17 used the HE equation while GP09 and AM10 used, respectively, $Y_X$ and $M_{gas}$ as a proxy). Nonetheless, we note that the flux-limited samples (GS17, AM10, and AM16) show flatter relations than all the other samples. The X-ray samples used by GP09 and AV09, which have properties of both a flux- and a volume-limited sample, have slopes for the derived scaling relations in better agreement with the results obtained from SZ selected clusters.  

\begin{figure*}[t!]
\figurenum{3}
\hbox{
\centering
\includegraphics[width=0.5\textwidth]{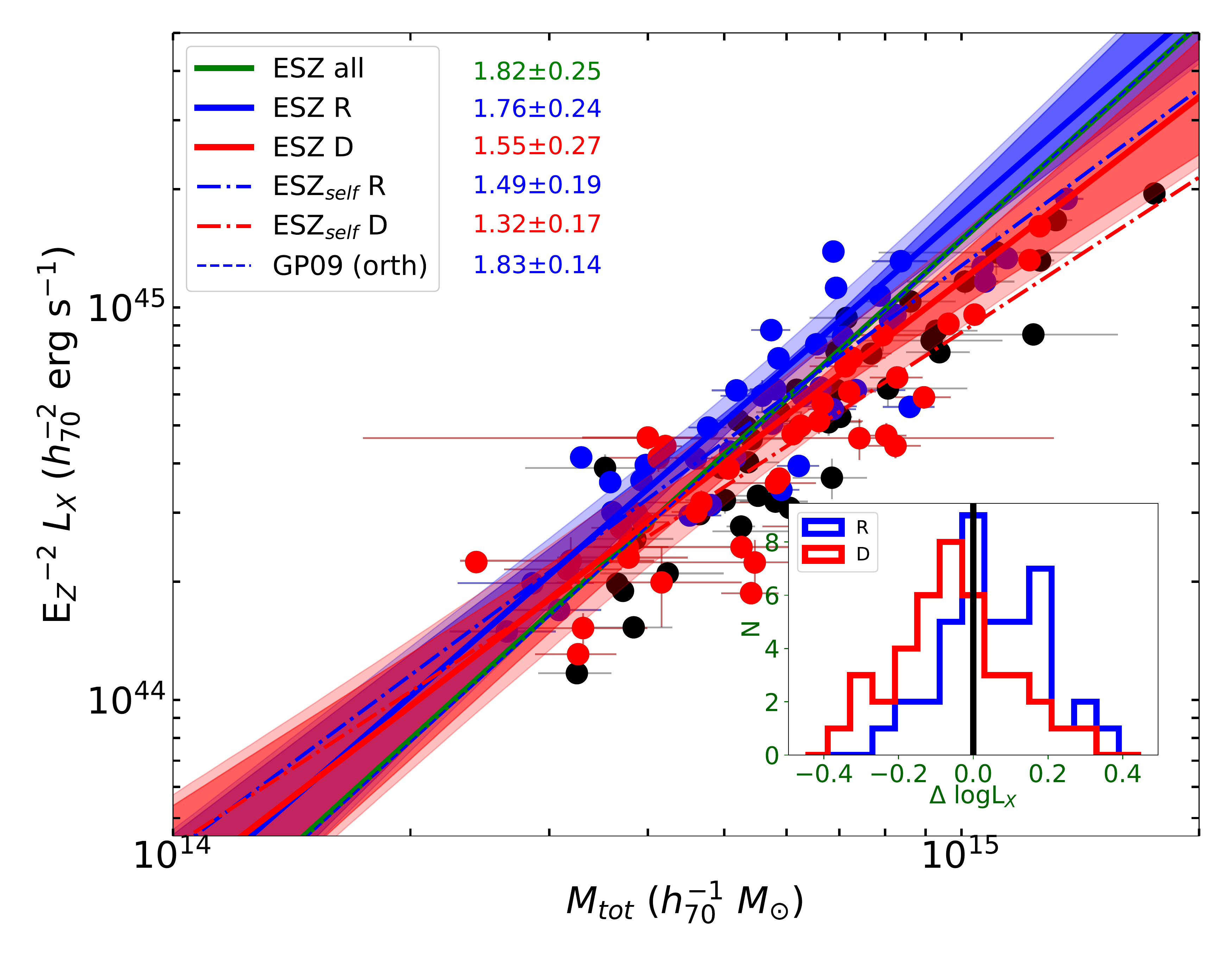}
\includegraphics[width=0.5\textwidth]{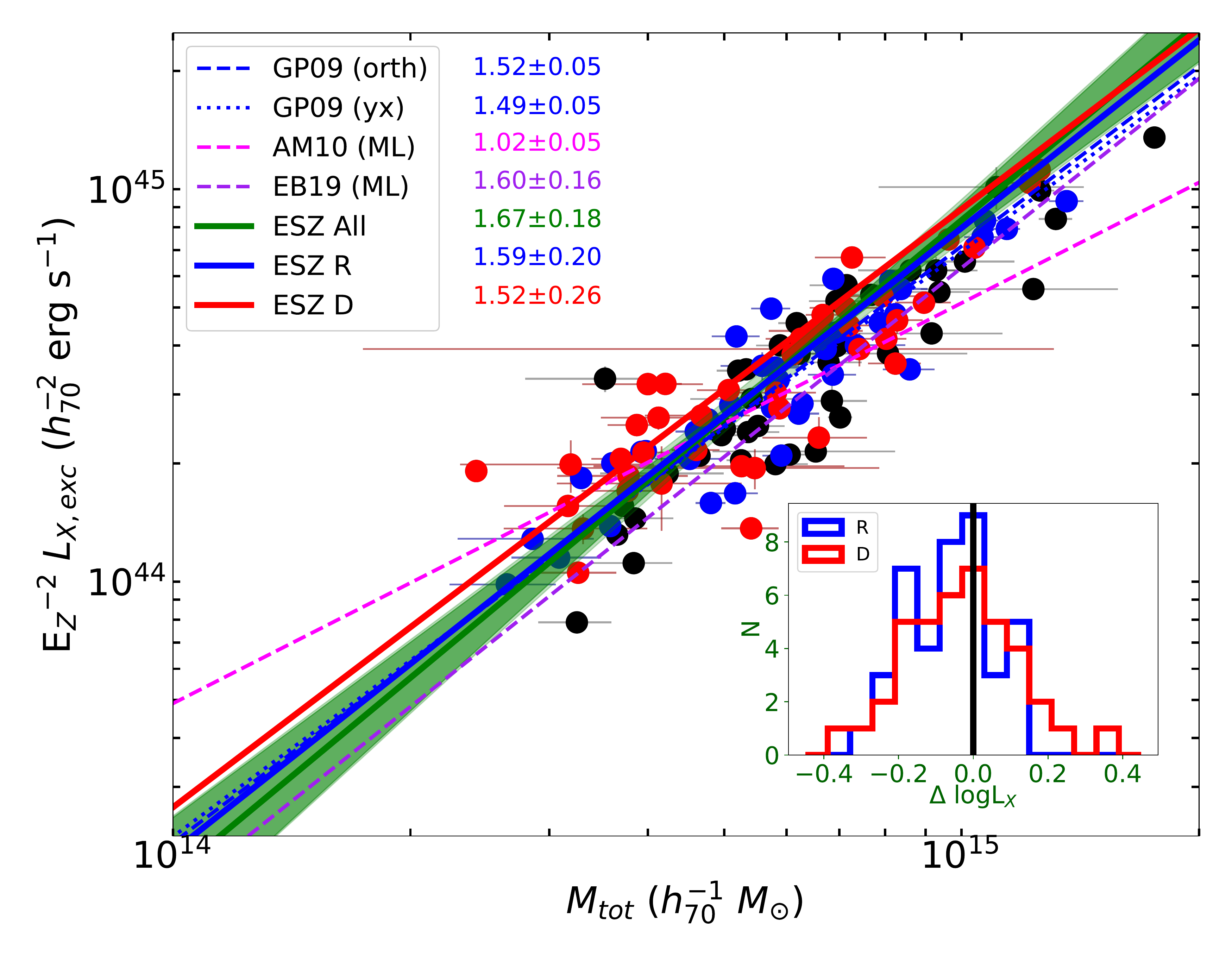}
}
\caption{{\it left:} $L_X$-$M_{tot}$ relations determined for relaxed (blue) and disturbed (red) systems independently. Clusters that are not classified in either of the two subsamples are shown in black.  The green line shows the fitted relation for all the clusters. The relations obtained with the redshift evolution frozen to the self-similar value, $\gamma$=2, are indicated with a {\it self} subscript.
In the inset plot we show the histogram of the log space residuals from the fitted $L_X$-$M_{tot}$ relation, derived with $\gamma$ free to vary. Relaxed clusters are, on average, above the relation, while disturbed clusters are, on average, below. {\it right:} comparison between the $L_{X,exc}$-$M_{tot}$ relation determined in this work with some of the relations available in the literature. Using the core-excised luminosities brings into better agreement the relations of relaxed and disturbed clusters. The histogram of the log space residuals in the inset plot shows that relaxed and disturbed systems are distributed around the fitted relation when core-excised luminosities are used. $L_X$ are soft band (0.1-2.4 keV) luminosities.  For the acronyms in the legend see Sect. \ref{sect:LM}.}
\label{fig:LMmor}
\end{figure*}

If we force the redshift evolution to be self-similar (i.e $\gamma$=2), similar to what done by, e.g., GS17, we find a flatter relation for the ESZ sample, more in agreement with GS17 results.  As discussed in Appendix  \ref{sect:fitcomp}, fixing the redshift evolution impacts our relations by changing both slope and normalization.

In Fig. \ref{fig:LMmor} ({\it top panel}), we show the distribution of the relaxed (in blue) and disturbed (in red) clusters, with respect to the fitted relation for the full sample (green line), along with the relations found in earlier studies.  We found that relaxed clusters have, on average, higher soft band (0.1-2.4 keV) luminosities $L_X$ than disturbed systems, confirming the finding by GP09.  Thus, when the relaxed and disturbed subsamples are fitted independently, we find that they have similar slopes (only slightly flatter for the disturbed clusters), but different intrinsic scatter and normalizations. The intrinsic scatter is only $\sim$16$\%$ for the relaxed systems, while it is significantly larger, $\sim$26$\%$, for the disturbed clusters. The relatively low scatter observed for the relaxed clusters is probably due to the fact that the dominant contribution to the scatter for these systems is the presence of a dense core which scatters $L_X$ always in the same direction (i.e. boost of $L_X$). On the contrary, in disturbed clusters there are many processes (e.g. non-thermal pressure, substructures and clumps, shocks and temperature inhomogeneities) playing a role, each of them acting in different directions and with a different magnitude. 

At $M_{tot}$=6$\times10^{14}M_{\odot}$ and $z$=0.2 the normalization  of the relation for the most relaxed clusters is $\sim$20$\%$ higher than the relation fitted if we included all the objects.  The normalization  of the relation for the most disturbed clusters is, instead, $\sim$10$\%$ lower than the relation fitted including all the objects.  This implies that, for a given total cluster mass the X-ray soft band luminosity of disturbed galaxy clusters is on average $\sim$30$\%$ lower than the luminosity of relaxed clusters. That means that if we do not take into account the dynamical state information for the X-ray selected samples, which are biased toward relaxed systems, we are not able to properly correct for all the selection biases. Indeed, this poses an issue for the eROSITA studies, because there will be too few counts to determine their cluster morphology or to derive accurate core-excised X-ray properties.  Assuming that SZ selected samples better represent the true cluster population, one can use them to correct the scaling relations derived with X-ray selected samples
and/or calibrate different mass proxies.

\begin{table*}
\caption{Fitted relations for the ESZ sample. The subsample of relaxed (disturbed) clusters have a surface brightness concentration higher (lower) than 0.18 and a centroid-shift lower (higher) than 0.0137. The morphological parameter values have been taken from \cite{2017ApJ...846...51L}. The definition of $C_{gof}$ is given in Eq. \ref{eq:chi2like}.}
\label{table:bestfit}
\centering
\begin{tabular}{c c c c c c c c}
\hline\hline
Relation (Y-X) & subsample & $\alpha$  & $\beta$ & $\gamma$ & $\sigma_{X|Z}$ & $\sigma_{Y|Z}$ & $C_{gof}$\\
\hline
$L_X$-$M_{tot}$  & all & 0.089$\pm$0.015 & 1.822$\pm$0.246 & 0.462$\pm$0.916 & 0.061$\pm$0.021 & 0.082$\pm$0.040 & 1.09 \\
	& relaxed & 0.162$\pm$0.022 & 1.756$\pm$0.243 & 0.323$\pm$1.058	& 0.061$\pm$0.022 & 0.066$\pm$0.040 & 1.04 \\
	& disturbed & 0.050$\pm$0.029 & 1.551$\pm$0.272 & 0.571$\pm$1.053	 & 0.049$\pm$0.024 & 0.102$\pm$0.041 & 1.01\\
\hline
$L_{X,exc}$-$M_{tot}$ & all & -0.091$\pm$0.013 & 1.668$\pm$0.183 & 1.325$\pm$0.804  & 0.063$\pm$0.015 & 0.034$\pm$0.029 & 1.03\\
	& relaxed & -0.092$\pm$0.018 & 1.589$\pm$0.196 & 1.357$\pm$1.154 & 0.055$\pm$0.016 & 0.038$\pm$0.026 & 1.06 \\
	& disturbed & -0.056$\pm$0.027 & 1.524$\pm$0.264 & 0.738$\pm$1.051 & 0.053$\pm$0.023 & 0.080$\pm$0.038 & 1.03 \\  
\hline	
$L_{bol}$-$M_{tot}$  & all & 0.174$\pm$0.016 & 2.079$\pm$0.230 & 0.541$\pm$0.904 & 0.050$\pm$0.020 & 0.098$\pm$0.042 & 1.10 \\
	& relaxed & 0.254$\pm$0.023 & 2.085$\pm$0.235 & 0.337$\pm$1.064	& 0.051$\pm$0.019 & 0.074$\pm$0.043 & 1.04 \\
	& disturbed & 0.134$\pm$0.031 & 1.868$\pm$0.265 & 0.375$\pm$1.025	 & 0.046$\pm$0.022 & 0.106$\pm$0.046 & 1.01 \\
\hline	
$L_{bol,exc}$-$M_{tot}$  & all & -0.006$\pm$0.014 & 1.921$\pm$0.189 & 1.561$\pm$0.848 & 0.052$\pm$0.016 & 0.050$\pm$0.034 & 1.07 \\
	& relaxed & 0.003$\pm$0.020 & 1.962$\pm$0.202 & 1.222$\pm$1.174	& 0.048$\pm$0.015 & 0.039$\pm$0.030 & 1.02 \\
	& disturbed & 0.027$\pm$0.028 & 1.787$\pm$0.264 & 0.814$\pm$1.097	 & 0.046$\pm$0.022 & 0.091$\pm$0.041 & 1.03 \\	
\hline
$L_X$-$T$  & all & -0.250$\pm$0.045 & 3.110$\pm$0.422 & 0.398$\pm$0.939  & 0.051$\pm$0.010 & 0.052$\pm$0.041 & 1.08  \\
	& relaxed & -0.133$\pm$0.045 & 2.703$\pm$0.380 & 0.114$\pm$1.044 & 0.040$\pm$0.013 & 0.079$\pm$0.041 & 1.02 \\
	& disturbed & -0.257$\pm$0.043 & 2.593$\pm$0.418 & 1.127$\pm$1.067 & 0.036$\pm$0.013 & 0.071$\pm$0.037 & 1.38 \\  
\hline
$L_{X,exc}$-$T_{exc}$  & all & -0.360$\pm$0.031 & 2.409$\pm$0.292 & 1.170$\pm$0.822  & 0.038$\pm$0.011 & 0.052$\pm$0.031 & 1.13 \\
	& relaxed & -0.365$\pm$0.026 & 2.350$\pm$0.194 & 1.166$\pm$0.838 & 0.026$\pm$0.007 & 0.030$\pm$0.018 & 1.13 \\
	& disturbed & -0.341$\pm$0.041 & 2.525$\pm$0.418 & 1.213$\pm$1.084 & 0.035$\pm$0.013 & 0.070$\pm$0.035 & 1.39 \\  
\hline
$L_{bol}$-$T$  & all & -0.209$\pm$0.044 & 3.464$\pm$0.400 & 0.661$\pm$0.964  & 0.044$\pm$0.010 & 0.061$\pm$0.045 & 1.10 \\
	& relaxed & -0.104$\pm$0.047 & 3.250$\pm$0.383 & 0.068$\pm$1.088 & 0.034$\pm$0.011 & 0.082$\pm$0.044 & 1.00 \\
	& disturbed & -0.241$\pm$0.042 & 3.134$\pm$0.386 & 1.042$\pm$1.043 & 0.032$\pm$0.011 & 0.066$\pm$0.039 & 1.35 \\  
\hline
$L_{bol,exc}$-$T_{exc}$  & all & -0.324$\pm$0.027 & 2.808$\pm$0.247 & 1.496$\pm$0.757  & 0.030$\pm$0.009 & 0.054$\pm$0.030 & 1.12 \\
	& relaxed & -0.341$\pm$0.027 & 2.924$\pm$0.198 & 0.920$\pm$0.839 & 0.021$\pm$0.006 & 0.030$\pm$0.019  & 1.04 \\
	& disturbed & -0.315$\pm$0.039 & 3.022$\pm$0.387 & 1.423$\pm$1.094 & 0.031$\pm$0.010 & 0.056$\pm$0.034 & 1.37 \\  	
\hline
$M_{tot}$-$T$  & all & -0.171$\pm$0.015 & 1.556$\pm$0.137 & 0.179$\pm$0.379  & 0.032$\pm$0.010 & 0.036$\pm$0.016 & 1.03 \\   
	& relaxed & -0.172$\pm$0.020 & 1.556$\pm$0.157 & -0.188$\pm$0.582  & 0.027$\pm$0.010 & 0.040$\pm$0.017 & 0.96  \\   
	& disturbed & -0.191$\pm$0.026 & 1.610$\pm$0.250 & 0.414$\pm$0.642  & 0.036$\pm$0.012 & 0.039$\pm$0.020 & 1.09 \\   
\hline
$M_{tot}$-$T_{exc}$  	& all & -0.165$\pm$0.014 & 1.508$\pm$0.126 & 0.235$\pm$0.370  & 0.026$\pm$0.009 & 0.043$\pm$0.015 & 1.02 \\  
			& relaxed & -0.178$\pm$0.019 & 1.536$\pm$0.138 & -0.243$\pm$0.549  & 0.025$\pm$0.009 & 0.037$\pm$0.015 & 0.97 \\  
			& disturbed & -0.180$\pm$0.026 & 1.616$\pm$0.259 & 0.329$\pm$0.633  & 0.035$\pm$0.013 & 0.046$\pm$0.022 & 1.10 \\      
\hline
$M_{tot}$-$M_{gas}$	& all & 0.073$\pm$0.007 & 0.802$\pm$0.049 & -0.317$\pm$0.307  & 0.028$\pm$0.015 & 0.043$\pm$0.011 & 1.04 \\   
			& relaxed & 0.080$\pm$0.008 & 0.864$\pm$0.047 & -0.801$\pm$0.411  & 0.025$\pm$0.010 & 0.027$\pm$0.009 & 1.32 \\   
			& disturbed & 0.064$\pm$0.022 & 0.800$\pm$0.127 & -0.063$\pm$0.664  & 0.054$\pm$0.027 & 0.055$\pm$0.021 & 1.01 \\      
\hline
$M_{tot}$-$Y_{X}$ 	& all & -0.010$\pm$0.005 & 0.540$\pm$0.030 & -0.292$\pm$0.287  & 0.039$\pm$0.023 & 0.039$\pm$0.011 & 1.00 \\  
			& relaxed & -0.010$\pm$0.007 & 0.561$\pm$0.034 & -0.635$\pm$0.428  & 0.034$\pm$0.019 & 0.031$\pm$0.010 & 1.09 \\  
			& disturbed & -0.019$\pm$0.013 & 0.538$\pm$0.069 & -0.043$\pm$0.568  & 0.066$\pm$0.037 & 0.049$\pm$0.019 & 0.96 \\  
\hline
$M_{tot}$-$Y_{X,exc}$ 	& all & -0.008$\pm$0.005 & 0.534$\pm$0.031 & -0.257$\pm$0.289  & 0.039$\pm$0.024 & 0.040$\pm$0.011 & 1.03 \\  
			& relaxed & -0.012$\pm$0.007 & 0.558$\pm$0.033 & -0.639$\pm$0.424  & 0.036$\pm$0.019 & 0.030$\pm$0.010 & 1.12 \\  
			& disturbed & -0.014$\pm$0.013 & 0.539$\pm$0.072 & -0.062$\pm$0.580  & 0.069$\pm$0.038 & 0.050$\pm$0.019 & 1.01 \\  			
\hline
\end{tabular}
\end{table*}

\subsection{L$_{X,exc}$-M$_{tot}$}\label{sect:LexcM}
Several studies have shown that using the core-excised luminosities helps to reduce the scatter of the $L_X$-$M_{tot}$ relation (e.g. \citealt{2009A&A...498..361P}, \citealt{2018MNRAS.473.3072M}). Therefore, we also computed the X-ray soft band luminosities $L_X$, excluding the core, corresponding to 0.15$R_{500}$. Indeed, these relations show a much lower scatter with respect to those derived using luminosities with the  cores included. The intrinsic scatter reduction is more significant for the most relaxed clusters, in agreement with the idea that the dominant contribution to the scatter for these systems is the dense and peaked core, while for disturbed systems, different processes, not associated with the core, are responsible for the scatter. The relations determined with core-excised luminosities are moderately flatter, but consistent within uncertainties, than the ones with the core-included ($\beta$=1.668$\pm$0.183 vs $\beta$=1.822$\pm$0.246), which may suggest a larger fraction of relaxed clusters at high masses. However, the same effect (i.e. flattening of the slope) is less obvious when fitting the data assuming the redshift self-similar evolution (i.e. $\gamma$=2). That could be explained if relaxed and disturbed systems evolve differently with redshift. 
Although the large uncertainties in $\gamma$, due to the limited redshift range of our sample, do not allow conclusive results, we have indeed hints of a slightly larger negative evolution for the most relaxed clusters (i.e. the fitted $\gamma$ for relaxed clusters tends to be systematically lower than the $\gamma$ derived for disturbed systems, see Table \ref{table:bestfit} and Fig. \ref{fig:slopes}). Moreover, when using the core-excised properties, we note a systematic trend to higher $\gamma$ values which is in better agreement with the self-similar predictions. 

Using the core-excised luminosities, we find that relaxed and disturbed systems share very similar relations. This agreement suggests that independently of the relative fraction of relaxed and disturbed systems in a sample, the core-excised luminosities can be used to obtain a universal relation that can be used for future surveys. However,
we note that core-excised temperatures usually cannot be measured for poor clusters or groups even in relatively deep surveys like XXL (e.g. \citealt{2016A&A...592A...1P}). Moreover, although the use of the  $L_{exc}$  helps to reduce the offset between the different relations for X-ray and SZ selected samples, there is still a normalization offset  which requires the determination of the true cluster mass scale (see the review by \citealt{2019SSRv..215...25P}) to obtain the universal $L_{X,exc}$-$M_{tot}$ relation (in the case of AM10 there is also a difference in the slope, which may depend on the different proxy used to estimated the total masses).  In Fig. \ref{fig:LMmor} ({\it right panel}) we show the $L_{X,exc}$-$M_{tot}$ relation showing good agreement between all the relations.  

\begin{figure*}[!ht]
\figurenum{4}
\hbox{
\centering
\includegraphics[scale=1.0, width=0.5\textwidth]{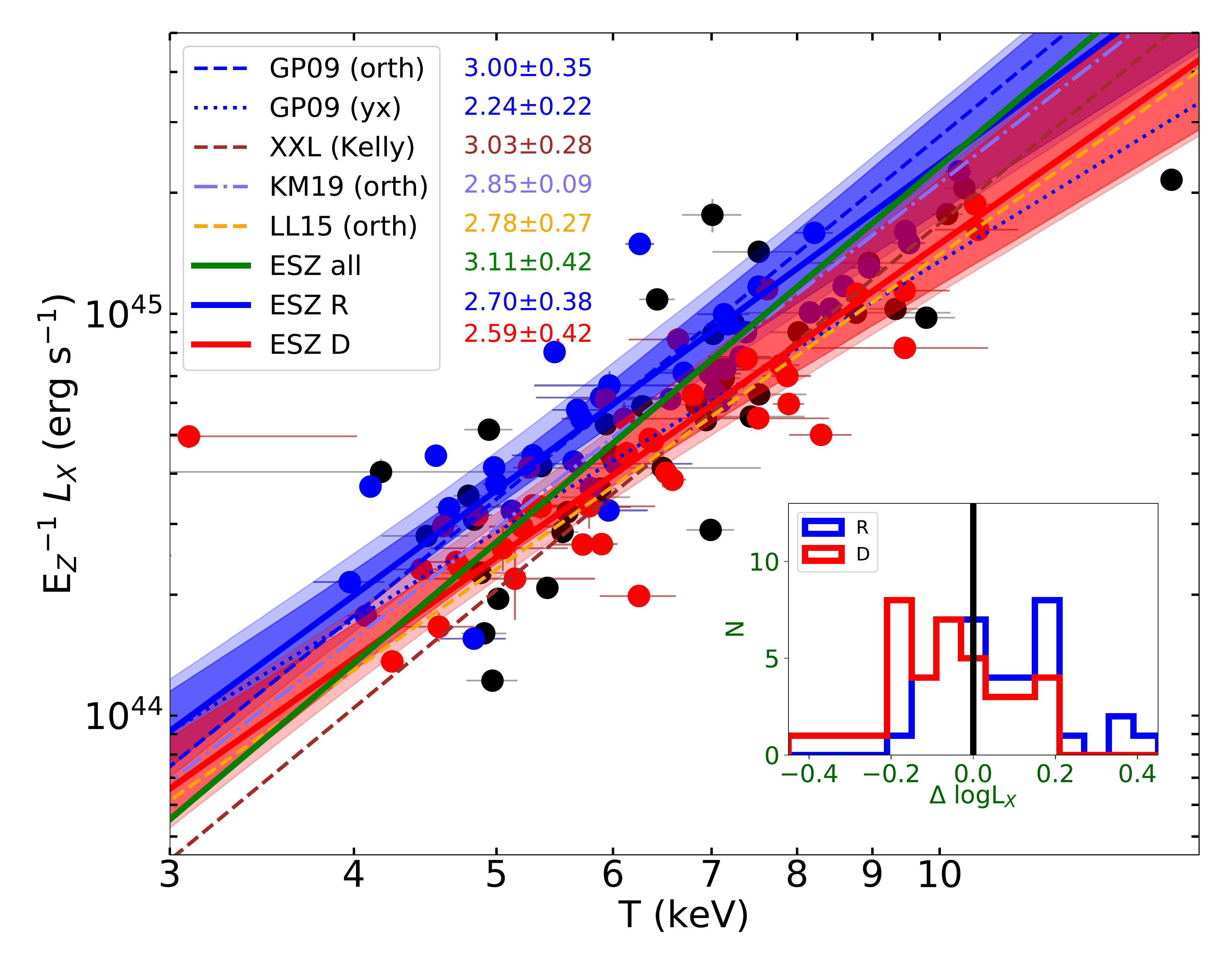}
\includegraphics[width=0.5\textwidth]{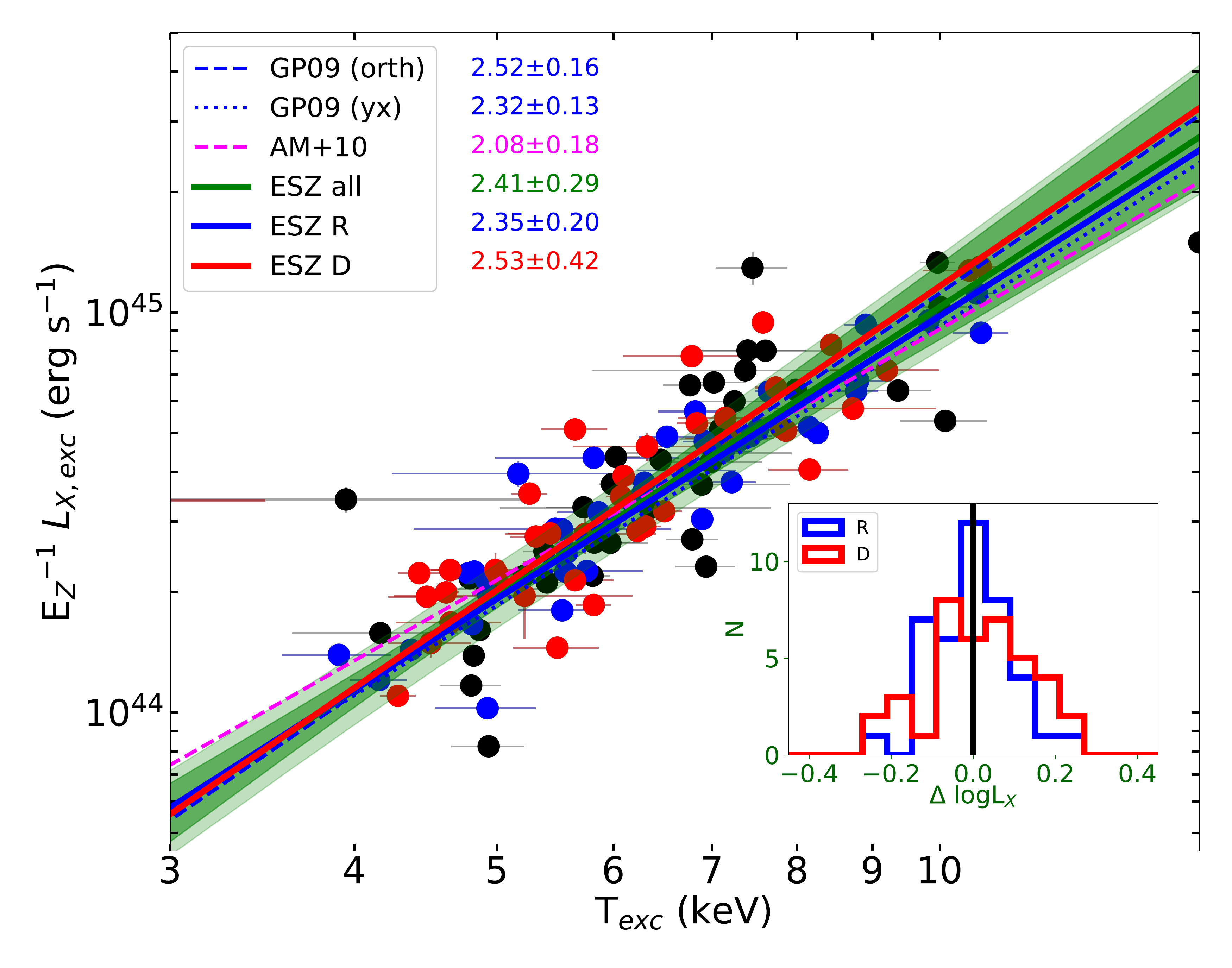}
}
\caption{{\it left:} comparison between the $L_X$-$T$ relation determined in this work with the best-fit relations obtained from other well-studied samples selected in the X-ray. $L_X$ are soft-band (0.1-2.4 keV) luminosities. The green line represents the fitted relation with $\gamma$ free to vary in the fit. The dark and light shaded areas represent the 1$\sigma$ statistical error and scatter, respectively. In blue and red we show the fitted relations determined for relaxed and disturbed systems, independently.  In the inset plot, we show the histogram of the log space residuals from the fitted $L_X$-$T$ relation. As for the $L_X$-$M_{tot}$ relation (see Fig. \ref{fig:LMmor}), relaxed objects are on average above the relation, while disturbed clusters are on average below.
{\it right:} comparison between the $L_{X,exc}$-$T_{exc}$ relation determined in this work with some of the relations available in the literature. The histogram of the log space residuals in the inset plot shows that relaxed and disturbed clusters distribute homogeneously around the fitted relation, when core-excised luminosities and temperatures are used. The acronyms in the legend are described in Sect. \ref{sect:LM} and Sect. \ref{sect:LT}.}
\label{fig:LT}
\end{figure*}

\subsection{L$_{bol}$-$M_{tot}$ and L$_{bol,exc}$-$M_{tot}$}
The relations of the ESZ sample determined using the bolometric luminosities (i.e. 0.01-100 keV) $L_{bol}$ are steeper  than those found using the luminosities in the soft X-ray band (i.e. 0.1-2.4 keV). The slope is steeper than the prediction from the self-similar scenario with a significance of more than 3$\sigma$, using both core-excised and core-included luminosities.
Similarly to the $L_{X}$-$M_{tot}$ relation, we observe hints of a negative evolution (i.e. $\gamma$=0.541$\pm$0.940) with respect to the self-similar scenario (i.e. $\gamma$=7/3).
The intrinsic scatter, larger than the one obtained using the soft band luminosities, is at the $\sim$25$\%$  level and it is comparable with others found in the literature (e.g. EB19).  Again, the scatter is larger for the disturbed clusters, and the reduction, after removing the cores, is larger for the most relaxed systems.

\subsection{$L_X$-$T$ {\rm and} $L_{X,exc}$-$T_{exc}$}\label{sect:LT}
Because of the different methods used to determine the total cluster mass (HE, WL, mass proxies, etc.), comparing the $L$-$M_{tot}$ relations from different studies is not always straightforward and can potentially bias our interpretation of the impact on the scaling relations of the different selection effects (e.g. the different fractions of relaxed/disturbed clusters). A more direct comparison can be done using the $L$-$T$ relation, although cross-calibration uncertainties between Chandra and XMM-Newton should be taken into account. While at low ICM temperatures, both observatories deliver similar results, the differences increase in the high temperature regime (see \citealt{2015A&A...575A..30S}, for more details) where most of our clusters reside.  

In Fig. \ref{fig:LT} (left panel), we compare our result with the finding by GP09, \citet[LL15]{2015A&A...573A.118L},  \citet[XXL]{2016A&A...592A...3G}, and Migkas\footnote{Migkas et al. used an  eeHIFLUGCS-like sample (see \citealt{2017AN....338..349R}) to determine the $L_X$-$T$ relation.  When  available, the Chandra data were used to determine the cluster temperature, while XMM-Newton observations have been used for clusters not observed with Chandra (i.e. roughly one third of the sample). 
For the comparison with our results, we converted the Chandra temperatures to XMM-like temperatures, using the equation determined by \cite{2015A&A...575A..30S}.} et al. (submitted, KM19).

The $L_X$-$T$ relation is significantly steeper than the value predicted by the self-similar scenario (i.e. $\beta$=3.110$\pm$0.422 vs $\beta$=1.5). Although the fit prefers a slightly smaller redshift evolution factor $\gamma$ (but consistent within the uncertainties with the self similar prediction), the impact on the slope is quite small.    
Similarly to the $L_X$-$M_{tot}$ relation, we observe quite good agreement for the  $L_X$-$T$ relation with other works for the slope, but a significant offset for the normalization. Again, the most relaxed clusters tend to have a higher luminosity for a given temperature (see Fig. \ref{fig:LT}, {\it left panel}). At 5 keV, relaxed clusters have, on average, a luminosity $\sim$50$\%$ higher than disturbed clusters. 

The use of the  core-excised luminosities brings into better agreement the best-fit relations for relaxed and disturbed clusters (see Fig.  \ref{fig:LT}, {\it right panel}). The $L_{X,exc}$-$T_{exc}$ relation, although shallower than the $L_X$-$T$ relation is still much steeper than what is predicted by the self-similar scenario ($\beta$=1.5). The slope determined by AM10 is only slightly flatter than other results and can be easily explained by the higher temperatures delivered by Chandra  used by AM10 compared to those from XMM-Newton used in this work and GP09.

The scatter of the temperature is smaller than the scatter of the total mass, indicating that the temperature is less sensitive than the total mass to the processes (e.g. presence of substructures) affecting the scatter. The results for the $L_X$-$T$ relation confirm that scatter in $L_X$ is basically driven by the boost in luminosity due to the peaked cores. In fact, we do not see any reduction of the scatter for the $L$-$T$ relation of the most disturbed clusters derived using the core-excised luminosities, neither in $\sigma_{X|Z}$ or $\sigma_{Y|Z}$. The lower scatter  of the L-T  relation can also be caused by positively correlated intrinsic scatter of luminosity and temperature at a given mass (\citealt{2019A&A...632A..54S}, \citealt{2016MNRAS.463.3582M}).

\subsection{$L_{bol}$-$T$ {\rm and} $L_{bol,exc}$-$T_{exc}$}
As was the case with soft X-ray luminosities, the $L$-$T$ relations obtained with the bolometric luminosities have a slope significantly steeper than the self-similar scenario (i.e.  $\beta$=3.464$\pm$0.400 vs $\beta$=2). Although the core-excised temperatures and luminosities help to flatten the slope  (i.e.  $\beta$=2.808$\pm$0.247), this is not sufficient to bring it into agreement with self-similar expectations (i.e. $\beta$=2).  In both cases, the redshift evolution of the full sample is in agreement with the predicted evolution  (i.e.  $\gamma$=1). However when using the core-included $L_{bol}$ and T, the evolution of the most relaxed objects deviates at $\sim$1$\sigma$ level from the prediction, while that is not the case for the disturbed systems. Again, using the core-excised luminosities helps to reduce the scatter for the subsample of relaxed clusters, but it has little effect on the subsample of disturbed systems (see Table \ref{table:bestfit}).

\subsection{$M_{tot}$-$M_{gas}$}\label{sect:MMg}
\begin{figure}[t!]
\figurenum{5}
\centering
\includegraphics[width=0.5\textwidth]{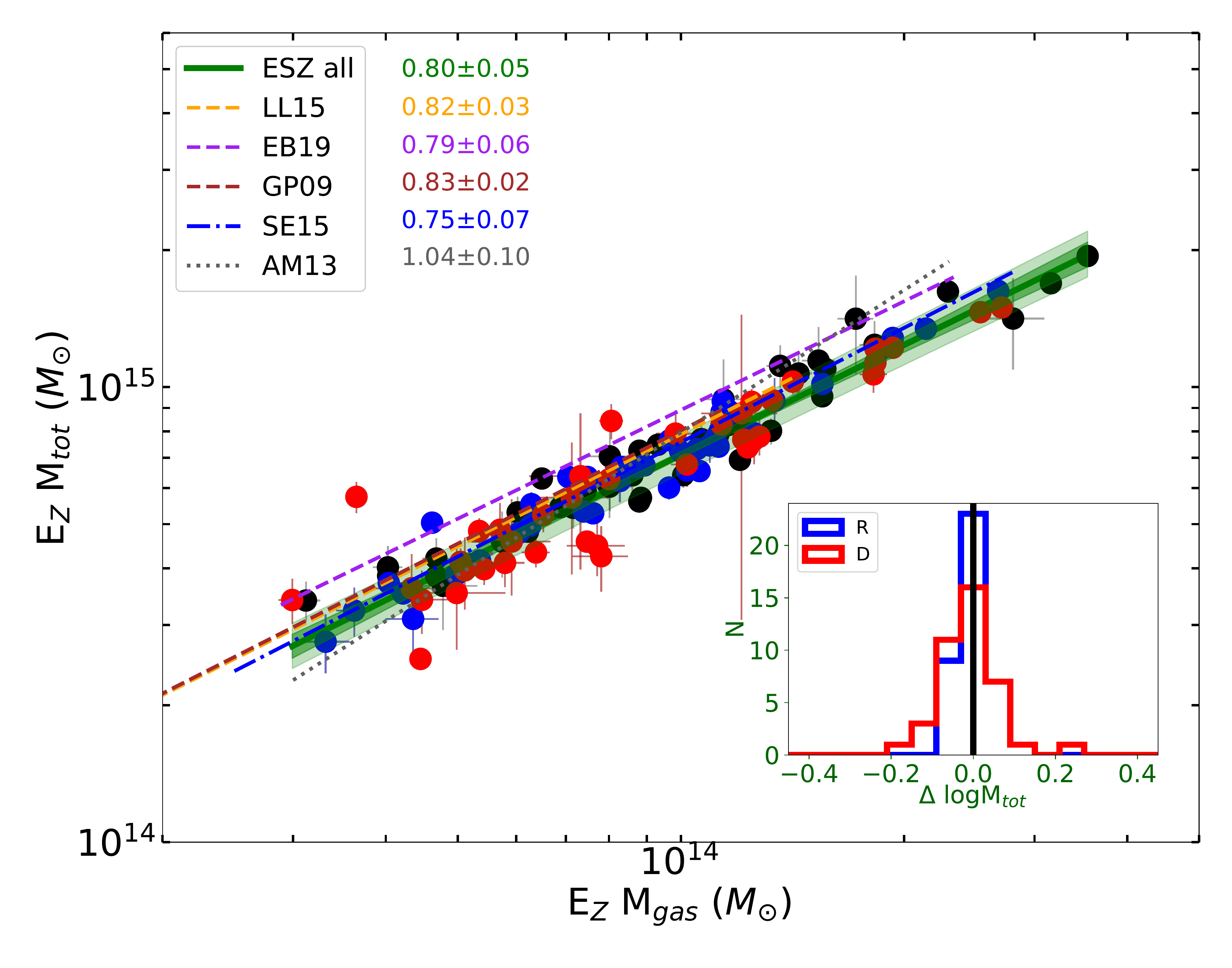}
\caption{$M_{tot}$-$M_{gas}$ relation for the ESZ clusters investigated in this work. The most relaxed and most disturbed clusters are shown in blue and red, respectively. The green line represents the fitted relation with $\gamma$ free to vary in the fit. The dark and light shaded areas represent the 1$\sigma$ statistical error and scatter, respectively. All the relations are plotted only in the mass range covered by the individual study. In the inset plot, we show the histogram of the log space residuals from the fitted $M_{tot}$-$M_{gas}$ relation. The agreement between the slopes obtained with samples having a different fraction of relaxed and disturbed systems, and between the subsamples of relaxed and disturbed clusters in the ESZ sample, suggests that this relation is insensitive to the dynamical state of the clusters. However, the offset observed in the normalization may suggest that the mass range of the investigated samples plays a role because of the increasing gas fraction for high mass systems. The acronyms in the legend are described in Sect. \ref{sect:LM}, Sect. \ref{sect:LT}, and Sect. \ref{sect:MMg}.}
\label{fig:MMgas}
\end{figure}

The mass of the ICM correlates very well with the total cluster mass with a relatively small intrinsic scatter (e.g. \citealt{2010ApJ...721..875O}, \citealt{2015A&A...573A.118L}, \citealt{2019arXiv190610455S}). 
Moreover, the $M_{gas}$ computed with Chandra and XMM-Newton within the same radius agree within a few percent (e.g. \citealt{2017A&A...608A..88B}).  In Fig. \ref{fig:MMgas}, we show that this is indeed also the case for the ESZ sample.  Moreover, we note that the slopes of fits to the different samples are in quite good agreement. Since the fraction of relaxed and disturbed systems in these samples is quite different, this implies that the slope of the $M_{tot}$-$M_{gas}$ relation is quite insensitive to the dynamical state of the clusters. This is indeed confirmed by the results for the subsamples of relaxed and disturbed systems, which show a good agreement in their slopes, although the intrinsic scatter for the disturbed clusters is larger than the one for relaxed clusters. The higher scatter observed in disturbed systems for both $M_{tot}$ and $M_{gas}$ is not surprising given the assumption of spherical symmetry. In fact, for these systems the presence of substructures and large scale inhomogeneities may bias the reconstruction of the clusters' properties  (e.g. \citealt{2013MNRAS.429..799V}, \citealt{2013MNRAS.428.3274Z}). 

There is an offset in the normalization, of the order of 5$\%$ at 10$^{14}M_{\odot}$ and $z$=0.2, between the relations obtained in this work and the results from \citet[SE15]{2015MNRAS.446.2629E}. The offset is even larger, of the order of 10$\%$ if compared with GP09 and LL15, and of the order of 20$\%$ with EB19. Part of this offset can be attributed to the nature of the different samples (i.e. ESZ contains more disturbed clusters), as shown by the better agreement between the ESZ relaxed clusters and the X-ray selected samples. This is because the normalization of the relaxed clusters is $\sim$4-5$\%$  higher than the one for the disturbed clusters. Moreover, the lower mass range covered by the other samples (see Fig. \ref{fig:MMgas}), with respect to the ESZ sample, may also play a role. In fact, low-mass systems have a lower gas fraction than the most massive ones (e.g. \citealt{2006ApJ...640..691V}, \citealt{2009A&A...498..361P},  \citealt{2015A&A...573A.118L}) implying a lower gas mass for a given cluster mass than what one would expect if the gas fraction were universal, and linearly related to the total mass. Thus, samples skewed towards massive systems, where the effect of baryonic processes and radiative cooling are expected to be relatively less impactful,  are expected to have a lower normalization in the $M_{tot}$-$M_{gas}$ plane. 
To support this interpretation, in Fig. \ref{fig:MMgas}, we only plot the relations in the mass range investigated in the individual papers and we can see that, with the exception of EB19, there is a shift toward lower normalizations in the $M_{tot}$-$M_{gas}$ plane when only massive clusters are considered. 
Apart from the best-fit relation from \citet[AM13]{2013ApJ...767..116M}, all the other relations point to a slope close to $\sim$0.8, therefore flatter than what is predicted if the gas fractions were the same on all mass scales.  The agreement between the slopes, but not in the normalization, suggests that this relation is almost independent of the fraction of relaxed and disturbed systems in the sample, but may depend on the mass range of the clusters that are investigated or on systematic differences in HE mass estimates..

\begin{figure}[t!]
\figurenum{6}
\centering
\includegraphics[width=0.5\textwidth]{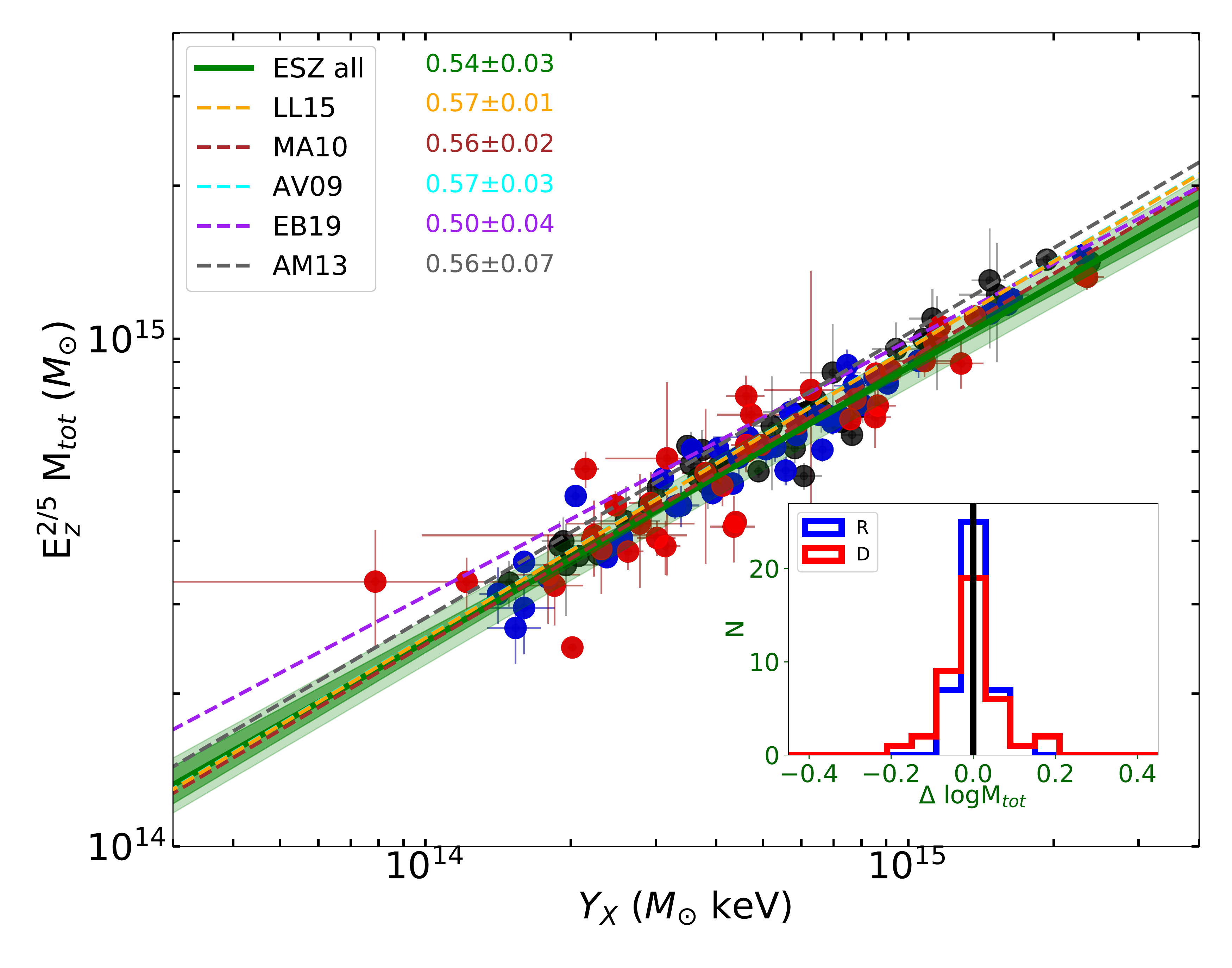}
\caption{$M_{tot}$-$Y_X$ relation for the ESZ clusters investigated in this work. The most relaxed and most disturbed clusters are shown in blue and red, respectively. The green line represents the fitted relation with $\gamma$ free to vary. The dark and light shaded areas represent the 1$\sigma$ statistical error and scatter, respectively. In the inset plot, we show the histogram of the log space residuals from the fitted $M_{tot}$-$Y_X$ relation. As for the $M_{tot}$-$M_{gas}$ relation, there is good agreement between the slopes obtained from the different samples, but an offset in the normalization. This, together with the good agreement between the relations obtained with the subsamples of relaxed and disturbed ESZ clusters, suggests that the $M_{tot}$-$Y_X$ relation  is insensitive to the dynamical state of the objects, but dependent on the mass range investigated. The effect is smaller than that observed  for the $M_{tot}$-$M_{gas}$. The acronyms in the legend are described in Sect. \ref{sect:LM}, Sect. \ref{sect:LT}, Sect. \ref{sect:MMg}, and Sect. \ref{sect:MY}.}
\label{fig:MY}
\end{figure}

\subsection{$M_{tot}$-$Y_{X}$}\label{sect:MY}
The $Y_X$ parameter (\citealt{2006ApJ...650..128K}), a measure of the total thermal energy in the ICM, is also a low scatter mass proxy (see Fig. \ref{fig:MY}).  
As for the other scaling relations, we observe quite good agreement in the slope derived from independent studies and an offset in the normalization, in particular in the best-fit derived by EB19 and AM13 ($\sim$8$\%$ lower normalization). The offset is smaller, on the order of $\sim$5$\%$, with respect to LL15 and AV09, while it is in perfect agreement with the result by \citet[MA10]{2010A&A...517A..92A}.   If the offset is caused by the lower $M_{gas}$ in low-mass systems, then we should observe a lower $M_{tot}$-$Y_{X}$ relation also for LL15 that includes systems with total masses down to $\sim$2$\times10^{13}M_{\odot}$. Instead the relation by LL15 has a lower normalization than the one from EB19, in particular in the low-mass regime. Unlike the other studies plotted in Fig.  \ref{fig:MY}, EB19 uses SZ derived masses, which may suggest a mass trend of the SZ signal with the total mass that would result in an offset in the X-ray observables and total mass relations. 

Almost all the studies find a slope of the $M_{tot}$-$Y_{X}$ relation shallower than what is expected from  the self-similar scenario (i.e. $\beta$=0.6). This may be caused by the increasing gas fractions for increasing total cluster masses. In fact, LL15 found that the slopes of the $M_{tot}$-$Y_{X}$ relation, derived in the low- and high-mass regimes agree well with the prediction of the self-similar scenario. However, the normalizations are quite different, with the galaxy groups having a $>$10$\%$ higher normalization than the clusters, due to their average lower gas fraction. This offset in the normalization leads to a slightly shallower relation, when fitting all the systems together. 

The relations, determined for relaxed and disturbed clusters separately, are in good agreement. Thus, as for the $M_{tot}$-$M_{gas}$ relation, the $M_{tot}$-$Y_{X}$ slope is insensitive to the dynamical state of the objects, but it may still have a small dependence on the mass range considered, which affects the normalization, although with a lower impact than what is seen in the $M_{tot}$-$M_{gas}$ relation. As for the $M_{tot}$-$M_{gas}$ relation there is a hint of a smaller scatter for relaxed clusters,  but the difference is within the statistical uncertainties. Using the core-excised temperatures to calculate the $Y_X$ parameters does not impact either the shape or the scatter of the relations. This confirms that the relation is basically insensitive to the influence of AGN feedback and/or star formation, as suggested by the numerical simulations (e.g. \citealt{2006ApJ...650..128K}, \citealt{2007ApJ...655...98N}). The similarity in the scatter of the $M_{tot}$-$M_{gas}$ and $M_{tot}$-$Y_{X}$ for both relaxed and disturbed systems, suggests that the scatter in the two relations is probably driven by the same processes.

\begin{figure}[t!]
\figurenum{7}
\centering
\includegraphics[width=0.5\textwidth]{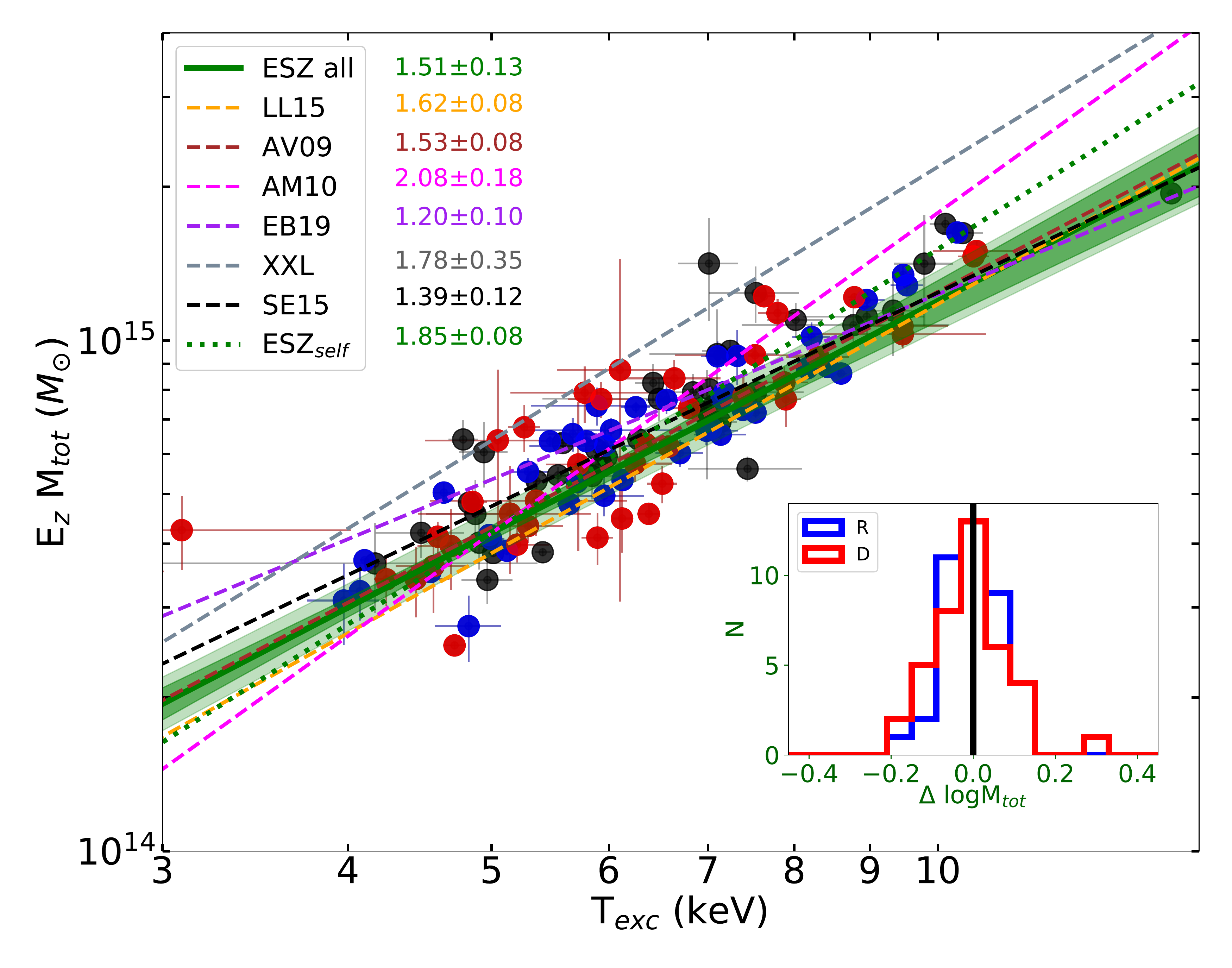}
\caption{$M_{tot}$-$T_{exc}$ relation for the ESZ clusters investigated in this work. The most relaxed and most disturbed clusters are shown in blue and red, respectively. Clusters that are not in these two subsamples are shown in black. The green line represents the fitted relation with the redshift evolution free to vary. The dark and light shaded areas represent the 1$\sigma$ statistical error and scatter, respectively. In the inset plot we show the histogram of the log space residuals from the fitted $M_{tot}$-$T_{exc}$ relation. The acronyms in the legend are described in Sections \ref{sect:LM}, \ref{sect:LT}, and \ref{sect:MMg}.}
\label{fig:MT}
\end{figure}

\subsection{$M_{tot}$-$T_{exc}$}
In Fig. \ref{fig:MT} we compare our $M_{tot}$-$T_{exc}$ relation with the relations available in the literature by  LL15, AV09,  AM10, XXL, EB19, and SE15. The slope for the ESZ sample, $\beta$=1.508$\pm$0.126, is in agreement with the self-similar expectations (i.e. $\beta$=1.5), but with a positive redshift  evolution at the $\sim$3$\sigma$ level. However, if the redshift evolution is forced to be self-similar, the fit prefers a much steeper relation (i.e. $\beta$=1.823$\pm$0.076), which is in better agreement with the result by AM10, who fitted jointly the luminosity, temperature, and total cluster mass, accounting for the selection biases, and assuming a self-similar redshift evolution. The $M_{tot}$-$T$  relation determined with core-excised temperatures is not significantly different from that obtained using the core-included temperatures. Relaxed and disturbed clusters share a similar relation, and also the scatter is not significantly different ($\sigma_{Y|Z}$=0.037$\pm$0.015 vs $\sigma_{Y|Z}$=0.046$\pm$0.022 for the relaxed and unrelaxed sub-samples, respectively), as instead was observed by \cite{2016A&A...592A...4L} for the XXL sample using WL masses, and XMM-Newton temperatures within a 300 kpc radius. A similar level of the scatter for the temperature suggests that the processes that alter the homogeneous temperature distribution have a relatively small impact on the scatter of the scaling relations.

\section{Discussion}
\begin{figure*}[!]
\figurenum{8}
\centering
\hbox
{
\includegraphics[width=0.5\textwidth]{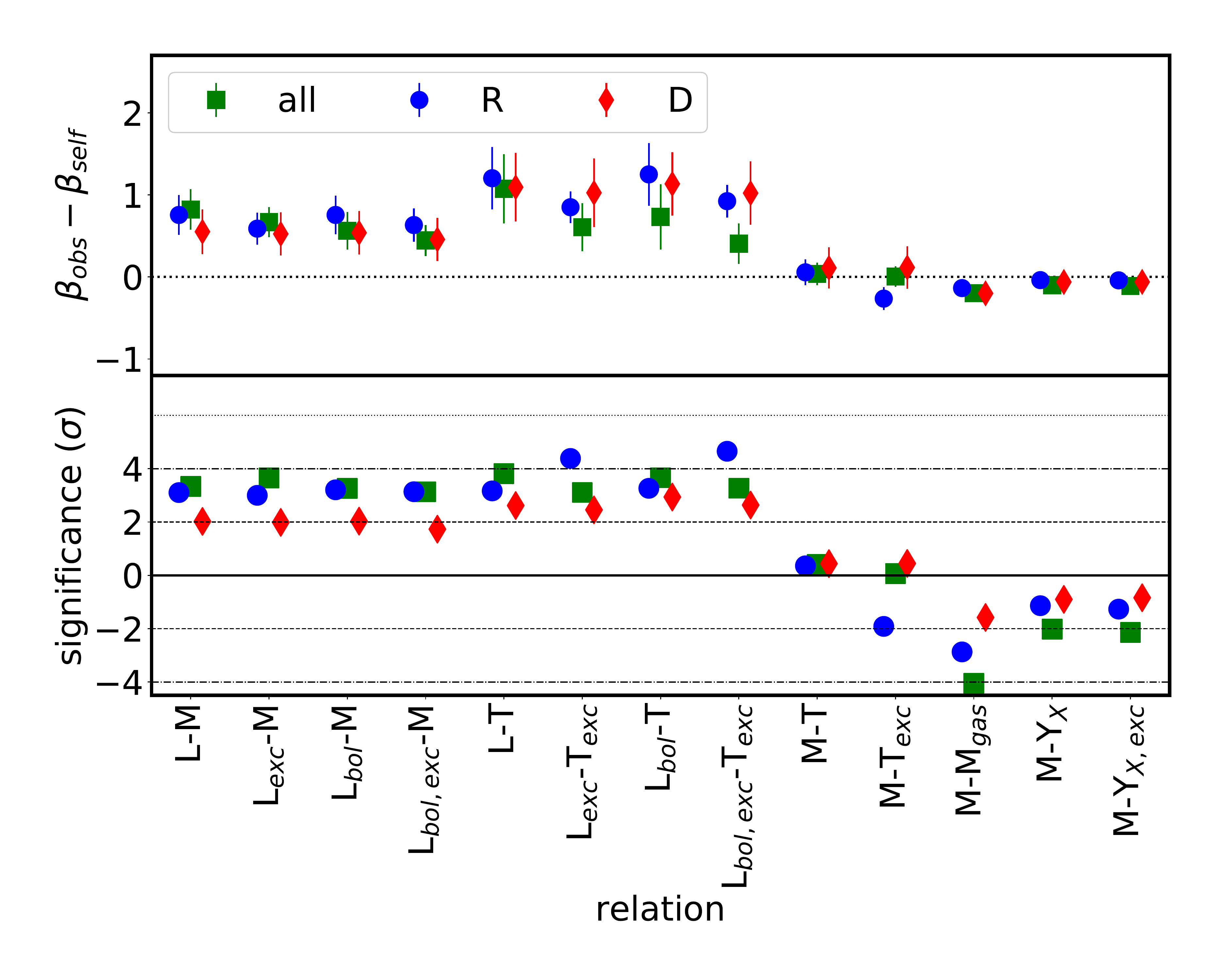}
\includegraphics[width=0.5\textwidth]{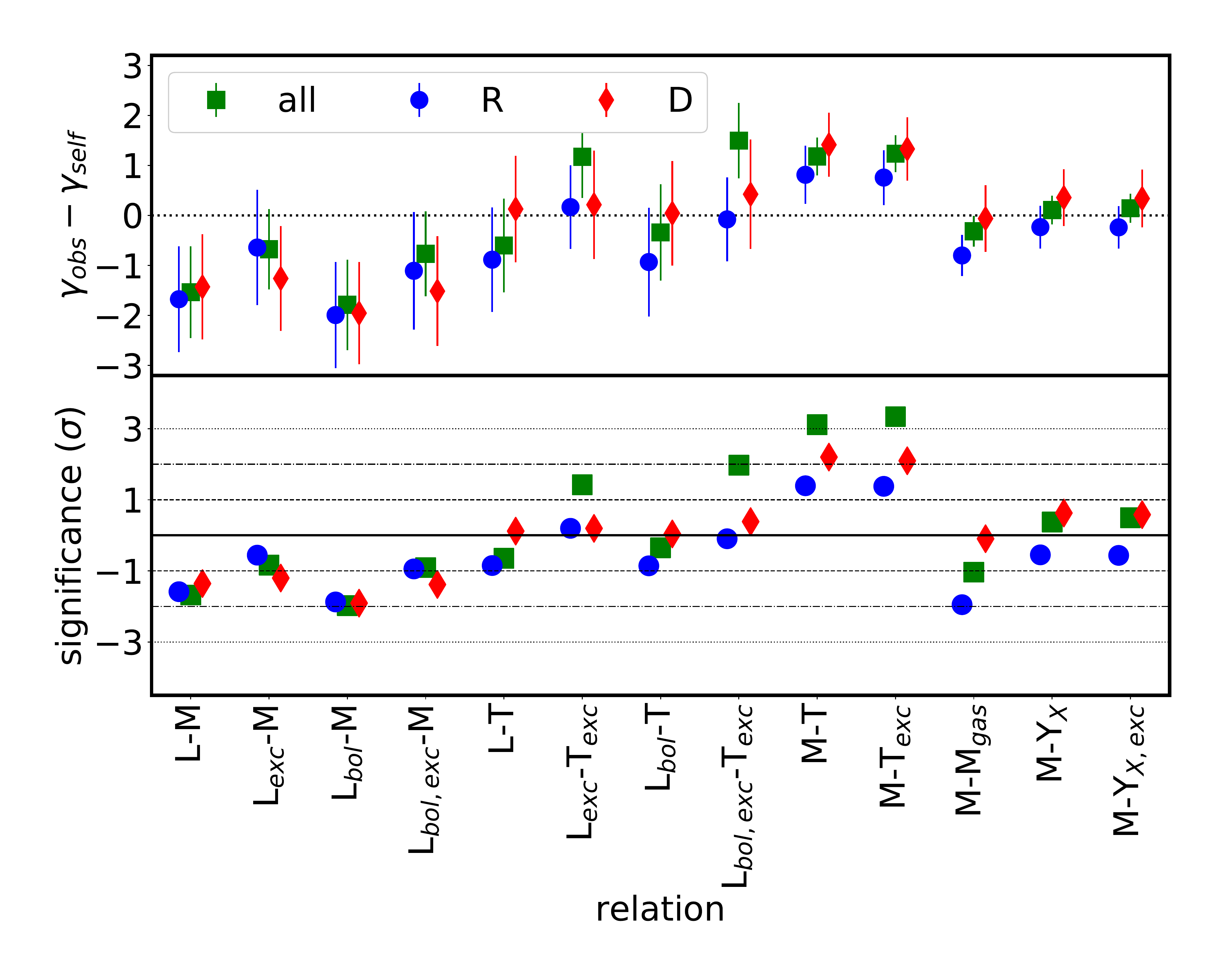}
}
\caption{Overview of the fitted slopes $\beta$ (left panel) and redshift evolution factors $\gamma$ (right panel) with respect to the self-similar predictions (see Table \ref{tab:pivot}). The results for the full sample are shown in green squares, while the results for the subsamples of relaxed (R) and disturbed (D) clusters are shown in blue circles and red diamonds, respectively. In the top panels, we show the difference between the observed and predicted values (i.e. if a relation behaves self-similarly the data point would be consistent with zero). In the bottom panels we show how significantly the parameters deviate from the prediction.}
\label{fig:slopes}
\end{figure*}

All the observed scaling relations, with the exception of the $M_{tot}$-$T_{exc}$, have a slope that deviates from the expectation of the self-similar scenario by more than 2$\sigma$ (see Fig. \ref{fig:slopes}, left panel), but are statistically consistent  with the results from the literature (within 1$\sigma$ if the same fitting method is used).   There are also hints that relaxed systems have steeper relations than disturbed clusters (see Table \ref{table:bestfit}), again with the exception of the $M_{tot}$-$T_{exc}$,  but that needs to be confirmed with a larger sample to strongly reduce the statistical uncertainties of the fits.

\begin{figure*}[!ht]
\figurenum{9}
\includegraphics[width=1\textwidth,height=0.25\textheight]{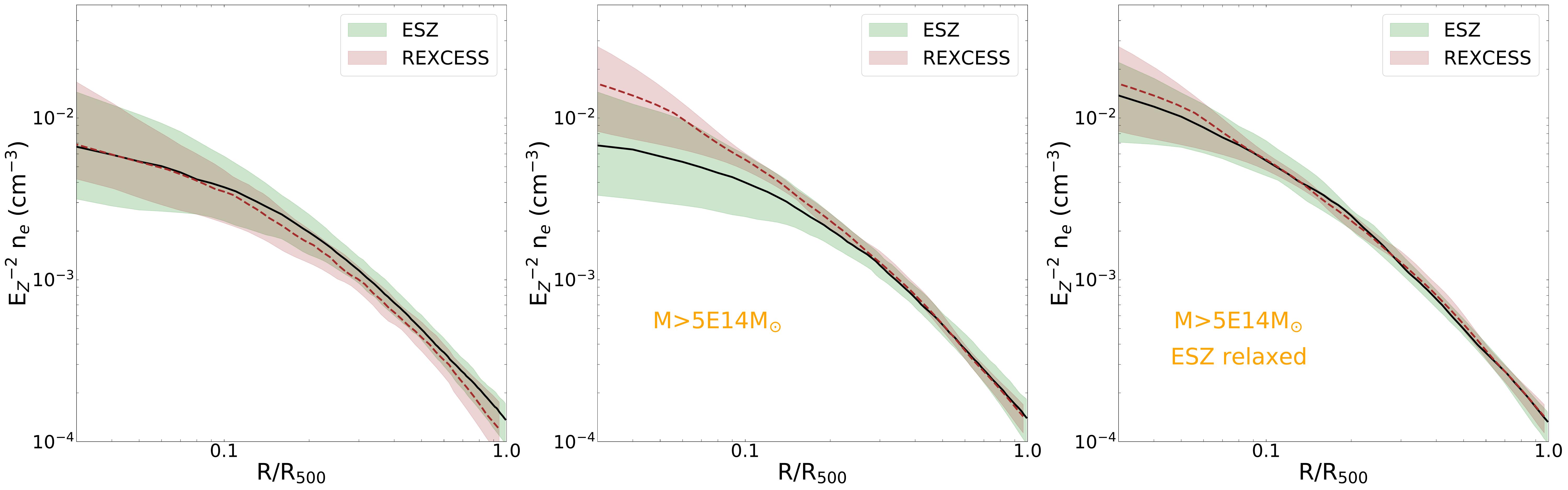}
\caption{{\it left:} comparison between the average density profiles for REXCESS and ESZ samples. {\it center:} as in the left panel but using for both samples only clusters with a $M_{tot}$$>$5$\cdot10^{14}M_{\odot}$.  {\it right:} as in the center panel but using only the relaxed subsample for the ESZ clusters. }
\label{fig:necomp}
\end{figure*}

In the case of the $L$-$M_{tot}$ relations, we observe quite large normalization offsets between the different studies. This is partially associated with the assumed fitting procedure and the choice of keeping  the redshift evolution frozen or free (see Appendix \ref{sect:fitcomp} for more details). Strongly contributing to the offset is also the different fractions of relaxed and disturbed systems in the investigated samples. In fact, on average, relaxed systems have a 30$\%$ higher X-ray soft band (0.1-2.4 keV) luminosity than disturbed clusters for the same mass. The origin of this difference is the lack of self-similarity in the gas density profiles of relaxed and disturbed clusters as discussed by \cite{2012MNRAS.421.1583M}. Disturbed clusters have flatter profiles, while relaxed clusters have more centrally concentrated gas density profiles. Moreover, \cite{2012MNRAS.421.1583M} found a temperature dependence  for the profiles of the disturbed systems with hotter clusters having higher densities than the cooler clusters. They did not find the same dependence for the relaxed clusters (see also \citealt{2016MNRAS.463.3582M}). The ESZ sample has a similar behavior, and we also find a temperature dependence for relaxed clusters in the low redshift (z$<$0.2) regime, while the most distant and relaxed clusters of our sample show similar profiles, independent of their temperature.  These results complicate the comparison of the $L$-$M_{tot}$ relation from different works. To illustrate why, we compare the average profile of the ESZ sample with the REXCESS sample (see left panel of Fig. \ref{fig:necomp}). The two samples have a similar electron density in the center, but in the outer regions, the ESZ clusters show a much higher density than the REXCESS clusters.  Since the two samples span a different mass range, in the middle panel of Fig. \ref{fig:necomp} we only compare the massive systems, which show much better agreement  in the outer regions, while showing a higher gas density in the core of the REXCESS clusters. However, since ESZ clusters are on average more disturbed than REXCESS clusters (e.g. \citealt{2017ApJ...846...51L}), in the right panel of Fig. \ref{fig:necomp}, we compare only the most relaxed systems in ESZ with all REXCESS clusters, and find that  the agreement is very good. This agreement reflects the good match between the relation found by GP09 and that from the most relaxed objects in the ESZ sample. However, the different shape of the electron density profiles for relaxed and disturbed systems, together with their temperature/redshift dependence and the relative fraction of relaxed/disturbed systems in the samples, has an impact on the observed slopes and normalizations, and also on the cluster redshift evolution.

To further complicate the comparison, we note that different studies compute  the total masses in different ways, which can easily result in an offset of 10-20$\%$ (e.g. \citealt{2015MNRAS.450.3633S}, \citealt{2019SSRv..215...25P}). The assumption of $N_H$ (LAB vs TOT), plasma model (e.g. mekal  vs apec v1.3.1 vs apec v3.0.9), and abundance tables (e.g. ASPL from \citealt{2009ARA&A..47..481A} vs ANGR from \citealt{1989GeCoA..53..197A}) in the spectral fitting and in the conversion of the total count rates from the surface brightness to the cluster fluxes also can cause an offset in the y-direction (i.e. in X-ray luminosity). Moreover, some relations have been obtained using the luminosities derived with RASS data, while others use the higher quality data of Chandra and XMM-Newton, which allow a more accurate point source detection.  
Similar arguments can be applied to the $L$-$T$ relation, with the complication that temperatures obtained with different detectors can vary significantly, with a larger deviation observed for hot clusters, which therefore can also lead to a different slope. 
Interestingly, both the $L$-$T$ and $L$-$M_{tot}$ relations from different studies tend to converge, when the core-excised luminosities are used, which points to the different fraction of relaxed and disturbed systems as the main contribution for the observed offset.  

The relativity good agreement between the $L_{X,exc}$-$T_{exc}$ relation of AM10 with the other relations may also indicate that their flatter slope for the  $L_{X,exc}$-$M_{tot}$ relation is probably related to a mass-dependent effect on the total mass estimation.  

The $M_{tot}$-$M_{gas}$ and $M_{tot}$-$Y_X$ relations are in good agreement between the different studies,  particularly if self-similar redshift evolution is assumed. One of the reasons is that, differently from the luminosity, $M_{gas}$ depends linearly on the gas density, reducing the impact of different selection methods. This is probably also due to the smaller offset between the relation obtained with the relaxed and disturbed cluster subsamples (i.e. only $\leq$5$\%$ difference).  Interestingly, both the ESZ and SPT samples suggest a slightly flatter $M_{tot}$-$Y_X$ relation than the relations derived from X-ray selected samples, although  at low significance. 
                                            
Special discussion is needed for the $M_{tot}$-$T$ relation. As can be seen in Fig. \ref{fig:MT}, there is some tension between the best-fit relations determined in different studies, both in terms of slope and normalization, with the latter easily connected to the method and/or proxy used for the total mass estimate. The slope can be as shallow as 1.25$\pm$0.16 as found by EB19 or as steep as 2.08$\pm$0.18 as found by AM10. The consistency of the $M_{tot}$-$T$ relation for galaxy groups and clusters (e.g. \citealt{2009ApJ...693.1142S} and \citealt{2015A&A...573A.118L}) excludes the possibility that the differences are related to the mass (temperature) range investigated. Moreover, the agreement of the $M_{tot}$-$T$ relation at low- and high-mass ends suggests that non-gravitational processes are not strongly impacting this relation. Therefore, the offset seen in Fig. \ref{fig:MT} may point to a possible bias introduced in the estimate of the total mass. For instance, EB19 used the SZ masses to derive the scaling relations and if the SPT mass estimates suffer from a mass-dependent bias as found by the Planck mass estimates  it could explain the shallower relation. 

All the fitted relations for the ESZ sample prefer different values of $\gamma$ with respect to the expectations of the self-similar scenario. However, given the relatively low redshift (i.e. z$_{med}$$\approx$0.2)  of the sample, only the value of $\gamma$ for the $M_{tot}$-$T$ relation lies more than 2$\sigma$ from the self similar value  (see Fig. \ref{fig:slopes}, right panel). The best-fits for the $L$-$M_{tot}$ and $L$-$T$ relations suggest a negative evolution of the scaling laws, in agreement with the finding by \cite{2011A&A...535A...4R} who combined several published  data sets to investigate the evolution of the X-ray scaling relations to $z$=1.46. The negative evolution estimated by \cite{2011A&A...535A...4R}  is more significant than the one estimated with the ESZ sample. This is likely due to the larger redshift range covered by their sample, but also because only X-ray selected clusters were used. In fact, interestingly, for the ESZ sample we found a systematic trend to have a larger evolution for the most relaxed objects than for the most disturbed clusters This indicates that relaxed and disturbed systems may evolve differently. 
We note that, given the large errors associated with $\gamma$, this result is not conclusive and a detailed investigation with a larger sample, which also includes more distant clusters,  should be performed. Nonetheless, the effect of the cores in the $L$-$M_{tot}$ and $L$-$T$ relations should be taken into account, not only for the effect on the scatter, but also the impact on the normalization and on the redshift evolution.

For example, the offset between the ESZ and SPT $L_X$-$M_{tot}$ relation decreases from $\sim$45$\%$ to $\sim$23$\%$ when self-similar redshift evolution (i.e. $\gamma$=2) is assumed. If relaxed and disturbed clusters evolve differently, and since the fraction of relaxed and disturbed clusters in the ESZ and SPT samples could be different, then forcing the same evolution may reduce the offset. However, that would be an extra effect on top of the offset caused by the different fraction of relaxed/disturbed systems (i.e. the larger the fraction of relaxed objects in the sample, the higher is the normalization). Unfortunately, we do not know the dynamical state for the SPT clusters analyzed by EB19.  

The evolution factor $\gamma$ determined for the global and core-excised luminosity relations is consistent in the case of  disturbed clusters, while it is different for relaxed clusters. In particular, using the core-excised luminosities leads to a $\gamma$ value consistent with the one derived for disturbed clusters and in better agreement with the expectations from the self-similar scenario. This suggests that either the peaked and relaxed clusters evolve differently from the disturbed clusters, or that the cool-cores mimic the evolution.

Given the higher normalization in the scaling relations of relaxed than disturbed systems and since flux selected samples have been shown to have a larger fraction of relaxed clusters, the lower normalization found by GS17, AM10, and AM16 must have a different explanation. For example, in this paper, we determined the masses as in GS17, but using XMM-Newton, which is known to deliver lower temperatures  than Chandra. It is possible that in the high mass regime, the masses obtained by GS17 are higher than the ones derived in this work because of the higher temperatures determined with Chandra. However, since the temperatures at large radii drops to values where Chandra and XMM-Newton agree better the difference maybe smaller. Indeed, \cite{2014MNRAS.443.2342M} obtained consistent results for the  total hydrostatic masses for the same clusters observed with both  Chandra- and XMM-Newton-based measurements. Therefore, a more detailed investigation is required to understand this difference.

The relation determined for the most relaxed clusters in our sample agrees well with the relation determined by GP09 for the REXCESS sample, which is dominated by centrally-peaked and relaxed systems. The significant overlap between the $n_e$ profiles from the REXCESS sample and our subsample of relaxed systems fully explains this good agreement.

The $M_{tot}$-$T$ relation shows a positive evolution with respect to the self-similar prediction. The evolution is detected at more than the 3$\sigma$ level (see Fig. \ref{fig:slopes}). The results by EB19 with the SPT clusters are also consistent with a positive evolution, although detected only at 1$\sigma$ level. Studies of X-ray selected samples, found seemingly contradictory results: e.g. \cite{2004A&A...417...13E} and \cite{2011A&A...535A...4R} found no evolution, while \cite{2016MNRAS.463.3582M} found a $\sim$2$\sigma$ positive evolution.
The fact that the positive evolution is observed more significantly with SZ selected samples is probably associated with the larger fraction of disturbed clusters than in the X-ray selected samples, and a better sampling of the full halo population. 
Indeed, the redshift evolution for the subsample of relaxed clusters is closer to the prediction of the self-similar scenario. This finding is consistent with the picture that clusters at higher redshift are on average more disturbed (as confirmed by the mild correlation of the centroid-shift with the redshift), therefore their temperatures are hotter than what one would expect from  self-similar evolution. Moreover, disturbed clusters tend to have a larger hydrostatic bias, which could potentially mimic the evolution. Although observationally this is the first time that the evolution was detected so significantly, a positive evolution was predicted by recent simulations.  For example, \cite{2017MNRAS.466.4442L} investigated a large population of groups and clusters obtained with the cosmo-OWLS suite in cosmological hydrodynamical simulations and found a positive evolution of the $M_{tot}$-$T$  relation, independent of the included ICM physics.  More recently,  \cite{2018MNRAS.474.4089T} also found that the normalization of the $M_{tot}$-$T$ varies only by $\sim$20$\%$ between z=0.6 and z=0 (roughly the same redshift range investigated in our paper) instead of the $\sim$40$\%$ predicted by the self-similar scenario. The evolution is somehow stronger when AGN feedback is included in the simulations. The reason for this good agreement between observations and simulations may be due to the characteristic SZ cluster selection which is approximately a mass selection. This is more representative of the cluster selection sampled by hydrodynamical simulations.

\begin{figure}[!]
\figurenum{10}
\centering
\includegraphics[width=0.5\textwidth]{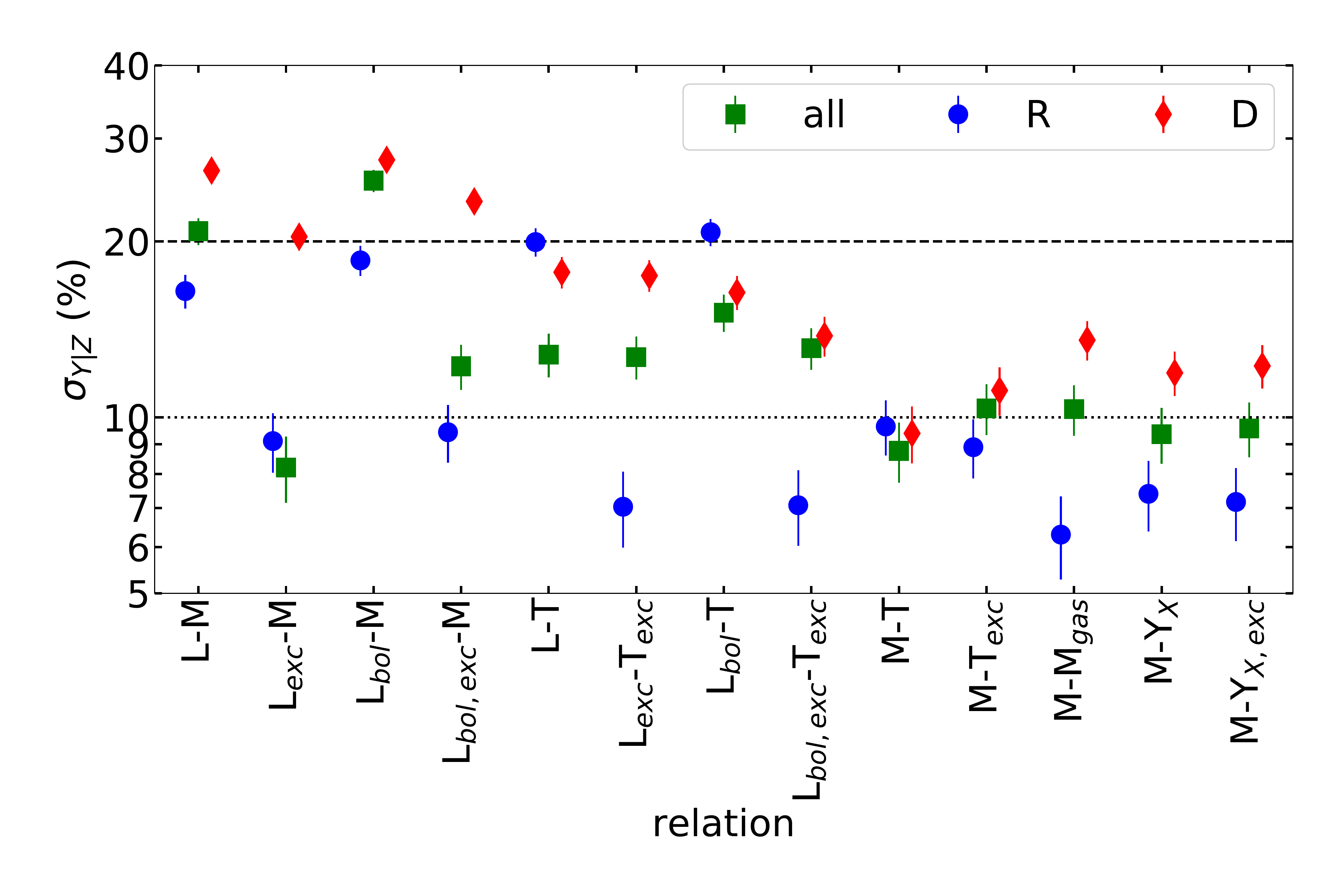}
\caption{Overview of the intrinsic scatter, $\sigma_{Y|Z}$ expressed in percent. The results for the full sample are shown in green squares, while the results for the subsamples of relaxed (R) and disturbed (D) clusters are shown in blue circles and red diamonds, respectively. }
\label{fig:scatter}
\end{figure}

The scatter in the scaling relations is the sum of different processes acting in different directions. In our study, apart from the $L_X$-$T$, $L_{bol}$-$T$,  and $M_{tot}$-$T$ relations, the subsample of relaxed clusters shows a lower scatter than the subsample of disturbed clusters (see Fig. \ref{fig:scatter}). Moreover, the scatter is clearly reduced by excluding the core regions to compute both luminosity and temperature, although the use of the core-excised temperatures has a much lower contribution in the scatter reduction than the core-excised luminosities. This suggests that  temperature inhomogeneities do not play a major role in the scatter of the scaling relations. As found in previous studies and by numerical simulations (e.g. \citealt{2017MNRAS.466.4442L}), we found that the $M_{tot}$-$M_{gas}$,  $M_{tot}$-$T$, and $M_{tot}$-$Y_X$ relations all have relatively low scatter. Moreover, the intrinsic scatter of the relations is further reduced, if only the relaxed clusters are considered. 

Depending on the orientation of the cluster, the estimated $M_{gas}$ can be easily incorrect  (i.e. typically the  $M_{gas}$ is overestimated if the cluster is elongated along the line-of-sight, while is underestimated if the cluster is elongated in the plane of the sky, see \citealt{2003A&A...398...41P} for more details). Since, the temperature structures have a small effect on the scatter, the determination of $M_{gas}$ by ignoring triaxiality, substructures, and clumps could be one of the major drivers for the scatter in these relations.

\section{Summary and Conclusion} 
We used archival XMM-Newton observations to determine the X-ray properties (i.e. $M_{tot}$, $M_{gas}$, k$T$, $L_X$, $L_{bol}$, $Y_X$) of a representative sample of 120 Planck Early Sunyaev-Zeldovich clusters to investigate the most common X-ray scaling relations: $L$-$M_{tot}$, $L$-$T$, $M_{tot}$-$M_{gas}$, $M$-$Y_X$, and $M_{tot}$-$T$.  We fit these relations, leaving free to vary the slope, normalization, redshift evolution, and  intrinsic scatter in both X and Y variables.  Our results are the following.

$\bullet$ The slopes of the scaling relations derived with SZ selected samples are in relatively good agreement with the relations derived from X-ray selected samples, particularly when the same fitting procedure is used. However, for most of the relations, there is some tension in the normalizations which can only partially be ascribed to the fitting algorithm.  Most differences come from the different fraction of relaxed and disturbed systems in the samples, which strongly depends on the different selection functions (e.g. SZ vs X-ray).  On top of that, differences also arise from different methods used to determine the global cluster properties (e.g. different mass proxies for the total mass) and the use of different instruments (e.g. gas temperatures from Chandra instead of XMM-Newton). Moreover, because of the mass dependence of the gas fraction, the range of masses considered in each sample has an impact on the slope, normalization, and probably also on the evolution of the different relations.  

$\bullet$ There is a hint for a different redshift evolution in relaxed and disturbed clusters. When the core regions are removed, the $\gamma$ values determined for the two subsamples tend to be in better agreement and also in better agreement with the self-similar predictions. 

$\bullet$ The positive redshift evolution of the $M_{tot}$-$T$ relation suggests an evolution of the kinetic-to-thermal energy ratio of the ICM in clusters. Higher redshift clusters are on average more disturbed, so that the contribution to the non-thermal pressure by large scale motions is larger. The $\gamma$ value obtained for relaxed clusters is in better agreement with the self-similar prediction than the one obtained for the most disturbed systems, in  support of this scenario.  

$\bullet$ The $M_{tot}$-$M_{gas}$ and in particular the $M$-$Y_X$ relations show the weakest dependence on the dynamical state of the systems, as suggested by numerical simulations. Both relations show consistent slopes (although shallower than the self-similar predictions) and normalizations. Moreover they also show a redshift evolution in relatively good agreement with the self-similar expectations.

$\bullet$ The intrinsic scatter of the relations derived for the relaxed cluster subsample is smaller than the one derived for the disturbed subsample. Moreover, removing the central regions of the clusters further reduces the scatter, particularly for the most relaxed systems.

\acknowledgments
The paper is based on observations obtained with XMM-Newton, an ESA science mission with instruments and contributions directly funded by ESA Member States and NASA. 
We thank the anonymous referee for his/her review of the manuscript.  We also acknowledge useful discussions with M. Arnaud. LL acknowledges support from NASA through contracts 80NSSCK0582 and 80NSSC19K0116. W.R.F. and C.J. are supported by the Smithsonian Institution. G.S. acknowledges support from NASA through contracts AR9-20013X.  SE and MS acknowledge financial contribution from the contracts ASI 2015-046-R.0 and ASI-INAF n.2017-14-H.0, and from INAF ``Call per interventi aggiuntivi a sostegno della ricerca di main stream di INAF". GWP acknowledges funding from the European Research Council under the European Union's Seventh Framework Programme (FP7/2007-2013)/ERC grant agreement No. 340519, and from the French space agency \hypertarget{foo}{CNES}.\\


\vspace{5mm}
\facilities{XMM-Newton}


\software{XSPEC (\citealt{1996ASPC..101...17A}), SAS (v16.0.0; \citealt{2004ASPC..314..759G})}          



\appendix

\section{Cluster properties}\label{sect:partable}
\hypertarget{foo}{All the X-ray properties used in this paper and calculated within $R_{500}$ are listed in Table} \hyperlink{foo}{A1}.

\renewcommand{\thetable}{A1}
\begin{longtable*}{c @{\hspace{0.3\tabcolsep}}  c @{\hspace{0.3\tabcolsep}} c @{\hspace{0.3\tabcolsep}} c @{\hspace{0.3\tabcolsep}} c @{\hspace{0.3\tabcolsep}} c @{\hspace{0.3\tabcolsep}} c @{\hspace{0.3\tabcolsep}} c @{\hspace{0.3\tabcolsep}} c @{\hspace{0.3\tabcolsep}} c @{\hspace{0.3\tabcolsep}} c @{\hspace{0.3\tabcolsep}} c @{\hspace{0.3\tabcolsep}} c @{\hspace{0.3\tabcolsep}} c }
\caption{Cluster Properties.  -- Col. (1): Planck Cluster Name as in the Planck Early Sunyaev-Zeldovich catalog (\citealt{2011A&A...536A...8P}). Col. (2): Cluster redshift. Cols. (3)-(4): total and gas mass within $R_{500}$. Cols. (5)-(6): core-included and core-excluded cluster temperature. Cols. (7)-(8): core-included and core-excluded cluster luminosity in the soft band (0.1-2.4 keV).  Cols. (9)-(10): core-included and core-excluded cluster bolometric luminosity (0.01-100 keV). Cols. (12): number of bins in the temperature profile. Col. (12): fraction of $R_{500}$ covered by the temperature profile. Col. (13): number of bins in the surface brightness profile. Col. (14): fraction of $R_{500}$ covered by the surface brightness profile.}\\

\hline\hline
Planck Name &  z & M$_{500}$ & M$_{g,500}$ & kT & kT$_{exc}$ & L$_X$ & L$_{X,exc}$ & L$_{bol}$ & L$_{bol,exc}$ & N$_{T}$ & f$_{T}$ & N$_{sb}$ & f$_{sb}$ \\
&   & [10$^{14}$M$_{\odot}$] & [10$^{13}$M$_{\odot}$] & [keV] & [keV] & [erg s$^{-1}$] & [erg s$^{-1}$] & [erg s$^{-1}$] & [erg s$^{-1}$] & & & &\\
\hline
\endfirsthead
\multicolumn{11}{c}%
{\tablename\ \thetable{} -- continued from previous page} \\
\hline\hline 
Planck Name & z & M$_{500}$ & M$_{g,500}$ & kT & kT$_{exc}$ & L$_X$ & L$_{X,exc}$ & L$_{bol}$ & L$_{bol,exc}$ & N$_{T}$ & f$_{T}$ & N$_{sb}$ & f$_{sb}$  \\
&   & [10$^{14}$M$_{\odot}$] & [10$^{13}$M$_{\odot}$] & [keV] & [keV] & [erg s$^{-1}$] & [erg s$^{-1}$] & [erg s$^{-1}$] & [erg s$^{-1}$] & & & & \\
\hline
\endhead
G000.44-41.83  & 0.165 & 5.01$^{+0.55}_{-0.48}$ & 6.61$^{+0.36}_{-0.33}$ & 5.85$^{+0.32}_{-0.32}$ & 5.82$^{+0.43}_{-0.43}$ & 3.79$^{+0.25}_{-0.25}$ & 2.88$^{+0.13}_{-0.13}$ & 8.36$^{+0.78}_{-0.78}$ & 6.36$^{+0.47}_{-0.47}$ & 6 & 1.00 & 20 & 1.21\\
G002.74-56.18 & 0.141 & 4.96$^{+0.43}_{-0.28}$ & 5.63$^{+0.21}_{-0.14}$ & 5.36$^{+0.12}_{-0.12}$ & 5.39$^{+0.18}_{-0.18}$ & 4.47$^{+0.16}_{-0.16}$ & 2.71$^{+0.07}_{-0.07}$ & 9.96$^{+0.49}_{-0.49}$ & 6.03$^{+0.23}_{-0.23}$ & 10 & 0.94 & 49 & 1.08 \\
G003.90-59.41 & 0.151 & 6.94$^{+0.19}_{-0.19}$ & 8.65$^{+0.08}_{-0.08}$ & 7.06$^{+0.13}_{-0.13}$ & 6.46$^{+0.14}_{-0.14}$ & 7.13$^{+0.24}_{-0.24}$ & 4.61$^{+0.17}_{-0.17}$ & 18.06$^{+0.67}_{-0.67}$ & 11.62$^{+0.43}_{-0.43}$ & 10 & 0.98 & 63 & 1.08\\
G006.70-35.54  & 0.089 & 2.42$^{+0.04}_{-0.03}$ & 4.27$^{+0.05}_{-0.08}$ & 4.72$^{+0.08}_{-0.08}$ & 4.62$^{+0.09}_{-0.09}$ & 2.44$^{+0.09}_{-0.09}$ & 2.08$^{+0.08}_{-0.08}$ & 4.67$^{+0.25}_{-0.25}$ & 3.98$^{+0.21}_{-0.21}$ & 15 & 1.14 &45 & 1.18\\
G006.78+30.46  & 0.203 & 17.56$^{+0.28}_{-0.27}$ & 32.00$^{+0.24}_{-0.22}$ & 14.37$^{+0.12}_{-0.12}$ & 15.01$^{+0.16}_{-0.16}$ & 23.84$^{+0.36}_{-0.36}$ & 16.55$^{+0.13}_{-0.13}$ & 86.79$^{+1.74}_{-1.74}$ & 60.27$^{+0.84}_{-0.84}$ & 19 & 1.26 & 171 & 1.62 \\
G008.44-56.35  & 0.149 & 3.61$^{+0.08}_{-0.07}$ & 4.64$^{+0.05}_{-0.04}$ & 5.12$^{+0.08}_{-0.08}$ & 4.91$^{+0.10}_{-0.10}$ & 3.47$^{+0.12}_{-0.12}$ & 2.31$^{+0.08}_{-0.08}$ & 6.78$^{+0.34}_{-0.34}$ & 4.54$^{+0.22}_{-0.22}$ & 8 & 1.10 & 41 & 1.24  \\
G008.93-81.23  & 0.307 & 10.39$^{+0.24}_{-0.23}$ & 15.64$^{+0.13}_{-0.13}$ & 8.78$^{+0.09}_{-0.09}$ & 8.43$^{+0.10}_{-0.10}$ & 13.12$^{+0.31}_{-0.31}$ & 9.72$^{+0.17}_{-0.17}$ & 36.54$^{+1.21}_{-1.21}$ & 27.09$^{+0.70}_{-0.70}$ & 8 & 0.97 & 78 & 1.03 \\
G021.09+33.25 & 0.151 & 6.88$^{+0.20}_{-0.18}$ & 10.45$^{+0.32}_{-0.35}$ & 6.25$^{+0.14}_{-0.14}$ & 8.77$^{+0.31}_{-0.31}$ & 16.07$^{+0.18}_{-0.18}$ & 6.84$^{+0.10}_{-0.10}$ & 38.83$^{+0.55}_{-0.55}$ & 16.51$^{+0.25}_{-0.25}$ & 10 & 1.21 & 86 & 1.24\\
G036.72+14.92$^{*}$  & 0.152 & 5.21$^{+0.32}_{-0.29}$ & 8.16$^{+0.29}_{-0.28}$ & 7.44$^{+0.66}_{-0.66}$ & 6.89$^{+1.02}_{-1.02}$ & 5.97$^{+0.27}_{-0.27}$ & 4.00$^{+0.17}_{-0.17}$ & 15.59$^{+1.11}_{-1.11}$ & 10.43$^{+0.71}_{-0.71}$ & 3 & 0.51 & 9 & 0.87\\
G039.85-39.98  & 0.176 & 3.77$^{+0.47}_{-0.17}$ & 5.32$^{+0.36}_{-0.17}$ & 5.89$^{+0.14}_{-0.14}$ & 5.82$^{+0.16}_{-0.16}$ & 2.91$^{+0.12}_{-0.12}$ & 2.02$^{+0.10}_{-0.10}$ & 6.54$^{+0.32}_{-0.32}$ & 5.69$^{+0.28}_{-0.28}$ & 7 & 1.22 & 71 & 1.43\\
G042.82+56.61 & 0.072 & 4.65$^{+0.15}_{-0.13}$ & 6.01$^{+0.08}_{-0.07}$ & 4.83$^{+0.07}_{-0.07}$ & 4.79$^{+0.09}_{-0.09}$ & 3.18$^{+0.08}_{-0.08}$ & 2.24$^{+0.05}_{-0.05}$ & 6.74$^{+0.22}_{-0.22}$ & 4.74$^{+0.14}_{-0.14}$ & 15 & 0.74 & 42 & 0.88\\
G046.08+27.18 & 0.389 & 6.26$^{+0.61}_{-0.48}$ & 9.90$^{+0.35}_{-0.30}$ & 5.93$^{+0.30}_{-0.30}$ & 5.65$^{+0.29}_{-0.29}$ & 7.51$^{+0.41}_{-0.41}$ & 6.26$^{+0.28}_{-0.28}$ & 17.12$^{+1.47}_{-1.47}$ & 14.28$^{+1.10}_{-1.10}$ & 3 & 1.08 & 15 & 1.17\\
G046.50-49.43  & 0.085 & 6.05$^{+0.28}_{-0.26}$ & 6.24$^{+0.09}_{-0.08}$ & 5.59$^{+0.12}_{-0.12}$ & 5.21$^{+0.14}_{-0.14}$ & 3.34$^{+0.12}_{-0.12}$ & 2.28$^{+0.07}_{-0.07}$ & 7.95$^{+0.39}_{-0.39}$ & 5.43$^{+0.23}_{-0.23}$ & 15 & 0.67 & 58 & 0.75\\
G049.20+30.86 & 0.164 & 5.86$^{+0.16}_{-0.16}$ & 6.51$^{+0.07}_{-0.07}$ & 5.47$^{+0.08}_{-0.08}$ & 6.28$^{+0.17}_{-0.17}$ & 8.70$^{+0.18}_{-0.18}$ & 3.83$^{+0.06}_{-0.06}$ & 22.46$^{+0.74}_{-0.74}$ & 9.89$^{+0.29}_{-0.29}$ & 8 & 0.99 & 24 & 0.99\\
G049.33+44.38  & 0.097 & 3.84$^{+0.46}_{-0.38}$ & 3.85$^{+0.18}_{-0.17}$ & 4.90$^{+0.17}_{-0.17}$ & 4.80$^{+0.23}_{-0.23}$ & 1.68$^{+0.08}_{-0.08}$ & 1.22$^{+0.04}_{-0.04}$ & 3.60$^{+0.23}_{-0.23}$ & 2.63$^{+0.13}_{-0.13}$ & 8 & 0.61 & 26 & 0.82 \\
G049.66-49.50  & 0.098 & 4.81$^{+0.21}_{-0.19}$ & 4.41$^{+0.08}_{-0.08}$ & 4.64$^{+0.08}_{-0.08}$ & 4.81$^{+0.14}_{-0.14}$ & 3.44$^{+0.12}_{-0.12}$ & 1.74$^{+0.05}_{-0.05}$ & 7.26$^{+0.33}_{-0.33}$ & 3.67$^{+0.15}_{-0.15}$ & 10 & 0.68 & 36 & 0.68 \\
G053.52+59.54 & 0.113 & 5.87$^{+0.20}_{-0.15}$ & 7.03$^{+0.10}_{-0.08}$ & 6.58$^{+0.14}_{-0.14}$ & 6.31$^{+0.16}_{-0.16}$ & 4.08$^{+0.11}_{-0.11}$ & 3.08$^{+0.04}_{-0.04}$ & 9.74$^{+0.40}_{-0.40}$ & 7.33$^{+0.20}_{-0.20}$ & 5 & 0.81 & 38 & 0.96 \\
G055.60+31.86  & 0.224 & 6.54$^{+0.18}_{-0.15}$ & 9.41$^{+0.12}_{-0.10}$ & 7.39$^{+0.11}_{-0.11}$ & 7.30$^{+0.14}_{-0.14}$ & 10.06$^{+0.30}_{-0.30}$ & 5.34$^{+0.21}_{-0.21}$ & 25.87$^{+0.85}_{-0.85}$ & 13.65$^{+0.56}_{-0.56}$ & 9 & 1.03 & 53 & 1.10 \\
G055.97-34.88 & 0.124 & 5.41$^{+0.45}_{-0.37}$ & 3.44$^{+0.09}_{-0.09}$ & 6.25$^{+0.37}_{-0.37}$ & 5.50$^{+0.37}_{-0.37}$ & 2.11$^{+0.08}_{-0.08}$ & 1.54$^{+0.07}_{-0.07}$ & 5.47$^{+0.32}_{-0.32}$ & 4.01$^{+0.22}_{-0.22}$ & 7 & 0.42 & 18 & 0.51 \\
G056.81+36.31 & 0.095 & 3.98$^{+0.08}_{-0.07}$ & 5.12$^{+0.04}_{-0.04}$ & 4.98$^{+0.05}_{-0.05}$ & 4.83$^{+0.07}_{-0.07}$ & 4.34$^{+0.12}_{-0.12}$ & 2.35$^{+0.06}_{-0.06}$ & 9.30$^{+0.31}_{-0.31}$ & 5.05$^{+0.17}_{-0.17}$ & 15 & 1.01 & 99 & 1.10\\
G056.96-55.07 & 0.447 & 9.62$^{+0.25}_{-0.25}$ & 15.23$^{+0.14}_{-0.14}$ & 7.63$^{+0.12}_{-0.12}$ & 7.58$^{+0.13}_{-0.13}$ & 14.61$^{+0.39}_{-0.39}$ & 11.98$^{+0.30}_{-0.30}$ & 38.36$^{+1.50}_{-1.50}$ & 31.48$^{+1.16}_{-1.16}$ & 5 & 0.91 & 48 & 0.94\\
G057.26-45.35 & 0.397 & 13.17$^{+0.64}_{-0.65}$ & 18.61$^{+0.29}_{-0.28}$ & 10.39$^{+0.33}_{-0.33}$ & 9.99$^{+0.44}_{-0.44}$ & 25.32$^{+0.73}_{-0.73}$ & 12.75$^{+0.27}_{-0.27}$ & 82.81$^{+4.14}_{-4.14}$ & 41.70$^{+1.75}_{-1.75}$ & 5 & 0.91 & 32 & 1.00\\
G058.28+18.59 & 0.065 & 3.72$^{+0.06}_{-0.07}$ & 4.54$^{+0.20}_{-0.05}$ & 5.01$^{+0.04}_{-0.04}$ & 4.87$^{+0.05}_{-0.05}$ & 2.01$^{+0.04}_{-0.04}$ & 1.66$^{+0.02}_{-0.02}$ & 4.03$^{+0.11}_{-0.11}$ & 3.31$^{+0.08}_{-0.08}$ & 19 & 1.01 & 65 & 1.07\\
G062.42-46.41 & 0.091 & 3.26$^{+0.39}_{-0.15}$ & 2.87$^{+0.13}_{-0.12}$ & 4.25$^{+0.08}_{-0.08}$ & 4.28$^{+0.12}_{-0.12}$ & 1.42$^{+0.05}_{-0.05}$ & 1.15$^{+0.04}_{-0.04}$ & 2.65$^{+0.13}_{-0.13}$ & 2.14$^{+0.10}_{-0.10}$ & 8 & 0.41 & 30 & 0.56\\
G067.23+67.46  & 0.171 & 8.11$^{+0.17}_{-0.18}$ & 10.44$^{+0.08}_{-0.12}$ & 8.16$^{+0.13}_{-0.13}$ & 7.65$^{+0.17}_{-0.17}$ & 10.96$^{+0.16}_{-0.16}$ & 6.90$^{+0.27}_{-0.27}$ & 30.38$^{+0.63}_{-0.63}$ & 19.11$^{+0.55}_{-0.55}$ & 14 & 1.03 & 71 & 1.08 \\
G071.61+29.79 & 0.157 & 4.24$^{+0.76}_{-0.59}$ & 5.32$^{+0.51}_{-0.43}$ & 4.88$^{+0.36}_{-0.36}$ & 4.93$^{+0.42}_{-0.42}$ & 2.44$^{+0.15}_{-0.15}$ & 2.20$^{+0.12}_{-0.12}$ & 4.58$^{+0.39}_{-0.39}$ & 4.13$^{+0.32}_{-0.32}$ & 4 & 0.95 & 15 & 1.05 \\
G072.63+41.46 & 0.228 & 10.71$^{+0.42}_{-0.39}$ & 16.51$^{+0.26}_{-0.25}$ & 8.95$^{+0.25}_{-0.25}$ & 8.91$^{+0.30}_{-0.30}$ & 14.61$^{+0.50}_{-0.50}$ & 10.44$^{+0.29}_{-0.29}$ & 45.16$^{+2.48}_{-2.48}$ & 32.30$^{+1.58}_{-1.58}$ & 9 & 0.97 & 45 & 0.97   \\
G072.80-18.72 & 0.143 & 5.33$^{+0.17}_{-0.12}$ & 8.25$^{+0.11}_{-0.08}$ & 5.93$^{+0.09}_{-0.09}$ & 5.99$^{+0.12}_{-0.12}$ & 5.69$^{+0.19}_{-0.19}$ & 3.99$^{+0.13}_{-0.13}$ & 13.32$^{+0.59}_{-0.59}$ & 9.35$^{+0.38}_{-0.38}$ & 12 & 0.93 & 45 & 0.93\\
G073.96-27.82 & 0.233 & 11.41$^{+0.45}_{-0.38}$ & 17.19$^{+0.28}_{-0.24}$ & 9.53$^{+0.24}_{-0.24}$ & 10.67$^{+0.47}_{-0.47}$ & 16.85$^{+0.36}_{-0.36}$ & 10.00$^{+0.20}_{-0.20}$ & 55.79$^{+1.80}_{-1.80}$ & 33.26$^{+0.90}_{-0.90}$ & 9 & 0.94 & 17 & 1.01 \\
G080.38-33.20  & 0.107 & 3.66$^{+0.07}_{-0.06}$ & 3.83$^{+0.03}_{-0.03}$ & 5.41$^{+0.07}_{-0.07}$ & 4.82$^{+0.08}_{-0.08}$ & 2.19$^{+0.04}_{-0.04}$ & 1.46$^{+0.02}_{-0.02}$ & 4.87$^{+0.12}_{-0.12}$ & 3.26$^{+0.06}_{-0.06}$ & 12 & 0.97 & 99 & 1.15\\
G080.99-50.90 & 0.300 & 6.62$^{+0.32}_{-0.23}$ & 9.55$^{+0.20}_{-0.17}$ & 7.15$^{+0.25}_{-0.25}$ & 6.92$^{+0.24}_{-0.24}$ & 8.49$^{+0.42}_{-0.42}$ & 5.55$^{+0.19}_{-0.19}$ & 23.28$^{+1.60}_{-1.60}$ & 15.18$^{+0.62}_{-0.62}$ & 6 & 0.94 & 20 & 1.19\\
G083.28-31.03 & 0.412 & 7.68$^{+0.47}_{-0.38}$ & 12.48$^{+0.30}_{-0.25}$ & 7.24$^{+0.31}_{-0.31}$ & 7.02$^{+0.34}_{-0.34}$ & 11.78$^{+0.40}_{-0.40}$ & 8.31$^{+0.17}_{-0.17}$ & 30.61$^{+1.90}_{-1.90}$ & 21.26$^{+0.83}_{-0.83}$ & 6 & 1.17 & 19 & 1.50 \\
G085.99+26.71 & 0.179 & 3.31$^{+0.68}_{-0.57}$ & 3.97$^{+0.47}_{-0.41}$ & 4.57$^{+0.26}_{-0.26}$ & 4.51$^{+0.29}_{-0.29}$ & 1.82$^{+0.14}_{-0.14}$ & 1.63$^{+0.12}_{-0.12}$ & 3.69$^{+0.41}_{-0.41}$ & 3.31$^{+0.35}_{-0.35}$ & 4 & 0.66 & 13 & 1.18\\
G086.45+15.29$^{*}$ & 0.260 & 5.74$^{+0.33}_{-0.25}$ & 9.30$^{+0.21}_{-0.18}$ & 7.14$^{+0.29}_{-0.29}$ & 6.82$^{+0.38}_{-0.38}$ & 11.38$^{+0.41}_{-0.41}$ & 6.45$^{+0.10}_{-0.10}$ & 27.51$^{+1.40}_{-1.40}$ & 17.61$^{+0.42}_{-0.42}$ & 7 & 1.10 & 22 & 1.16\\
G092.73+73.46 & 0.228 & 6.18$^{+0.32}_{-0.28}$ & 10.72$^{+0.29}_{-0.26}$ & 7.13$^{+0.19}_{-0.19}$ & 7.09$^{+0.21}_{-0.21}$ & 7.74$^{+0.31}_{-0.31}$ & 5.72$^{+0.15}_{-0.15}$ & 19.00$^{+1.08}_{-1.08}$ & 10.90$^{+0.41}_{-0.41}$ & 9 & 1.17 & 27 & 1.26\\
G093.91+34.90 & 0.081 & 4.68$^{+0.71}_{-0.50}$ & 5.49$^{+0.44}_{-0.33}$ & 5.36$^{+0.81}_{-0.81}$ & 5.31$^{+0.21}_{-0.21}$ & 3.43$^{+0.24}_{-0.24}$ & 2.86$^{+0.19}_{-0.19}$ & 7.57$^{+0.66}_{-0.66}$ & 5.46$^{+0.36}_{-0.36}$ & 7 & 0.85 & 44 & 1.07\\
G096.87+24.21 & 0.300 & 5.47$^{+2.40}_{-1.39}$ & 6.28$^{+1.08}_{-0.89}$ & 5.05$^{+0.54}_{-0.54}$ & 4.99$^{+0.60}_{-0.60}$ & 3.04$^{+0.32}_{-0.32}$ & 2.65$^{+0.23}_{-0.23}$ & 6.80$^{+0.97}_{-0.97}$ & 5.94$^{+0.74}_{-0.74}$ & 2 & 0.63 & 17 & 0.95\\
G097.73+38.11 & 0.171 & 4.21$^{+0.21}_{-0.20}$ & 6.87$^{+0.17}_{-0.16}$ & 6.38$^{+0.11}_{-0.11}$ & 6.07$^{+0.14}_{-0.14}$ & 5.24$^{+0.13}_{-0.13}$ & 3.77$^{+0.05}_{-0.05}$ & 12.26$^{+0.43}_{-0.43}$ & 8.84$^{+0.19}_{-0.19}$ & 9 & 1.00 & 18 & 1.07\\
G098.95+24.86 & 0.093 & 3.25$^{+0.35}_{-0.15}$ & 2.99$^{+0.14}_{-0.07}$ & 4.97$^{+0.20}_{-0.20}$ & 4.94$^{+0.28}_{-0.28}$ & 1.28$^{+0.06}_{-0.06}$ & 0.86$^{+0.03}_{-0.03}$ & 2.77$^{+0.19}_{-0.19}$ & 1.87$^{+0.11}_{-0.11}$ & 8 & 0.72 & 22 & 0.74 \\
G106.73-83.22 & 0.292 & 5.18$^{+0.36}_{-0.29}$ & 8.31$^{+0.26}_{-0.22}$ & 6.70$^{+0.25}_{-0.25}$ & 6.53$^{+0.28}_{-0.28}$ & 8.28$^{+0.30}_{-0.30}$ & 5.68$^{+0.10}_{-0.10}$ & 18.64$^{+1.27}_{-1.27}$ & 12.80$^{+0.50}_{-0.50}$ & 4 & 0.51 & 14 & 0.78 \\
G107.11+65.31 & 0.292 & 7.26$^{+0.75}_{-0.67}$ & 6.94$^{+0.30}_{-0.28}$ & 6.64$^{+0.49}_{-0.49}$ & 6.78$^{+0.69}_{-0.69}$ & 10.02$^{+0.56}_{-0.56}$ & 9.03$^{+0.42}_{-0.42}$ & 27.55$^{+3.06}_{-3.06}$ & 24.83$^{+2.38}_{-2.38}$ & 6 & 0.8 & 25 & 1.00\\
G113.82+44.35 & 0.225 & 3.87$^{+0.32}_{-0.21}$ & 5.70$^{+0.28}_{-0.22}$ & 5.29$^{+0.30}_{-0.30}$ & 5.43$^{+0.34}_{-0.34}$ & 3.73$^{+0.22}_{-0.22}$ & 3.14$^{+0.14}_{-0.14}$ & 7.95$^{+0.68}_{-0.68}$ & 6.68$^{+0.45}_{-0.45}$ & 4 & 0.64 & 19 & 0.89 \\
G124.21-36.48 & 0.197 & 5.81$^{+0.57}_{-0.62}$ & 7.81$^{+0.29}_{-0.33}$ & 4.78$^{+0.11}_{-0.11}$ & 5.81$^{+0.16}_{-0.16}$ & 3.89$^{+0.14}_{-0.14}$ & 2.42$^{+0.06}_{-0.06}$ & 7.91$^{+0.34}_{-0.34}$ & 4.94$^{+0.16}_{-0.16}$ & 9 & 0.81 & 23 & 1.23\\
G125.70+53.85 & 0.302 & 6.78$^{+0.69}_{-0.65}$ & 9.64$^{+0.43}_{-0.43}$ & 6.84$^{+0.44}_{-0.44}$ & 6.99$^{+0.58}_{-0.58}$ & 6.94$^{+0.39}_{-0.39}$ & 4.93$^{+0.14}_{-0.14}$ & 18.07$^{+1.81}_{-1.81}$ & 12.02$^{+0.70}_{-0.70}$ & 5 & 0.96 & 14 & 1.15\\
G139.19+56.35 & 0.322 & 7.42$^{+5.68}_{-1.68}$ & 10.23$^{+1.42}_{-0.80}$ & 6.10$^{+0.57}_{-0.57}$ & 6.32$^{+0.69}_{-0.69}$ & 6.47$^{+0.56}_{-0.56}$ & 5.45$^{+0.38}_{-0.38}$ & 16.18$^{+1.83}_{-1.83}$ & 13.66$^{+1.07}_{-1.07}$ & 4 & 0.39 & 15 & 0.63 \\
G149.73+34.69 & 0.182 & 7.12$^{+0.71}_{-0.63}$ & 11.68$^{+0.57}_{-0.55}$ & 7.39$^{+0.42}_{-0.42}$ & 7.15$^{+0.51}_{-0.51}$ & 8.46$^{+0.36}_{-0.36}$ & 5.96$^{+0.15}_{-0.15}$ & 22.10$^{+1.79}_{-1.79}$ & 15.61$^{+0.81}_{-0.81}$  & 6 & 0.59 & 11 & 0.72 \\
G157.43+30.33 & 0.450 & 6.23$^{+0.52}_{-0.42}$ & 9.64$^{+0.35}_{-0.31}$ & 7.54$^{+0.58}_{-0.58}$ & 7.06$^{+0.59}_{-0.59}$ & 8.01$^{+0.27}_{-0.27}$ & 6.15$^{+0.23}_{-0.23}$ & 21.05$^{+1.57}_{-1.57}$ & 16.26$^{+1.14}_{-1.14}$ & 3 & 1.00 & 13 & 1.13 \\
G159.85-73.47 & 0.206 & 6.73$^{+0.64}_{-0.70}$ & 9.88$^{+0.41}_{-0.47}$ & 5.89$^{+0.57}_{-0.57}$ & 5.82$^{+0.83}_{-0.83}$ & 6.86$^{+0.27}_{-0.27}$ & 4.80$^{+0.11}_{-0.11}$ & 16.15$^{+0.90}_{-0.90}$ & 11.29$^{+0.37}_{-0.37}$ & 10 & 1.09 & 34 & 1.09\\
G164.18-38.89$^{*}$ & 0.074 & 5.07$^{+0.45}_{-0.22}$ & 6.30$^{+0.28}_{-0.15}$ & 6.52$^{+0.15}_{-0.15}$ & 6.50$^{+0.18}_{-0.18}$ & 4.16$^{+0.19}_{-0.19}$ & 3.30$^{+0.13}_{-0.13}$ & 8.83$^{+0.47}_{-0.47}$ & 7.01$^{+0.34}_{-0.34}$ & 13 & 0.70 & 66 & 1.01 \\
G166.13+43.39  & 0.217 & 6.86$^{+0.48}_{-0.41}$ & 8.68$^{+0.22}_{-0.19}$ & 6.56$^{+0.21}_{-0.21}$ & 6.30$^{+0.29}_{-0.29}$ & 6.83$^{+0.22}_{-0.22}$ & 4.18$^{+0.11}_{-0.11}$ & 16.37$^{+0.51}_{-0.51}$ & 10.02$^{+0.37}_{-0.37}$ & 6 & 0.76 & 15 &0.83\\
G167.65+17.64$^{*}$ & 0.174 & 5.88$^{+0.40}_{-0.30}$ & 9.25$^{+0.26}_{-0.21}$ & 6.28$^{+0.20}_{-0.20}$ & 6.02$^{+0.23}_{-0.23}$ & 6.41$^{+0.25}_{-0.25}$ & 4.74$^{+0.11}_{-0.11}$ & 13.84$^{+0.76}_{-0.76}$ & 10.24$^{+0.36}_{-0.36}$ & 11 & 0.77 & 29 & 0.97\\
G171.94-40.65 & 0.270 & 12.33$^{+3.45}_{-2.12}$ & 15.02$^{+0.96}_{-0.75}$ & 9.79$^{+0.45}_{-0.45}$ & 9.37$^{+0.49}_{-0.49}$ & 11.21$^{+0.36}_{-0.36}$ & 7.32$^{+0.18}_{-0.18}$ & 32.74$^{+2.28}_{-2.28}$ & 21.47$^{+0.26}_{-0.26}$ & 4 & 0.34 & 11 & 0.54\\
G180.24+21.04 & 0.546 & 12.58$^{+0.53}_{-0.68}$ & 23.47$^{+0.36}_{-0.50}$ & 10.12$^{+0.25}_{-0.25}$ & 9.96$^{+0.27}_{-0.27}$ & 23.81$^{+0.78}_{-0.78}$ & 17.95$^{+0.47}_{-0.47}$ & 72.74$^{+3.61}_{-3.61}$ & 54.55$^{+2.36}_{-2.36}$ & 5 & 0.95 & 33 & 1.58 \\
G182.44-28.29 & 0.088 & 6.93$^{+0.17}_{-0.18}$ & 9.58$^{+0.11}_{-0.20}$ & 7.53$^{+0.12}_{-0.12}$ & 7.13$^{+0.32}_{-0.32}$ & 12.18$^{+0.11}_{-0.11}$ & 4.69$^{+0.01}_{-0.01}$ & 33.55$^{+0.44}_{-0.44}$ & 12.89$^{+0.08}_{-0.08}$ & 26 & 0.99 & 65 & 1.10 \\
G182.63+55.82  & 0.206 & 4.77$^{+0.26}_{-0.16}$ & 6.88$^{+0.15}_{-0.10}$ & 5.71$^{+0.11}_{-0.11}$ & 5.54$^{+0.15}_{-0.15}$ & 6.06$^{+0.20}_{-0.20}$ & 3.18$^{+0.07}_{-0.07}$ & 13.90$^{+0.65}_{-0.65}$ & 7.28$^{+0.25}_{-0.25}$ & 8 & 0.97 & 40 & 1.34\\
G186.39+37.25  & 0.282 & 9.28$^{+1.19}_{-0.89}$ & 12.49$^{+0.61}_{-0.52}$ & 8.77$^{+1.40}_{-1.40}$ & 7.38$^{+1.58}_{-1.58}$ & 11.61$^{+0.60}_{-0.60}$ & 8.27$^{+0.50}_{-0.50}$ & 30.83$^{+2.56}_{-2.56}$ & 22.24$^{+1.51}_{-1.51}$ & 1 & 0.57 & 4 & 0.84\\
G195.62+44.05  & 0.295 & 5.82$^{+0.72}_{-0.49}$ & 8.77$^{+0.49}_{-0.36}$ & 5.26$^{+0.13}_{-0.13}$ & 5.26$^{+0.15}_{-0.15}$ & 4.82$^{+0.21}_{-0.21}$ & 4.10$^{+0.15}_{-0.15}$ & 10.30$^{+0.58}_{-0.58}$ & 8.77$^{+0.43}_{-0.43}$ & 6 & 0.97 & 32 & 1.06 \\
G195.77-24.30 & 0.203 & 6.67$^{+0.32}_{-0.22}$ & 11.14$^{+0.24}_{-0.16}$ & 6.80$^{+0.18}_{-0.18}$ & 6.84$^{+0.21}_{-0.21}$ & 6.95$^{+0.24}_{-0.24}$ & 5.84$^{+0.19}_{-0.19}$ & 17.45$^{+1.03}_{-1.03}$ & 14.67$^{+0.81}_{-0.81}$ & 12 & 1.03 & 47 & 1.26\\
G218.85+35.50  & 0.175 & 3.86$^{+0.45}_{-0.41}$ & 4.30$^{+0.20}_{-0.19}$ & 4.48$^{+0.31}_{-0.31}$ & 4.17$^{+0.54}_{-0.54}$ & 3.05$^{+0.17}_{-0.17}$ & 1.72$^{+0.06}_{-0.06}$ & 6.52$^{+0.46}_{-0.46}$ & 3.68$^{+0.18}_{-0.18}$ & 5 & 0.44 & 26 & 1.14 \\
G225.92-19.99 & 0.460 & 11.07$^{+3.23}_{-4.48}$ & 21.96$^{+2.86}_{-5.42}$ & 7.01$^{+0.33}_{-0.33}$ & 7.46$^{+0.42}_{-0.42}$ & 22.54$^{+1.69}_{-1.69}$ & 16.54$^{+1.24}_{-1.24}$ & 56.88$^{+5.50}_{-5.50}$ & 41.61$^{+3.16}_{-3.16}$ & 4 & 0.90 & 34 & 1.07 \\
G226.17-21.91 & 0.099 & 4.61$^{+0.32}_{-0.23}$ & 5.10$^{+0.11}_{-0.09}$ & 4.85$^{+0.11}_{-0.11}$ & 4.65$^{+0.14}_{-0.14}$ & 3.31$^{+0.12}_{-0.12}$ & 2.38$^{+0.07}_{-0.07}$ & 6.95$^{+0.34}_{-0.34}$ & 4.99$^{+0.20}_{-0.20}$ & 14 & 0.94 & 46 & 0.99 \\
G226.24+76.76 & 0.143 & 6.28$^{+0.12}_{-0.09}$ & 8.34$^{+0.06}_{-0.05}$ & 7.03$^{+0.06}_{-0.06}$ & 6.89$^{+0.08}_{-0.08}$ & 6.83$^{+0.10}_{-0.10}$ & 3.26$^{+0.02}_{-0.02}$ & 17.63$^{+0.35}_{-0.35}$ & 8.40$^{+0.07}_{-0.07}$ & 10 & 1.11 & 58 & 1.31\\
G228.15+75.19 & 0.545 & 7.94$^{+0.96}_{-0.60}$ & 13.54$^{+0.78}_{-0.57}$ & 9.46$^{+0.69}_{-0.69}$ & 9.20$^{+0.78}_{-0.78}$ & 15.35$^{+0.78}_{-0.78}$ & 9.64$^{+0.42}_{-0.42}$ & 43.85$^{+3.43}_{-3.43}$ & 34.24$^{+1.67}_{-1.67}$ & 3 & 1.07 & 17 & 1.16\\
G228.49+53.12 & 0.143 & 5.16$^{+0.35}_{-0.42}$ & 5.87$^{+0.17}_{-0.21}$ & 5.29$^{+0.17}_{-0.17}$ & 5.54$^{+0.37}_{-0.37}$ & 4.76$^{+0.16}_{-0.16}$ & 1.93$^{+0.03}_{-0.03}$ & 10.23$^{+0.46}_{-0.46}$ & 4.16$^{+0.11}_{-0.11}$ & 7 & 0.94 & 29 & 1.26\\
G229.21-17.24 & 0.171 & 5.26$^{+1.84}_{-1.19}$ & 6.56$^{+0.84}_{-0.68}$ & 5.72$^{+0.28}_{-0.28}$ & 5.65$^{+0.35}_{-0.35}$ & 2.90$^{+0.17}_{-0.17}$ & 2.33$^{+0.06}_{-0.06}$ & 6.39$^{+0.52}_{-0.52}$ & 5.14$^{+0.26}_{-0.26}$ & 8 & 0.75 & 20 & 0.81\\
G229.94+15.29  & 0.070 & 7.01$^{+0.25}_{-0.26}$ & 8.51$^{+0.12}_{-0.13}$ & 6.94$^{+0.15}_{-0.15}$ & 6.79$^{+0.28}_{-0.28}$ & 5.62$^{+0.09}_{-0.09}$ & 2.80$^{+0.02}_{-0.02}$ & 14.67$^{+0.35}_{-0.35}$ & 7.30$^{+0.11}_{-0.11}$ & 17 & 1.17 & 163 & 1.43\\
G236.95-26.67  & 0.148 & 5.91$^{+0.32}_{-0.33}$ & 6.96$^{+0.14}_{-0.14}$ & 5.79$^{+0.13}_{-0.13}$ & 5.57$^{+0.17}_{-0.17}$ & 3.96$^{+0.14}_{-0.14}$ & 2.42$^{+0.05}_{-0.05}$ & 9.20$^{+0.43}_{-0.43}$ & 5.61$^{+0.20}_{-0.20}$ & 10 & 0.83 & 52 & 0.79\\
G241.74-30.88 & 0.271 & 5.80$^{+0.42}_{-0.33}$ & 7.59$^{+0.25}_{-0.22}$ & 6.98$^{+0.35}_{-0.35}$ & 6.75$^{+0.52}_{-0.52}$ & 8.14$^{+0.28}_{-0.28}$ & 4.62$^{+0.11}_{-0.11}$ & 19.94$^{+1.15}_{-1.15}$ & 11.32$^{+0.52}_{-0.52}$ & 4 & 0.91 & 15 & 1.03 \\
G241.77-24.00 & 0.139 & 3.29$^{+0.06}_{-0.06}$ & 3.95$^{+0.04}_{-0.04}$ & 4.55$^{+0.06}_{-0.06}$ & 4.93$^{+0.12}_{-0.12}$ & 4.74$^{+0.14}_{-0.14}$ & 2.10$^{+0.07}_{-0.07}$ & 9.89$^{+0.32}_{-0.32}$ & 4.43$^{+0.15}_{-0.15}$ & 9 & 0.84 & 48 & 1.01 \\
G241.97+14.85 & 0.169 & 4.13$^{+0.64}_{-0.38}$ & 7.10$^{+0.69}_{-0.45}$ & 6.12$^{+0.10}_{-0.10}$ & 6.23$^{+0.11}_{-0.11}$ & 4.88$^{+0.33}_{-0.33}$ & 3.08$^{+0.18}_{-0.18}$ & 10.54$^{+0.82}_{-0.82}$ & 6.66$^{+0.47}_{-0.47}$ & 14 & 1.15 & 33 & 1.51 \\
G244.34-32.13 & 0.284 & 6.94$^{+0.54}_{-0.50}$ & 11.46$^{+0.39}_{-0.38}$ & 7.02$^{+0.29}_{-0.29}$ & 7.25$^{+0.41}_{-0.41}$ & 10.30$^{+0.39}_{-0.39}$ & 6.93$^{+0.21}_{-0.21}$ & 24.97$^{+1.74}_{-1.74}$ & 16.90$^{+1.06}_{-1.06}$ & 5 & 0.90 & 11 & 0.94\\
G244.69+32.49 & 0.153 & 3.70$^{+0.31}_{-0.19}$ & 5.04$^{+0.17}_{-0.12}$ & 5.20$^{+0.26}_{-0.26}$ & 5.01$^{+0.29}_{-0.29}$ & 3.18$^{+0.13}_{-0.13}$ & 2.39$^{+0.08}_{-0.08}$ & 6.54$^{+0.46}_{-0.46}$ & 4.91$^{+0.30}_{-0.30}$ & 5 & 0.63 & 11 & 0.70\\
G247.17-23.32 & 0.152 & 3.17$^{+0.54}_{-0.35}$ & 4.16$^{+0.37}_{-0.26}$ & 4.45$^{+0.24}_{-0.24}$ & 4.65$^{+0.38}_{-0.38}$ & 2.49$^{+0.14}_{-0.14}$ & 1.81$^{+0.08}_{-0.08}$ & 4.84$^{+0.40}_{-0.40}$ & 3.51$^{+0.25}_{-0.25}$ & 7 & 1.01 & 17 & 1.01\\
G249.87-39.86 & 0.165 & 2.86$^{+0.56}_{-0.40}$ & 4.02$^{+0.36}_{-0.28}$ & 3.97$^{+0.22}_{-0.22}$ & 3.91$^{+0.33}_{-0.33}$ & 2.33$^{+0.10}_{-0.10}$ & 1.51$^{+0.04}_{-0.04}$ & 3.91$^{+0.24}_{-0.24}$ & 2.53$^{+0.10}_{-0.10}$ & 5 & 0.51 & 15 & 0.84\\
G250.90-36.25 & 0.200 & 5.36$^{+0.51}_{-0.36}$ & 6.74$^{+0.27}_{-0.20}$ & 5.98$^{+0.23}_{-0.23}$ & 5.97$^{+0.36}_{-0.36}$ & 4.91$^{+0.18}_{-0.18}$ & 2.93$^{+0.08}_{-0.08}$ & 11.18$^{+0.66}_{-0.66}$ & 6.67$^{+0.31}_{-0.31}$ & 6 & 0.83 & 21 & 0.83  \\
G252.96-56.05 & 0.075 & 3.58$^{+0.04}_{-0.05}$ & 3.90$^{+0.03}_{-0.03}$ & 4.10$^{+0.03}_{-0.03}$ & 4.37$^{+0.08}_{-0.08}$ & 3.85$^{+0.04}_{-0.04}$ & 1.49$^{+0.01}_{-0.01}$ & 7.42$^{+0.10}_{-0.10}$ & 2.88$^{+0.01}_{-0.01}$ & 15 & 0.60 & 133 & 0.77 \\
G253.47-33.72 & 0.191 & 4.52$^{+0.44}_{-0.48}$ & 5.71$^{+0.30}_{-0.33}$ & 5.96$^{+0.38}_{-0.38}$ & 5.76$^{+0.52}_{-0.52}$ & 3.56$^{+0.18}_{-0.18}$ & 2.48$^{+0.11}_{-0.11}$ & 8.59$^{+0.70}_{-0.70}$ & 5.98$^{+0.43}_{-0.43}$ & 6 & 1.00  & 16 & 1.07\\
G256.45-65.71 & 0.220 & 5.42$^{+0.89}_{-0.76}$ & 7.18$^{+0.57}_{-0.54}$ & 4.94$^{+0.19}_{-0.19}$ & 5.73$^{+0.33}_{-0.33}$ & 5.74$^{+0.25}_{-0.25}$ & 3.63$^{+0.12}_{-0.12}$ & 11.96$^{+0.72}_{-0.72}$ & 7.54$^{+0.38}_{-0.38}$ & 7 & 1.01 & 22 & 1.17\\
G257.34-22.18 & 0.203 & 3.19$^{+0.88}_{-0.51}$ & 4.51$^{+0.82}_{-0.51}$ & 1.67$^{+1.34}_{-1.34}$ & 1.40$^{+1.33}_{-1.33}$ & 2.76$^{+0.34}_{-0.34}$ & 2.43$^{+0.31}_{-0.31}$ & 6.09$^{+1.06}_{-1.06}$ & 5.37$^{+0.94}_{-0.94}$ & 2 & 0.35 & 11 & 0.82\\
G260.03-63.44 & 0.284 & 7.15$^{+0.73}_{-0.73}$ & 10.03$^{+0.33}_{-0.29}$ & 6.43$^{+0.18}_{-0.18}$ & 6.76$^{+0.28}_{-0.28}$ & 12.55$^{+0.26}_{-0.26}$ & 7.60$^{+0.01}_{-0.01}$ & 28.88$^{+1.13}_{-1.13}$ & 17.53$^{+0.26}_{-0.26}$ & 5 & 0.72 & 22 & 1.17 \\
G262.25-35.36 & 0.295 & 6.59$^{+0.90}_{-0.70}$ & 10.80$^{+0.78}_{-0.67}$ & 7.90$^{+0.19}_{-0.19}$ & 7.86$^{+0.20}_{-0.20}$ & 6.94$^{+0.37}_{-0.37}$ & 5.89$^{+0.28}_{-0.28}$ & 17.22$^{+1.60}_{-1.60}$ & 14.66$^{+1.26}_{-1.26}$ & 5 & 0.68 & 12 & 0.95\\
G262.71-40.91 & 0.420 & 9.16$^{+2.11}_{-1.59}$ & 12.29$^{+0.78}_{-0.69}$ & 9.33$^{+0.39}_{-0.39}$ & 10.08$^{+0.68}_{-0.68}$ & 12.84$^{+0.44}_{-0.44}$ & 6.69$^{+0.16}_{-0.16}$ & 36.85$^{+2.32}_{-2.32}$ & 19.17$^{+0.10}_{-0.10}$ & 3 & 0.44 & 12 & 0.75\\
G263.16-23.41 & 0.227 & 7.07$^{+0.35}_{-0.28}$ & 11.22$^{+0.22}_{-0.18}$ & 7.18$^{+0.16}_{-0.16}$ & 7.43$^{+0.26}_{-0.26}$ & 10.55$^{+0.34}_{-0.34}$ & 5.49$^{+0.12}_{-0.12}$ & 27.27$^{+1.28}_{-1.28}$ & 14.18$^{+0.54}_{-0.54}$ & 8 & 0.97 & 45 & 1.06\\
G263.66-22.53  & 0.164 & 8.59$^{+0.65}_{-0.45}$ & 10.52$^{+0.29}_{-0.21}$ & 7.10$^{+0.18}_{-0.18}$ & 7.22$^{+0.28}_{-0.28}$ & 6.55$^{+0.23}_{-0.23}$ & 4.08$^{+0.11}_{-0.11}$ & 16.83$^{+0.88}_{-0.88}$ & 10.48$^{+0.47}_{-0.47}$ & 10 & 1.05 & 34 & 1.12\\
G266.03-21.25 & 0.296 & 12.56$^{+0.34}_{-0.34}$ & 21.79$^{+0.22}_{-0.23}$ & 10.57$^{+0.26}_{-0.26}$ & 10.66$^{+0.34}_{-0.34}$ & 21.79$^{+0.68}_{-0.68}$ & 15.16$^{+0.38}_{-0.38}$ & 65.46$^{+2.90}_{-2.90}$ & 45.55$^{+1.78}_{-1.78}$ & 10 & 1.03 & 59 & 1.32\\
G269.31-49.87 & 0.085 & 2.65$^{+0.41}_{-0.23}$ & 3.19$^{+0.26}_{-0.16}$ & 4.82$^{+0.25}_{-0.25}$ & 4.93$^{+0.39}_{-0.39}$ & 1.62$^{+0.08}_{-0.08}$ & 1.07$^{+0.03}_{-0.03}$ & 3.34$^{+0.25}_{-0.25}$ & 2.20$^{+0.14}_{-0.14}$ & 7 & 0.73 & 34 & 0.83 \\
G271.19-30.96 & 0.370 & 8.38$^{+0.68}_{-0.53}$ & 12.80$^{+0.49}_{-0.52}$ & 8.22$^{+0.24}_{-0.24}$ & 8.80$^{+0.46}_{-0.46}$ & 19.29$^{+0.67}_{-0.67}$ & 8.19$^{+0.03}_{-0.03}$ & 52.15$^{+3.21}_{-3.21}$ & 22.16$^{+0.49}_{-0.49}$ & 3 & 0.37 & 17 & 0.54\\
G271.50-56.55 & 0.300 & 8.07$^{+2.11}_{-1.51}$ & 9.79$^{+0.75}_{-0.61}$ & 7.10$^{+0.71}_{-0.71}$ & 7.02$^{+0.91}_{-0.91}$ & 8.44$^{+0.62}_{-0.62}$ & 5.18$^{+0.35}_{-0.35}$ & 24.85$^{+1.97}_{-1.97}$ & 16.30$^{+1.21}_{-1.21}$ & 4 & 0.57 & 36 & 0.83 \\
G272.10-40.15  & 0.059 & 6.11$^{+0.09}_{-0.08}$ & 7.79$^{+0.05}_{-0.05}$ & 6.35$^{+0.04}_{-0.04}$ & 6.10$^{+0.04}_{-0.04}$ & 5.02$^{+0.08}_{-0.08}$ & 4.01$^{+0.06}_{-0.06}$ & 12.62$^{+0.29}_{-0.29}$ & 10.10$^{+0.21}_{-0.21}$ & 33 & 0.89 & 372 & 0.95\\
G277.75-51.73 & 0.440 & 8.96$^{+0.73}_{-0.59}$ & 14.49$^{+0.54}_{-0.47}$ & 7.80$^{+0.23}_{-0.23}$ & 7.74$^{+0.25}_{-0.25}$ & 9.41$^{+0.38}_{-0.38}$ & 8.20$^{+0.29}_{-0.29}$ & 25.18$^{+1.51}_{-1.51}$ & 21.96$^{+1.21}_{-1.21}$ & 6 & 1.01 & 25 & 1.35 \\
G278.60+39.17 & 0.307 & 9.37$^{+0.87}_{-0.81}$ & 13.38$^{+0.56}_{-0.55}$ & 8.02$^{+0.35}_{-0.35}$ & 7.98$^{+0.46}_{-0.46}$ & 10.53$^{+0.38}_{-0.38}$ & 7.50$^{+0.26}_{-0.26}$ & 26.14$^{+1.23}_{-1.23}$ & 18.64$^{+0.80}_{-0.80}$ & 6 & 0.93 & 17 & 1.04\\
G280.19+47.81  & 0.156 & 6.53$^{+1.70}_{-1.09}$ & 7.44$^{+0.61}_{-0.47}$ & 6.99$^{+0.26}_{-0.26}$ & 6.94$^{+0.32}_{-0.32}$ & 3.13$^{+0.16}_{-0.16}$ & 2.50$^{+0.06}_{-0.06}$ & 7.48$^{+0.55}_{-0.55}$ & 5.99$^{+0.27}_{-0.27}$ & 8 & 0.66 & 22 & 0.89 \\
G282.49+65.17 & 0.077 & 5.25$^{+0.22}_{-0.20}$ & 6.64$^{+0.11}_{-0.10}$ & 5.54$^{+0.14}_{-0.14}$ & 5.41$^{+0.18}_{-0.18}$ & 2.97$^{+0.05}_{-0.05}$ & 2.19$^{+0.02}_{-0.02}$ & 6.82$^{+0.18}_{-0.18}$ & 5.02$^{+0.11}_{-0.11}$ & 16 & 1.05 & 141 & 1.12\\
G283.16-22.93 & 0.450 & 7.34$^{+1.14}_{-0.97}$ & 10.53$^{+0.65}_{-0.61}$ & 7.32$^{+0.36}_{-0.36}$ & 7.52$^{+0.49}_{-0.49}$ & 9.94$^{+0.39}_{-0.39}$ & 6.47$^{+0.27}_{-0.27}$ & 26.47$^{+1.30}_{-1.30}$ & 17.22$^{+0.77}_{-0.77}$ & 3 & 0.99 & 9 & 1.02\\
G284.46+52.43 & 0.441 & 10.63$^{+0.55}_{-0.48}$ & 16.94$^{+0.34}_{-0.31}$ & 9.48$^{+0.14}_{-0.14}$ & 9.83$^{+0.21}_{-0.21}$ & 20.29$^{+0.43}_{-0.43}$ & 12.07$^{+0.38}_{-0.38}$ & 63.47$^{+1.53}_{-1.53}$ & 37.68$^{+1.10}_{-1.10}$ & 7 & 0.98 & 59 & 1.51\\
G284.99-23.70$^{*}$ & 0.390 & 10.10$^{+1.57}_{-1.23}$ & 14.88$^{+0.83}_{-0.70}$ & 7.53$^{+0.53}_{-0.53}$ & 7.61$^{+0.78}_{-0.78}$ & 17.51$^{+0.83}_{-0.83}$ & 9.85$^{+0.22}_{-0.22}$ & 41.13$^{+2.89}_{-2.89}$ & 23.17$^{+1.06}_{-1.06}$ & 3 & 0.36 & 16 & 0.62\\
G285.63-17.24$^{*}$ & 0.350 & 6.59$^{+1.00}_{-1.17}$ & 8.20$^{+0.57}_{-0.78}$ & 5.78$^{+0.63}_{-0.63}$ & 5.74$^{+0.68}_{-0.68}$ & 3.98$^{+0.40}_{-0.40}$ & 3.35$^{+0.30}_{-0.30}$ & 7.80$^{+1.09}_{-1.09}$ & 6.56$^{+0.85}_{-0.85}$ & 1 & 0.86 & 14 & 0.86\\
G286.58-31.25  & 0.210 & 5.52$^{+0.45}_{-0.26}$ & 7.18$^{+0.24}_{-0.15}$ & 5.88$^{+0.15}_{-0.15}$ & 5.87$^{+0.19}_{-0.19}$ & 4.08$^{+0.13}_{-0.13}$ & 3.07$^{+0.08}_{-0.08}$ & 9.28$^{+0.44}_{-0.44}$ & 7.00$^{+0.29}_{-0.29}$ & 7 & 1.01 & 30 & 1.21\\
G286.99+32.91 & 0.390 & 12.20$^{+0.76}_{-0.70}$ & 22.08$^{+0.62}_{-0.57}$ & 10.62$^{+0.69}_{-0.69}$ & 10.47$^{+0.73}_{-0.73}$ & 19.86$^{+0.77}_{-0.77}$ & 15.63$^{+0.53}_{-0.53}$ & 62.75$^{+4.77}_{-4.77}$ & 49.38$^{+3.45}_{-3.45}$ & 5 & 1.03 & 12 & 1.09  \\
G288.61-37.65 & 0.127 & 4.00$^{+0.70}_{-0.48}$ & 7.35$^{+0.67}_{-0.50}$ & 3.09$^{+0.93}_{-0.93}$ & 2.38$^{+1.10}_{-1.10}$ & 5.26$^{+0.32}_{-0.32}$ & 3.60$^{+0.19}_{-0.19}$ & 12.77$^{+1.00}_{-1.00}$ & 8.75$^{+0.61}_{-0.61}$ & 5 & 1.03 & 33 & 1.06\\
G292.51+21.98  & 0.300 & 8.03$^{+0.49}_{-0.44}$ & 11.39$^{+0.29}_{-0.27}$ & 7.53$^{+0.88}_{-0.88}$ & 7.22$^{+0.50}_{-0.50}$ & 6.41$^{+0.36}_{-0.36}$ & 5.65$^{+0.30}_{-0.30}$ & 15.97$^{+1.25}_{-1.25}$ & 14.08$^{+1.06}_{-1.06}$ & 6 & 0.69 & 33 & 1.05 \\
G294.66-37.02  & 0.274 & 7.20$^{+0.59}_{-0.39}$ & 9.88$^{+0.36}_{-0.26}$ & 7.88$^{+0.30}_{-0.30}$ & 7.80$^{+0.39}_{-0.39}$ & 8.05$^{+0.44}_{-0.44}$ & 5.93$^{+0.28}_{-0.28}$ & 22.32$^{+2.01}_{-2.01}$ & 16.46$^{+1.37}_{-1.37}$ & 4 & 0.97 & 14 & 1.09  \\
G304.67-31.66 & 0.193 & 4.16$^{+1.10}_{-0.79}$ & 5.38$^{+0.76}_{-0.61}$ & 5.15$^{+0.69}_{-0.69}$ & 5.22$^{+0.96}_{-0.96}$ & 2.41$^{+0.46}_{-0.46}$ & 2.15$^{+0.43}_{-0.43}$ & 4.54$^{+1.07}_{-1.07}$ & 4.02$^{+1.00}_{-1.00}$ & 2 & 0.67 & 5 & 0.90\\
G304.84-41.42 & 0.410 & 8.28$^{+0.64}_{-0.60}$ & 11.41$^{+0.38}_{-0.35}$ & 9.47$^{+1.31}_{-1.31}$ & 8.73$^{+1.22}_{-1.22}$ & 10.21$^{+0.26}_{-0.26}$ & 7.14$^{+0.35}_{-0.35}$ & 23.76$^{+1.16}_{-1.16}$ & 16.56$^{+0.91}_{-0.91}$ & 2 & 0.95 & 6 & 0.95\\
G306.68+61.06 & 0.085 & 3.93$^{+0.13}_{-0.31}$ & 5.03$^{+0.07}_{-0.17}$ & 5.00$^{+0.10}_{-0.10}$ & 4.78$^{+0.15}_{-0.15}$ & 3.93$^{+0.07}_{-0.07}$ & 2.32$^{+0.01}_{-0.01}$ & 8.36$^{+0.19}_{-0.19}$ & 4.93$^{+0.05}_{-0.05}$ & 22 & 1.02 & 144 & 1.02\\
G306.80+58.60 & 0.085 & 4.60$^{+0.26}_{-0.16}$ & 5.82$^{+0.14}_{-0.09}$ & 5.64$^{+0.11}_{-0.11}$ & 5.58$^{+0.15}_{-0.15}$ & 4.46$^{+0.07}_{-0.07}$ & 2.61$^{+0.04}_{-0.04}$ & 10.62$^{+0.22}_{-0.22}$ & 6.23$^{+0.10}_{-0.10}$ & 17 & 0.98 & 72 & 1.04\\
G308.32-20.23$^{*}$  & 0.480 & 8.61$^{+1.22}_{-0.98}$ & 10.52$^{+0.57}_{-0.48}$ & 8.95$^{+0.84}_{-0.84}$ & 7.40$^{+0.70}_{-0.70}$ & 17.31$^{+1.14}_{-1.14}$ & 10.39$^{+0.43}_{-0.43}$ & 67.99$^{+7.98}_{-7.98}$ & 40.90$^{+3.79}_{-3.79}$ & 3 & 0.66 & 21 & 0.66\\
G313.36+61.11  & 0.183 & 7.87$^{+0.10}_{-0.09}$ & 10.53$^{+0.07}_{-0.06}$ & 8.60$^{+0.08}_{-0.08}$ & 8.26$^{+0.11}_{-0.11}$ & 12.82$^{+0.13}_{-0.13}$ & 5.47$^{+0.14}_{-0.14}$ & 37.61$^{+0.68}_{-0.68}$ & 16.14$^{+0.13}_{-0.13}$ & 9 & 0.98 & 80 & 1.01 \\
G313.87-17.10 & 0.153 & 8.24$^{+0.25}_{-0.23}$ & 10.78$^{+0.13}_{-0.13}$ & 8.43$^{+0.15}_{-0.15}$ & 8.15$^{+0.24}_{-0.24}$ & 11.10$^{+0.33}_{-0.33}$ & 5.57$^{+0.15}_{-0.15}$ & 31.84$^{+1.60}_{-1.60}$ & 15.91$^{+0.77}_{-0.77}$ & 11 & 0.91 & 61 & 0.91 \\
G318.13-29.57 & 0.217 & 5.59$^{+0.64}_{-0.57}$ & 7.43$^{+0.40}_{-0.38}$ & 5.96$^{+0.66}_{-0.66}$ & 5.17$^{+0.93}_{-0.93}$ & 7.39$^{+0.56}_{-0.56}$ & 4.40$^{+0.29}_{-0.29}$ & 22.10$^{+3.07}_{-3.07}$ & 13.15$^{+1.67}_{-1.67}$ & 2 & 0.83 & 4 & 1.11\\
G321.96-47.97  & 0.094 & 3.95$^{+0.29}_{-0.23}$ & 4.83$^{+0.17}_{-0.14}$ & 4.60$^{+0.11}_{-0.11}$ & 4.43$^{+0.14}_{-0.14}$ & 3.10$^{+0.09}_{-0.09}$ & 2.33$^{+0.05}_{-0.05}$ & 6.51$^{+0.23}_{-0.23}$ & 4.91$^{+0.15}_{-0.15}$ & 19 & 0.84 & 52 & 1.12\\
G324.49-44.97  & 0.095 & 3.09$^{+0.40}_{-0.19}$ & 3.47$^{+0.20}_{-0.10}$ & 4.08$^{+0.12}_{-0.12}$ & 4.16$^{+0.18}_{-0.18}$ & 1.85$^{+0.08}_{-0.08}$ & 1.26$^{+0.04}_{-0.04}$ & 3.46$^{+0.20}_{-0.20}$ & 2.36$^{+0.11}_{-0.11}$  & 10 & 0.99 & 37 & 1.08\\
G332.23-46.36 & 0.098 & 5.09$^{+0.15}_{-0.10}$ & 7.06$^{+0.08}_{-0.06}$ & 6.13$^{+0.09}_{-0.09}$ & 5.95$^{+0.12}_{-0.12}$ & 4.71$^{+0.11}_{-0.11}$ & 3.09$^{+0.06}_{-0.06}$ & 11.37$^{+0.34}_{-0.34}$ & 7.48$^{+0.20}_{-0.20}$ & 17 & 1.04 & 102 & 1.08\\
G332.88-19.28 & 0.147 & 6.22$^{+0.38}_{-0.33}$ & 7.74$^{+0.21}_{-0.19}$ & 6.02$^{+0.77}_{-0.77}$ & 5.48$^{+1.09}_{-1.09}$ & 4.55$^{+0.19}_{-0.19}$ & 3.09$^{+0.11}_{-0.11}$ & 11.06$^{+0.82}_{-0.82}$ & 7.51$^{+0.50}_{-0.50}$ & 6 & 0.47 & 13 & 0.58\\
G335.59-46.46  & 0.076 & 3.53$^{+0.74}_{-0.56}$ & 4.61$^{+0.54}_{-0.44}$ & 4.17$^{+1.20}_{-1.20}$ & 3.95$^{+1.36}_{-1.36}$ & 4.19$^{+0.32}_{-0.32}$ & 3.53$^{+0.25}_{-0.25}$ & 10.13$^{+0.94}_{-0.94}$ & 8.55$^{+0.75}_{-0.75}$ & 22 & 0.92 & 23 & 0.92\\
G336.59-55.44 & 0.097 & 3.78$^{+0.71}_{-0.52}$ & 4.88$^{+0.46}_{-0.36}$ & 4.69$^{+0.24}_{-0.24}$ & 4.48$^{+0.26}_{-0.26}$ & 2.53$^{+0.10}_{-0.10}$ & 2.04$^{+0.05}_{-0.05}$ & 5.36$^{+0.26}_{-0.26}$ & 4.32$^{+0.15}_{-0.15}$ & 20 & 1.05 & 23 & 1.19\\
G337.09-25.97 & 0.260 & 5.75$^{+0.50}_{-0.41}$ & 8.94$^{+0.34}_{-0.30}$ & 5.67$^{+0.22}_{-0.22}$ & 5.86$^{+0.36}_{-0.36}$ & 6.57$^{+0.25}_{-0.25}$ & 3.61$^{+0.06}_{-0.06}$ & 14.33$^{+0.80}_{-0.80}$ & 7.87$^{+0.28}_{-0.28}$ & 5 & 0.56 & 21 & 1.16\\
G342.31-34.90 & 0.232 & 6.85$^{+0.74}_{-0.62}$ & 9.49$^{+0.45}_{-0.40}$ & 6.48$^{+1.07}_{-1.07}$ & 6.35$^{+1.33}_{-1.33}$ & 4.64$^{+0.44}_{-0.44}$ & 3.64$^{+0.35}_{-0.35}$ & 13.17$^{+2.16}_{-2.16}$ & 10.34$^{+1.71}_{-1.71}$ & 2 & 0.59 & 13 & 0.92 \\
G347.18-27.35 & 0.237 & 8.24$^{+0.63}_{-0.73}$ & 11.07$^{+0.43}_{-0.50}$ & 8.31$^{+0.40}_{-0.40}$ & 8.16$^{+0.51}_{-0.51}$ & 5.63$^{+0.31}_{-0.31}$ & 4.56$^{+0.19}_{-0.19}$ & 15.15$^{+1.21}_{-1.21}$ & 12.25$^{+0.81}_{-0.81}$ & 8 & 0.96 & 24 & 1.21\\
G349.46-59.94 & 0.347 & 13.59$^{+0.68}_{-0.65}$ & 22.38$^{+0.44}_{-0.42}$ & 10.30$^{+0.22}_{-0.22}$ & 10.60$^{+0.31}_{-0.31}$ & 27.09$^{+0.38}_{-0.38}$ & 13.37$^{+0.03}_{-0.03}$ & 86.74$^{+2.35}_{-2.35}$ & 42.79$^{+0.39}_{-0.39}$ & 7 & 1.00 & 47 & 1.14\\ 
\hline
 \label{tab:properties} 
\end{longtable*}

\section{Comparisons between different fitting methods}\label{sect:fitcomp}
\begin{figure*}[!tb]
\figurenum{B1}
\includegraphics[width=1\textwidth]{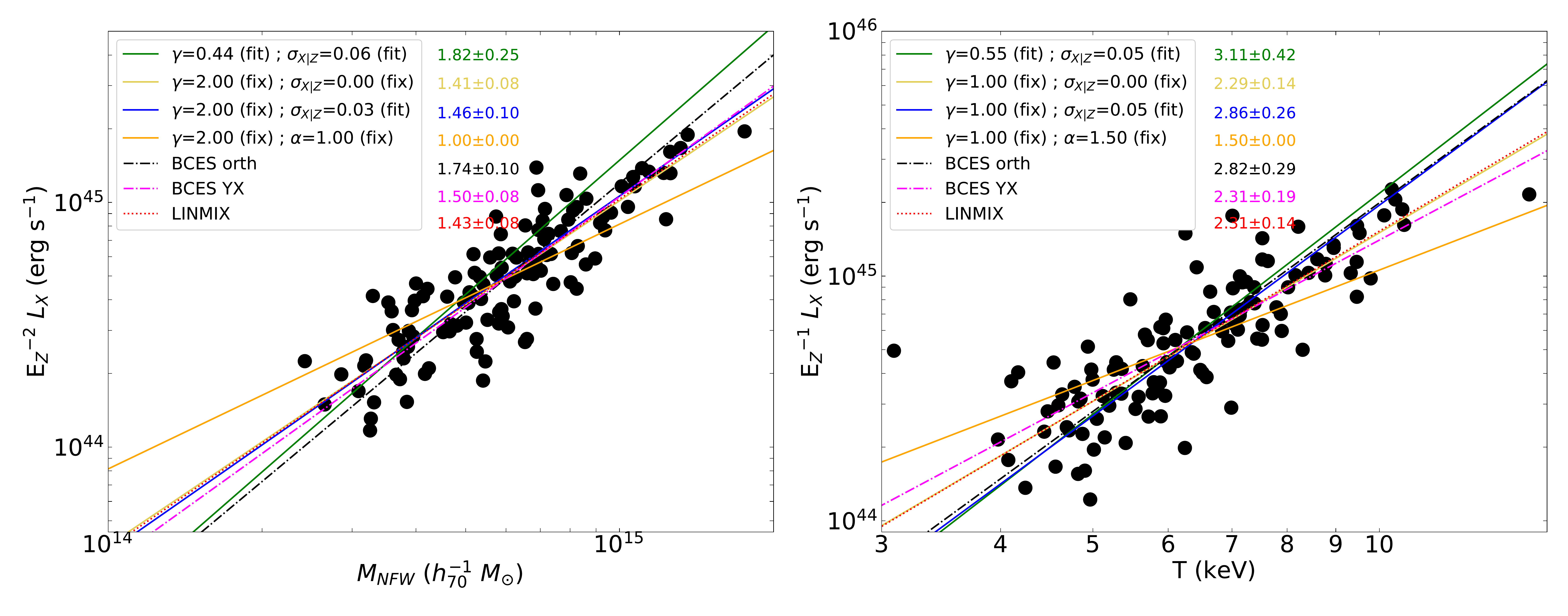}
\caption{{\it left panel}: lines show comparison of the fitted results for the $L_X$-$M_{tot}$ relation obtained using different fitting methods. In green we show the fitted relation obtained using LIRA  allowing all the parameters to vary (our reference model in the paper). In blue (intrinsic scatter in both variable) and yellow (intrinsic scatter only in the Y-axis) we show the best-fit results obtained assuming the self-similar redshift evolution (i.e $\gamma$=2). In orange we show the best-fit obtained assuming that both the slope and the redshift evolution are self-similar. In black and magenta, we show the result from BCES, while in red we show the result using the LINMIX algorithm. In both cases the redshift evolution was chosen to be self-similar.  {\it right panel}: the same as in the top panel but for the $L_X$-$T$ relation.}
\label{fig:LMfit}
\end{figure*}

In this paper we compared our results with the best-fit relations from papers that used different linear regression techniques for their analysis. This complicates the comparison and the interpretation of the different results because they are affected by how one treats the measurement errors, which may be heteroscedastic  and correlated, and the intrinsic scatter. On top of that, the treatment of the selection effects that bias some cluster samples also has a non-negligible effect on the regression results. Many methods have been proposed to account for these effects (e.g. see \citealt{2007ApJ...665.1489K}, \citealt{2016MNRAS.457.1279M}, \citealt{2016MNRAS.455.2149S}, and references therein), each with their advantages and disadvantages.  Here we compare the results from LIRA \citep{2016MNRAS.455.2149S}, the technique used in this paper, with BCES \citep{1996ApJ...470..706A} and MLINMIX \citep{2007ApJ...665.1489K}, two  linear regression  techniques, publicly available, and widely used when fitting scaling relations. We summarize the results in Table \hyperlink{foo2}{B1}, and we plot the $L_X$-$M_{tot}$ and the $L_X$-$T$ relations for illustration in Fig. \ref{fig:LMfit}.  For both relations, the slope obtained with LIRA leaving all the parameters free to vary (in green)  is the steepest and is in fairly good agreement with the result of the orthogonal method with BCES (in black). However in the latter case, the redshift evolution is forced to be self-similar in contrast to the negative redshift evolution determined in the fit with LIRA. Fixing the redshift evolution either to the self-similar value or to zero (i.e. redshift independent relation) impacts the shape of the relations: the larger the $\gamma$ factor, the flatter  the relation (see Table \hyperlink{foo2}{B2}). For all the relations, but the $M_{tot}$-$T$, the fit prefers a $\gamma$ value smaller than the self-similar prediction (although consistent within $\sim$1$\sigma$). Although, the significance for each relation is small, the systematic trend for all relations suggests that it is probably a real effect. Most relaxed clusters, which are thought to be less affected by processes such as gas motions, inhomogeneities, clumps, and shocks, show a redshift evolution more in agreement with the self-similar prediction, which strengthens our argument.  However, if we assume that gravity is the only force driving structure formation, then we could fix the redshift evolution to the predictions, and check if we can recover the predicted slopes. Fixing the redshift evolution tightens the relations, by breaking the degeneracy between $\alpha$, $\beta$, and $\gamma$. In this case, the slope of $L_X$-$M_{tot}$ and the $L_X$-$T$ relations would be slightly flatter but still significantly steeper than  the self-similar predictions. Also the slopes of the  $M_{tot}$-$M_{gas}$ and $M_{tot}$-$Y_X$ relations become flatter, but in this case the deviation from the self-similar prediction is even larger. The slope of the $M_{tot}$-$T$ relation gets steeper than the self-similar prediction. Since the temperature of a cluster is only determined by the depth of its potential well, a deviation from self-similarity would require a mass bias which is temperature dependent.
Considering the scatter in both variables can also play a significant role, depending on the distribution of the data points on the X-axis and on their errors and intrinsic scatter. In Fig.  \ref{fig:LMfit} we show in blue and yellow the best-fit results obtained by fitting or not the intrinsic scatter in the X-axis. In the case of the $L_X$-$T$, the determined slopes are significantly different.  When we set the scatter on X  to zero, we find good agreement  between LIRA and LINMIX.

\section{Reproducibility of the results}\hypertarget{foo3}
The R-package LIRA is a very powerful tool with many parameters that can be frozen or left free to vary depending on the analyses of interest (see \citealt{2016MNRAS.455.2149S} for all the details). In this paper, we consider  two main cases. In the first case, both the variables $X$ and $Y$ are treated as scattered proxy of an underlying quantity $Z$, e.g. the true mass or a rescaled version of the true mass. Here, $\sigma_{X|Z}$ and $\sigma_{Y|Z}$ are the intrinsic scatters of $X$ and $Y$ for a fixed value of $Z$. In the second case, we consider only the scatter in the $Y$ variable, and $\sigma_{Y|Z=X}$ is the intrinsic scatter of $Y$ for a given value of $X$.

To allow the full reproducibility of our results, below, we provide the commands used in the different cases.  Let \texttt{x} and \texttt{y}, \texttt{delta.x} and \texttt{delta.y}, \texttt{covariance.xy}, and \texttt{z} be the vectors storing the values of the observed $\mathbfit{x}$ and $\mathbfit{y}$, their uncertainties $\mathbfit{\delta_x}$ and $\mathbfit{\delta_y}$, the uncertainty covariance $\mathbfit{\delta_{xy}}$, and the redshifts $\mathbfit{z}$, respectively. If not stated otherwise, priors and parameter values are set to default.

\begin{itemize}
\item For regressions without scatter on the $X$ variable, the analysis was performed with the command

\noindent \texttt{> mcmc <- lira (x, y, delta.x = delta.x, delta.y = delta.y, covariance.xy = covariance.xy, z = z, z.ref = 0.2, gamma.mu.Z.Fz=0.0, gamma.sigma.Z.D=$^\prime$dt$'$, n.chains = 4, n.adapt = 2$\times$10$^3$, n.iter = 2$\times$10$^4$)},

\noindent where the covariate distribution is modelled as a Gaussian function with redshift evolving width (\texttt{gamma.sigma.Z.Fz=$'$dt$'$}). Each of the \texttt{n.chains = 4} chain was \texttt{n.iter = 2$\times$10$^4$} long, and the number of iterations for inizialization was set to \texttt{n.adapt = 2$\times$10$^3$}.
 
\item For regressions with scatter on the $X$ variable, the analysis was performed with the command

\noindent \texttt{> mcmc <- lira (x, y, delta.x = delta.x, delta.y = delta.y, covariance.xy = covariance.xy, z = z, z.ref = 0.2, sigma.XIZ.0 = $'$prec.dgamma$'$, gamma.mu.Z.Fz=0.0, gamma.sigma.Z.D=$'$dt$'$, n.chains = 4, n.adapt = 2$\times$10$^3$, n.iter = 2$\times$10$^4$)},

\noindent where the argument (\texttt{sigma.XIZ.0 = $'$prec.dgamma$'$}) makes the scatter in $X$ a parameter to be fitted.

\item For regressions with fixed time evolution, e.g $\gamma=2$, the analysis was performed with the command

\noindent \texttt{> mcmc <- lira (x, y, delta.x = delta.x, delta.y = delta.y, covariance.xy = covariance.xy, z = z, z.ref = 0.2, gamma.YIZ=2.0,  gamma.mu.Z.Fz=0.0, gamma.sigma.Z.D=$'$dt$'$, n.chains = 4, n.adapt = 2$\times$10$^3$, n.iter = 2$\times$10$^4$)},

\noindent where the argument \texttt{gamma.YIZ=2} freezes $\gamma$ to a fixed value.

\item For regressions with fixed time evolution and fixed slope, e.g $\beta=1$ and $\gamma=2$, the analysis was performed with the command

\noindent \texttt{> mcmc <- lira (x, y, delta.x = delta.x, delta.y = delta.y, covariance.xy = covariance.xy, z = z, z.ref = 0.2, beta.YIZ=1.0, gamma.YIZ=2.0, sigma.XIZ.0 = $'$prec.dgamma$'$, gamma.mu.Z.Fz=0.0, gamma.sigma.Z.D=$'$dt$'$, n.chains = 4, n.adapt = 2$\times$10$^3$, n.iter = 2$\times$10$^4$)},

\noindent where the values of \texttt{beta.YIZ} and \texttt{gamma.YIZ} were frozen (\texttt{beta.YIZ=1.0} and \texttt{gamma.YIZ=2}).

\end{itemize}

\renewcommand{\thetable}{B2}
\begin{longtable*}{c @{\hspace{0.3\tabcolsep}}  c @{\hspace{0.3\tabcolsep}} c @{\hspace{0.3\tabcolsep}} c @{\hspace{0.3\tabcolsep}} c @{\hspace{0.3\tabcolsep}} c @{\hspace{0.3\tabcolsep}} c @{\hspace{0.3\tabcolsep}} c   }
\caption{Comparison between the different fitting methods. In the third column we indicate the parameters that were left free to vary with the exception of $\alpha$ and $\sigma_{Y|Z}$ which were always free when appropriate.}\\

\hline\hline
\hypertarget{foo2}Relation (Y-X) & estimator & fitted & $\alpha$  & $\beta$ & $\gamma$ & $\sigma_{X|Z}$ & $\sigma_{Y|Z}$ \\
\hline
\endfirsthead
\multicolumn{8}{c}%
{\tablename\ \thetable{} -- continued from previous page} \\
\hline\hline 
Relation (Y-X) & estimator & fitted & $\alpha$  & $\beta$ & $\gamma$ & $\sigma_{X|Z}$ & $\sigma_{Y|Z}$ \\
\hline
\endhead
$L_X$-$M_{tot}$  	& LIRA & $\beta$, $\gamma$, $\sigma_{X|Z}$ & 0.089$\pm$0.015 & 1.822$\pm$0.246 & 0.462$\pm$0.916  & 0.061$\pm$0.021 & 0.082$\pm$0.040 \\
				& LIRA & $\beta$,  $\sigma_{X|Z}$ & 0.081$\pm$0.013 & 1.455$\pm$0.098 & [2]  & 0.025$\pm$0.015 & 0.123$\pm$0.014 \\
				& LIRA & $\beta$ & 0.080$\pm$0.013 & 1.409$\pm$0.077 & [2]		 & [0]  & 0.129$\pm$0.010 \\
				& LIRA & - & 0.076$\pm$0.014 & [1]		  & [2]		 & 0.015$\pm$0.008 & 0.149$\pm$0.011 \\	
				& LIRA & $\beta$, $\sigma_{X|Z}$ & 0.094$\pm$0.015 & 1.949$\pm$0.127		& [0]		 & 0.069$\pm$0.010 & 0.049$\pm$0.031 \\
      				& BCES YX & $\beta$ & -0.009$\pm$0.013 & 1.500$\pm$0.080 & [2] &  [0]	 &  [0]	\\ 
       				& BCES orth & $\beta$ & -0.008$\pm$0.014 & 1.740$\pm$0.094 & [2] &  [0]	 &  [0]	 \\ 
       				& LINMIX & $\beta$ & -0.001$\pm$0.013 & 1.427$\pm$0.077 & [2] & [0]  & 0.129$\pm$0.051\\ 
\hline
L$_{exc}$-M 		& LIRA & $\beta$, $\gamma$, $\sigma_{X|Z}$ & -0.091$\pm$0.013 & 1.668$\pm$0.183 & 1.325$\pm$0.804  & 0.063$\pm$0.015 & 0.034$\pm$0.029 \\
				& LIRA & $\beta$,  $\sigma_{X|Z}$ & -0.095$\pm$0.011 & 1.525$\pm$0.113 &  [2] & 0.052$\pm$0.016 & 0.063$\pm$0.028 \\
				& LIRA & $\beta$ & -0.097$\pm$0.011 & 1.357$\pm$0.064 & [2]		 &  [0] & 0.104$\pm$0.008 \\
				& LIRA & - & -0.100$\pm$0.012 &  [1]	  &  [2]		 & 0.018$\pm$0.008 & 0.123$\pm$0.009 \\	
				& LIRA & $\beta$, $\sigma_{X|Z}$ & -0.087$\pm$0.013 & 1.872$\pm$0.089	&  [0]	 & 0.067$\pm$0.006 & 0.022$\pm$0.014 \\
      				& BCES YX & $\beta$ & -0.201$\pm$0.011 & 1.390$\pm$0.077 &  [2] & [0]	&  [0]	 \\ 
       				& BCES orth & $\beta$ & -0.200$\pm$0.011 & 1.540$\pm$0.073 &  [2] &  [0]	 &  [0]	 \\ 
           			& LINMIX & $\beta$ & -0.182$\pm$0.011 & 1.375$\pm$0.064 &  [2]	 &  [0]	 & 0.104$\pm$0.042\\    
\hline
L-T  				& LIRA & $\beta$, $\gamma$, $\sigma_{X|Z}$ & -0.250$\pm$0.045 & 3.110$\pm$0.422 & 0.398$\pm$0.939  & 0.051$\pm$0.010 & 0.052$\pm$0.041 \\
				& LIRA & $\beta$,  $\sigma_{X|Z}$ & -0.228$\pm$0.032 & 2.862$\pm$0.261 &  [1]	 & 0.047$\pm$0.009 & 0.080$\pm$0.036 \\
				& LIRA & $\beta$ & -0.168$\pm$0.021 & 2.288$\pm$0.137 & [1]	 & - 				& 0.150$\pm$0.011 \\
				& LIRA & - & -0.084$\pm$0.016 & [1.5]	 & [1]	 & 0.016$\pm$0.007 & 0.174$\pm$0.012 \\	
				& LIRA & $\beta$, $\sigma_{X|Z}$ & -0.264$\pm$0.028 & 3.237$\pm$0.207	& [0]	 & 0.053$\pm$0.005 & 0.066$\pm$0.027 \\
      				& BCES YX & $\beta$ & -0.176$\pm$0.033 & 2.070$\pm$0.209 & [1] & [0]	 & [0]	 \\ 
       				& BCES orth & $\beta$ & -0.253$\pm$0.037 & 2.830$\pm$0.240 & [1] & [0]	& [0]	  \\ 
       				& LINMIX & $\beta$ & -0.242$\pm$0.020 & 2.156$\pm$0.132 & [1] &[0]	 & 0.145$\pm$0.056\\ 
\hline
L$_{exc}$-T$_{exc}$  & LIRA & $\beta$, $\gamma$, $\sigma_{X|Z}$ & -0.360$\pm$0.031 & 2.409$\pm$0.292 & 1.170$\pm$0.822  & 0.038$\pm$0.011 & 0.052$\pm$0.031 \\
				 & LIRA & $\beta$,  $\sigma_{X|Z}$ & -0.390$\pm$0.020 & 2.732$\pm$0.143 & [1]	 & 0.043$\pm$0.004 & 0.030$\pm$0.018 \\
				 & LIRA & $\beta$ & -0.347$\pm$0.016 & 2.292$\pm$0.106 & [1]		 & [0]				& 0.111$\pm$0.009 \\
				& LIRA & - & -0.259$\pm$0.014 &[1.5]	  & [1]	 & 0.015$\pm$0.006 & 0.145$\pm$0.011 \\	
				& LIRA & $\beta$, $\sigma_{X|Z}$ & -0.410$\pm$0.020 & 2.949$\pm$0.147	& [0]		 & 0.045$\pm$0.004 & 0.023$\pm$0.014 \\
      				& BCES YX & $\beta$ & -0.331$\pm$0.038 & 1.920$\pm$0.273 & [1] & [0]	 & [0]	\\ 
       				& BCES orth & $\beta$ & -0.387$\pm$0.034 & 2.560$\pm$0.226 & [1] & [0]	 & [0]	 \\ 
      	 			& LINMIX & $\beta$ & -0.416$\pm$0.015 & 2.135$\pm$0.101 & [1] & [0]	 & 0.106$\pm$0.043\\ 
\hline
M-T  				& LIRA & $\beta$, $\gamma$, $\sigma_{X|Z}$ & -0.171$\pm$0.015 & 1.556$\pm$0.137 & 0.179$\pm$0.379  & 0.032$\pm$0.010 & 0.036$\pm$0.016 \\
				& LIRA & $\beta$,  $\sigma_{X|Z}$ & -0.199$\pm$0.012 & 1.843$\pm$0.084 & [-1]	 & 0.040$\pm$0.004 & 0.016$\pm$0.008 \\
				& LIRA & $\beta$ & -0.173$\pm$0.010 & 1.591$\pm$0.067 &  [-1]	 & [0]			& 0.068$\pm$0.006 \\
				& LIRA & - & -0.162$\pm$0.007 & [1.5] &  [-1] & 0.031$\pm$0.009 & 0.050$\pm$0.012 \\	
				& LIRA & $\beta$, $\sigma_{X|Z}$ & -0.176$\pm$0.011 & 1.606$\pm$0.085	&[0]		 & 0.035$\pm$0.007 & 0.028$\pm$0.014 \\
      				& BCES YX & $\beta$ & -0.110$\pm$0.018 & 1.500$\pm$0.116 &  [-1] & [0] & [0]\\ 
       				& BCES orth & $\beta$ & -0.119$\pm$0.018 & 1.610$\pm$0.116 &  [-1] & [0] & [0] \\ 
       				& LINMIX & $\beta$ & -0.130$\pm$0.011 & 1.607$\pm$0.067 &  [-1] &[0] & 0.064$\pm$0.028\\        
\hline
M-M$_{gas}$ 		& LIRA & $\beta$, $\gamma$, $\sigma_{X|Z}$ & 0.073$\pm$0.007 & 0.802$\pm$0.049 & -0.317$\pm$0.307  & 0.028$\pm$0.015 & 0.043$\pm$0.011 \\ 
				& LIRA & $\beta$,  $\sigma_{X|Z}$ & 0.057$\pm$0.006 & 0.669$\pm$0.027 & [0] & 0.014$\pm$0.007 & 0.052$\pm$0.004 \\
				& LIRA & $\beta$ & 0.057$\pm$0.007 & 0.620$\pm$0.030 & [0]& [0]			& 0.160$\pm$0.011 \\
				& LIRA & $\sigma_{X|Z}$ & 0.097$\pm$0.010 & [1]) & [0]	& 0.104$\pm$0.008 & 0.123$\pm$0.009 \\	
	     			& BCES YX & $\beta$ & 0.081$\pm$0.005 & 0.790$\pm$0.025 & [0] & [0] & [0]\\ 
       				& BCES orth & $\beta$ & 0.081$\pm$0.005 & 0.079$\pm$0.024 & [0] & [0] & [0] \\ 
	      			& LINMIX & $\beta$ & 0.080$\pm$0.005 & 0.778$\pm$0.023 & [0] & [0] & 0.047$\pm$0.005\\ 
\hline
M-Y$_{X}$ 		& LIRA & $\beta$, $\gamma$, $\sigma_{X|Z}$ & -0.010$\pm$0.005 & 0.540$\pm$0.030 & -0.292$\pm$0.287  & 0.039$\pm$0.023 & 0.039$\pm$0.011 \\ 
				& LIRA & $\beta$,  $\sigma_{X|Z}$ & -0.010$\pm$0.005 & 0.549$\pm$0.019 & [-0.4] & 0.047$\pm$0.022  & 0.037$\pm$0.010 \\
				& LIRA & $\beta$ & -0.005$\pm$0.006 & 0.516$\pm$0.018 & [-0.4] & - 	& 0.235$\pm$0.016 \\
				& LIRA & $\sigma_{X|Z}$ & -0.003$\pm$0.006 & 0.6 (fix) & [-0.4] & 0.097$\pm$0.010 & 0.207$\pm$0.015 \\	
				& LIRA & $\beta$, $\sigma_{X|Z}$ & -0.011$\pm$0.005 & 0.517$\pm$0.082	& [0]	 & 0.024$\pm$0.017 & 0.042$\pm$0.006 \\
				& BCES YX & $\beta$ & 0.015$\pm$0.005 & 0.544$\pm$0.016 & [-0.4] & [0]	 & [0]	\\ 
       				& BCES orth & $\beta$ & 0.015$\pm$0.005 & 0.540$\pm$0.016 & [-0.4] & [0]	 & [0]	\\ 
	 			& LINMIX & $\beta$ & 0.006$\pm$0.005 & 0.538$\pm$0.016 & [-0.4] & - & 0.043$\pm$0.019\\ 
\hline
 \label{tab:properties} 
\end{longtable*}

\section{Test the underlying assumptions of our regression method}
The \textsc{LIRA} software and its underlying assumptions and methods have been extensively tested with data and simulations (see, e.g., \citealt{2016MNRAS.455.2149S} or \citealt{ser+19_hscxxl}). Accurate sampling of the parameter posterior probability distribution is crucial. Here we compare the LIRA sampling which relies on Gibbs sampling exploiting the \textsc{JAGS} (Just Another Gibbs sampler) library\footnote{JAGS by M. Plummer is publicly available at \url{http://mcmc-jags.sourceforge.net}.}, with an alternative method where the original data-set is perturbed proportional to the observed measurement errors, the fitting procedure is repeated for each random data extraction, and the parameter posterior is built as the distribution of the central momenta.

\begin{figure*}[!tb]
\figurenum{D1}
\includegraphics[width=1\textwidth]{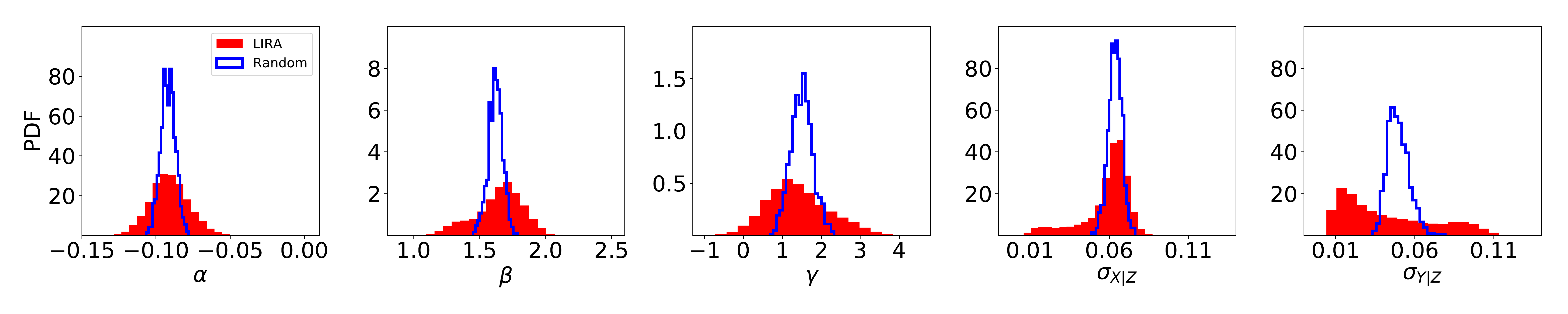}
\caption{Comparison between the distributions of the parameters obtained using the Gibbs sampling used in LIRA (in blue) and the distribution obtained by perturbing the original data-sets proportional to the observed measurement errors (in red).}
\label{fig:gibbs}
\end{figure*}

Let us consider the mass versus core excised soft luminosity for the full sample in the more general case, e.g. time-evolution and scatter in the $X$ variable. We extracted a collection of $10^3$ simulated data-sets, where the random pairs $\{ M_{500}-L_X\}$ pair are extracted from bi-variate Gaussians centered on the actual pair and with the same measured uncertainty covariance matrix. For each regression, we collect the posterior mean. As can be seen from Fig. \ref{fig:gibbs}, there is good agreement between the two methods. The peaks of the posterior distributions are  located well within the statistical uncertainty. The Gibbs sampling proves to be better suited to fully explore the full parameter space, with posterior generally broader.

\bibliography{plckscaling}{}
\bibliographystyle{aasjournal}



\end{document}